\documentclass[useAMS,usenatbib]{mn2e}
\usepackage{epsfig}
\usepackage{placeins}
\def\gtsima{$\; \buildrel > \over \sim \;$}
\def\ltsima{$\; \buildrel < \over \sim \;$}
\def\gsim{\lower.5ex\hbox{\gtsima}}
\def\lsim{\lower.5ex\hbox{\ltsima}}
\def\kms{km\,s$^{-1}$}
\def\zabs{z_{\rm abs}}
\def\zem{z_{\rm em}}

\def\Lya{Ly$\alpha$~}

\newcommand{\CIV}{\mbox{C\,{\sc iv}}}

\newcommand{\CII}{\mbox{C\,{\sc ii}}}

\newcommand{\OI}{\mbox{O\,{\sc i}}}
\newcommand{\MgI}{\mbox{Mg\,{\sc i}}}
\newcommand{\MgII}{\mbox{Mg\,{\sc ii}}}
\newcommand{\AlII}{\mbox{Al\,{\sc ii}}}
\newcommand{\AlIII}{\mbox{Al\,{\sc iii}}}
\newcommand{\SiIV}{\mbox{Si\,{\sc iv}}}

\newcommand{\SiII}{\mbox{Si\,{\sc ii}}}

\newcommand{\FeII}{\mbox{Fe\,{\sc ii}}}
\newcommand{\ZnII}{\mbox{Zn\,{\sc ii}}}
\newcommand{\HI}{\mbox{H\,{\sc i}}}
\newcommand{\HeII}{\mbox{He\,{\sc ii}}}

\title[Metals in the IGM with X-shooter at the VLT]{Metals in the IGM approaching the re-ionization epoch:
  results from X-shooter at the VLT\thanks{Based on observations collected at 
  the  European  Southern  Observatory  Very  Large  Telescope,  Cerro
  Paranal,  Chile -- Programs 069.A-0529, 079.A-0226, 084.A-0390,
  084.A-0550, 085.A-0299, 086.A-0162, 087.A-0607 and 268.A-5767}}
\author[V. D'Odorico, et al.]{V. D'Odorico$^{1}$\thanks{E-mail:
dodorico@oats.inaf.it}, G. Cupani$^{1}$, S. Cristiani$^{1,2}$,  R. Maiolino$^{3,4,5}$, P. Molaro$^{1}$, M. Nonino$^{1}$, \and
M. Centuri\'on${^1}$, A. Cimatti$^6$, S. di Serego Alighieri$^7$, F. Fiore$^3$,
A. Fontana$^3$, \and 
S. Gallerani$^8$, E. Giallongo$^3$, F. Mannucci$^7$, A. Marconi$^9$,
L. Pentericci$^3$, M. Viel$^{1,2}$, \and G. Vladilo$^{1}$ \\
$^1$ INAF-OATS, Via Tiepolo 11, 34143 Trieste, Italy \\
$^2$ INFN/National Institute of Nuclear Physics, via Valerio 2, 34127 Trieste, Italy \\
$^3$ INAF-OAR, via di Frascati 33, 00040 Monte Porzio Catone, Italy\\
$^4$ Cavendish Laboratory, University of Cambridge, 19 JJ Thomson Avenue, Cambridge
CB3 0HE, UK \\
$^5$ Kavli Institute for Cosmology, Madingley Road, Cambridge CB3 0HA,
UK \\
$^6$ Dipartimento di Astronomia, Universit\`a di Bologna Via Ranzani
1, I-40127 Bologna, Italy \\
$^7$INAF-OAArcetri, Largo Enrico Fermi 5, 50125 Firenze. Italy \\
$^8$Scuola Normale Superiore di Pisa, Piazza dei Cavalieri 7, 56126,
Pisa, Italy \\
$^9$Dipartimento di Fisica e Astronomia, Universit\`a di Firenze, via
G. Sansone 1, Sesto Fiorentino (Firenze), Italy} 

\begin{document}

\date{}

\pagerange{\pageref{firstpage}--\pageref{lastpage}} \pubyear{}

\maketitle

\label{firstpage}

\begin{abstract}
We present the results of observations taken with the X-shooter  spectrograph
devoted to the study of quasars at $z\sim6$. This paper focuses on the
properties of metals  at high redshift traced, in particular, by the
\CIV\ doublet absorption systems. Six objects were observed with
resolutions $\simeq27$ and 34 \kms\ in the visual, and 37.5 and 53.5
\kms\ in the near-infrared.  
We detected 102 \CIV\ lines in the range: $4.35 < z < 6.2$ of which 27
are above $z\sim5$. Thanks to the characteristics of resolution and
spectral coverage of X-shooter, we could also detect 25
\SiIV\ doublets associated with the \CIV\ at $z\gsim5$.  The column
density distribution function of the \CIV\ line sample is
observed to evolve in redshift for $z\gsim 5.3$,  with respect to the
normalization defined by low redshift ($1.5 < z <4$) \CIV\ lines. This
behaviour is reflected in the redshift evolution of the \CIV\ cosmic
mass density, $\Omega_{\rm CIV}$,  of lines with column density in the
range $13.4 < \log N($\CIV$) < 15$, which is consistent with a drop of
a factor of $\sim2$ for $z\gsim 5.3$.  
%At $z<5.3$, the mass density stays flat to $z \simeq 3$ and it rises
%  again by a factor $\sim4$ to $z\simeq 1.5$.  
Considering only the stronger \CIV\ lines ($13.8 < \log N($\CIV$) <
15$), $\Omega_{\rm CIV}$ gently rises by a factor of $\sim 10$ between
$z\simeq 6.2$ and $z\simeq 1.5$ with a possible flattening towards
$z\sim 0$.  The increase is well fitted by a power law: $\Omega_{\rm
  CIV} = (2\pm1)\times10^{-8} [(1+z)/4]^{-3.1\pm0.1}$. An insight into
the properties of the \CIV\ absorbers and their evolution with
redshift is obtained by comparing the observed column densities of
associated \CIV, \SiIV\ and \CII\ absorptions with the output of a set
of {\small CLOUDY} photo-ionization models.  As already claimed by
cosmological simulations, we find that \CIV\ is a good tracer of the
metallicity in the low density intergalactic medium gas at $z\sim5-6$
while at $z\sim3$ 
it arises in gas with over-density $\delta\sim100$.    
\end{abstract}


\section{Introduction}

The properties of stars, galaxies and quasars in the local and early
Universe can be investigated through their impact on the intergalactic
medium (IGM). In particular, the radiation emitted and the metals
ejected from these objects re-ionized and polluted the IGM. 
As a consequence, the detailed understanding of these mechanisms has
the potential to significantly constrain models for the formation and
evolution of galaxies and quasars, and the re-ionization history of the
Universe.
The IGM is mainly studied through the absorption signature it leaves
in the spectra of   
bright high-redshift sources, quasars and gamma-ray bursts. The
highest redshift quasars have been detected mainly by the 
Sloan Digital Sky Survey (SDSS) at $z\sim6$ \citep[e.g.][]{fan2001}, corresponding
to $\sim1$ billion years after the big bang. 
This sample of $\sim 20$ objects has been used to investigate several
topics, in particular the ionization and chemical status of the IGM at
these high redshifts \citep[see][for a review]{fan_review}.

The same high-$z$ quasars can be used to put an indirect constraint on
the epoch of re-ionization by investigating the redshift evolution of
metal abundances traced by ionic 
absorption lines. 
%The two main questions that need to be answered are:
%when most of the enrichment took place and how metals were able to
%travel to large distances from their production sites. 
%Two possible
%scenarios have emerged. The first postulates that metals are produced
%by star formation within the galaxies observed at $z\sim3$ and ejected in
%low-density regions via super-winds from supernova explosions. This
%scenario is supported by the strong correlation between the
%\CIV\ absorbers and the Lyman Break Galaxies (LBGs) at those 
%redshifts (e.g. Adelberger et al. 2005; Steidel et al. 2010). An
%alternative scenario predicts that metals are produced at much higher
%redshift, in an earlier stage of metal production that has uniformly
%enhanced the IGM metallicity. The latter process, often referred to as
%preenrichment, is generally attributed to primeval galaxies and/or
%PopIII stars at $z>6$, when shallow potential wells allow winds to
%distribute metals over large comoving volumes, thus producing a quite
%uniform metallicity distribution (e.g. Madau et al. 2001). 
The investigation of the regime beyond $z\sim5$ is essential since in this
redshift range the comoving star formation rate density appears to
decline with redshift \citep[e.g.][]{mannucci07,gonzalez10}. If a
similar behaviour is observed for the mass density of metals in the
IGM, then a scenario where winds from in situ star-forming galaxies
pollute the IGM with metals would be favoured.   On the other hand, if
the mass density of metals is observed to remain constant, this would 
point to an epoch of very early enrichment of the IGM, presumably by
massive stars (e.g. PopIII stars) in mini-haloes \citep[M~$\sim 10^6$
M$_{\odot}$; e.g.][]{choudhury}, when shallow potential wells 
allow winds to distribute metals over large comoving volumes, thus
producing a quite uniform metallicity distribution
\citep[e.g.][]{madau01}. 
These same massive stars could have been the main sources of ionizing
photons allowing for an earlier reionization epoch. 
%%% Simulazioni
%Due to the fact that the abundance 
%of \CIV\ relative to C is highly sensitive to the physical conditions
%of the IGM, sophisticated cosmological simulations are needed in order
%to explore this relation and to link the \CIV\ and 
%the global metallicity evolution (e.g. Oppenheimer & Davé 2008;
%%Tescari et al. 2010). 

Recent estimates of the evolution with redshift of the
\CIV\ cosmic mass density \citep[see definition in
  Section 5.2,][]{dodorico10,cooksey12} show a decrease of the
\CIV\ content from redshift $\sim0$ to 3, and then a flat behaviour in the
range $z\simeq 3-5$ and a possible downturn at $z>5$.  
Searching for intergalactic \CIV\ absorption lines at $z>5$ becomes
challenging because the \CIV\ doublet moves into the near-infrared spectral
region, absorption line spectroscopy is much more difficult in this
regime because of increased detector noise, OH emission from the sky
and more severe telluric absorption. 

The first results at $z\gsim5.5$ were based on low/intermediate resolution QSO
spectra and a few detections of \CIV\ absorptions indicating a
possible decrease of the value of $\Omega_{\rm CIV}$
\citep{ryanweber09} and a significant decline of the \CIV\ column density
distribution function \citep[CDDF;][]{becker09}. 
In a recent work,  \citet[][hereafter Simcoe11]{simcoe11}
claim the presence of a downturn in the \CIV\ abundance at $\langle
z\rangle = 5.66$ by a factor of 4.1 relative to its value at $\langle
z\rangle = 4.96$. The result is based on the spectra (at resolution $\delta v\simeq 50$ \kms) of seven quasars obtained with the new spectrograph FIRE \citep{simcoe10} at the
Magellan telescope, coupled with six observations of northern objects
taken from the literature, of which four \citep[see][]{ryanweber09} have
three times lower resolution and lower signal-to-noise ratio than the FIRE
sample. The northern sample shows only one \CIV\ absorption system, while
almost doubling the surveyed redshift path, the effect is a decrease
of the value of $\Omega_{\rm CIV}$, which however could be due to the
lower quality of those spectra.   

In this work, we present the spectra obtained in a guaranteed time
observation (GTO) programme
(PI: V. D'Odorico) carried out with the X-shooter spectrograph at
the European Southern Observatory (ESO) Very Large Telescope (VLT)
between 2010 January and 2011 June. The scientific aim of this project was to observe the brightest
(J$_{\rm Vega}\lsim 19$) quasars known with $z_{\rm em}\gsim5.7$ and
observable from Paranal to derive constraints both on the abundance of
neutral hydrogen and on the abundance of metals at high redshift. Here the results on the
metal lines are reported, the \HI\ abundance will be the topic of a
forthcoming paper.      
The X-shooter spectrograph \citep{xshooter}, with its high sensitivity, extended
spectral coverage (from 3000 \AA\ to 2.5 $\mu$m) and intermediate
resolution, appeared to be the ideal instrument to allow significant
steps forward in this research field.   
Seven objects satisfy the chosen selection criteria. However, only
four of them were observed due to bad weather downtime. 
%We added to the
%sample also the spectrum of J1306+0356 obtained during the 3rd
%commissioning run in March 2009.  
Luckily, two of the objects we could not observe are present in the
ESO X-shooter archive and we have retrieved and analysed their
spectra. 
%The sample has been increased with the addition of two more objects:
%J1306+0356 and J1319+0950, whose spectra were retrieved from the ESO
%archive.  

Preliminary results on a sample of three objects (J1306+0356, J0818+1722,
J1509-1749) were reported in \citet[][hereafter Paper I]{dodorico11}. Here, we describe the
whole sample and carry out a more detailed analysis. 

\begin{table*}
\begin{center}
\caption{X-shooter observations. Spectra of J0818+1722, J1509-1749 and
  J1030+0524 were obtained using the combination of slits 1.0/0.9/0.9
  arcsec in the UVB/VIS/NIR arms, while for the frames of J0836+0054,
  J1306+0356 and J1319+0950 the combination 0.8/0.7/0.6 arcsec was
  adopted. $\Delta$~X is the redshift absorption path defined in
  Section~5.1 and the SNR is computed per resolution element.}
\begin{minipage}{150mm}
\label{tab_obs}
\begin{tabular}{@{} l l l r r c c c c c}
\hline  
QSO & $z_{\rm em}$ & $J_{\rm mag}$ & $T_{\rm VIS}$ & $T_{\rm NIR}$ &
SNR @
& SNR @& $z_{\rm CIV}$ & $\Delta\,X_{\rm CIV}$ &  $z_{\rm SiIV}$ \\
& & & (s) & (s) & (9000\,\AA)& (10350\,\AA) & & & \\
\hline
 J0818+1722$^1$ & 6.002 & 18.54 & $22910$ & $23390$ & 85 & 24 & $4.516-5.886$ & 6.646 & $5.128-5.886$\\
 J0836+0054$^2$ & 5.810 & 17.9 & 8160 & 8400 & 65 & 10 & $4.365-5.697$ & 6.371 & $4.960-5.697$\\
 J1030+0524$^2$ & 6.308 & 18.87 & 26860 & 26000 & 125 & 7 & $4.757-6.187$ & 7.091 & $5.395-6.187$\\
 J1306+0356$^2$ & 6.016 & 18.77 & $46800$ & $46800$ & 55 & 12 & $4.527-5.900$ & 6.666 & $5.140-5.900$\\
 J1319+0950$^3$ & 6.13 & 18.8 & $36000$ & $36000$ & 95 & 15 & $4.617-6.012$ & 6.831 & $5.240-6.012$\\
 J1509-1749$^4$ & 6.118 & 18.78 & $23360$ & $24000$ & 50 & 8 & $4.608-6.000$ & 6.814 & $5.229-6.000$ \\
\hline
\end{tabular}
(1) \citet{fan2006}; (2) \citet{fan2001}; (3) \citet{mortlock09}; (4) \citet{willott07}
\end{minipage}
\end{center}
\end{table*}

The rest of the paper is organized as follows. Data reduction and
analysis are presented in \S~2. In \S~3 the spectra of each object in
the sample are described in details. \S~4 is dedicated to the
\CIV\ lines statistics: CDDF and
cosmic mass density.  The results are discussed in \S~6.   
Throughout this paper, we assume $\Omega_{\rm m} = 0.26$,
$\Omega_{\Lambda} = 0.74$ and $h \equiv H_0/(100 {\rm km\ s}^{-1} {\rm
  Mpc}^{-1}) =0.72$.           
     
\section{Observations and data reduction}

%The QSO spectra presented in this work were obtained with the new ESO
%spectrograph X-shooter \citep{xshooter} in operation at the VLT since
%early 2009. 
%In particular, the spectrum of J1306+0356 was obtained during the 3rd
%commissioning run in March 2009, while the spectra of the other
%objects were obtained during the Guaranteed Time of Observation (GTO) 
%assigned to Italy  (proposal P.I. V. D'Odorico) between January and June
%2011.
%The details of the observations are reported in Table~\ref{tab_obs}.

The spectra were acquired with a binning of two pixels in the
dispersion direction and adopting two sets of slit apertures for the
blue, visual (VIS) and near-infra red (NIR) arms respectively: 1.0/0.9/0.9
and 0.8/0.7/0.6 arcsec depending on the seeing conditions at the moment of the
observations.  
The journal of observations is reported in Table~\ref{tab_obs}. 

All the raw frames were reduced with the public release of the
X-shooter pipeline \citep{xshpipe}. The pipeline
reduction proceeds along the following steps. Pixels in the
2D frames are first mapped to wavelength space using calibration
frames. Sky emission lines are subtracted before any re-sampling
using the method developed by \citet{kelson}. The different orders of
the echelle spectrum are then extracted, rectified, wavelength
calibrated and merged, with a weighted average used in the overlapping
regions. The final product is a 1D, background-subtracted spectrum and
the corresponding error file. The spectra are rebinned to 0.3 (narrow
slit) or 0.4 \AA\ (broad slit) in the VIS and
to 0.6 \AA\ in the NIR arm, following the prescription of the pipeline
manual.   

We followed the standard procedure with the exception of the
extraction of the 1D spectrum from the 2D merged spectrum which
was carried out with the EXTRACT/LONG command in {\small
  MIDAS} reduction package (using a predefined aperture). The
extraction within {\small MIDAS} results in a better signal-to-noise
ratio at least for those objects which are faint and whose spectrum is
strongly absorbed in the \Lya\ forest region.   
%% Flux calibration
The instrument response curve was obtained reducing with the specific
pipeline recipe the standard flux stars observed the same night of the
scientific observations. 
Each extracted frame was then flux calibrated by dividing for the
response curve.  
Finally, the set of 1D spectra obtained for each object was added with a
weighted sum to obtain the final spectrum. 

%% Continuum det
The continuum level in the region red-wards of the \Lya\ emission was
determined interpolating with a spline polynomial of third degree the
portions of the spectrum free from absorption lines. The same approach
cannot be applied to the heavily absorbed \Lya\ forest. The continuum
in this spectral range was obtained by extrapolating the power law
which fits the red region cleaned from the intrinsic emission lines.   

%% Telluric correction
Finally, the VIS and NIR spectra were corrected for telluric
absorption dividing by the normalized spectrum of  standard
spectroscopic stars observed with the same instrumental set-up as the
quasars in our sample, using the command {\it telluric} in {\small
  IRAF}.   

\subsection{Identification of metal absorptions}

Metal absorptions red-wards of the \Lya\ emission have been identified
mainly by eye. First, we looked for the most common ion doublets and
multiplets (e.g. \CIV, \MgII, \FeII\ and \SiIV)  and fitted them to confirm
their nature and determine a precise redshift. Then, the lines without
an identification were processed with automatic routines which try to 
associate those lines with the identified systems or with known ions on
the basis of their wavelength ratios.   

In the VIS region of the spectrum, the section inspected for metals
extends from the observed wavelength of the \Lya\ emission to
$\lambda\ =10200$ \AA, while in the NIR 
spectra it goes from this wavelength to $\sim 2$ $\mu$m, with the known
gaps due to atmospheric absorption.  At $\lambda\ >2$ $\mu$m telluric
lines are severely affecting the spectrum preventing the detection of
other absorption lines.  
The redshift range of the \CIV\ and \SiIV\ forests reported in Table~\ref{tab_obs} goes from the \Lya\ emission $+ 1000$ \kms, to avoid blending with \Lya lines, to the \CIV\ and \SiIV\ emission respectively $- 5000$ \kms, to avoid proximity effects. 

%% Measure of equivalent width and fit of lines
The analysis of high resolution spectra (full width at half-maximum $\simeq 6-7$ \kms) shows
that metal lines are characterized by Doppler velocity widths of the
order of $5-10$ \kms\ \citep[e.g.][]{tescari}, which are not resolved
in the X-shooter spectra. In spite of this, we have fitted the detected
metal lines (in particular, \CIV\ and \SiIV) with Voigt profiles using
the context LYMAN of the {\small MIDAS} reduction package
\citep{font:ball}. None of the fitted doublets shows saturation
(the strength ratio in the doublet would not be preserved in that
case), so we are confident that we are not underestimating their true column
densities. 

For comparison with previous works, we have also computed the rest frame
equivalent widths, $W_0$, of \CIV\ and \SiIV\ doublets by direct summation of the
continuum-normalized flux of the spectral pixels. The errors on $W_0$
were computed as the square root of the quadratic sum of the flux
standard deviations. The sums have been
performed over the velocity range extending between the two pixels at
which the normalized profile returned to unity.
In the case of blending with other transitions, we report either a lower
or an upper limit to the equivalent width depending on whether only part
of the line profile is blended with a nearby absorption or the
considered line falls within other strong absorptions (e.g. telluric
line). In the case of complex velocity profiles if components are
blended, we measured the total equivalent width.
Tables~\ref{tab_CIV_0818}-\ref{tab_CIV_1509} report
the equivalent width and 
the measured column density of all the \CIV\ and \SiIV\ lines
detected in our spectra.  

The level of completeness of our sample was computed by generating
a set of mock \CIV\ absorption doublets with column densities in the
range $13.3 - 13.8$ and random redshifts in the interval covered by the
observed spectra.Doppler parameters were derived randomly from a list
of measured Doppler parameters for the X-shooter observed
\CIV\ sample.  The correct velocity profiles for the lines were
obtained with the procedure COMPUTE/LYMAN in {\small MIDAS},
and the lines were then superposed to the observed spectra and
detected by eye. 
Our X-shooter spectra result to be complete to almost 100
per cent for \CIV\ doublets with $\log $N(CIV)$ \ge 13.6$ for the
whole inspected redshift range. Considering the redshift range $z<
5.3$, the sample is complete to above 95 per cent for $\log $N(CIV)$
= 13.5$ and 13.4, and at 85 per cent for  $\log $N(CIV)$ = 13.3$. 
Above $z=5.3$, the completeness decreases significantly with
\CIV\ lines with column density in log equal to 13.5 detected 85
per cent of the time, 13.4 detected 70 per cent of the time and 13.3 in 60
per cent of the cases.
 
\section{Notes on individual objects}
In this section, all the objects in the sample are briefly
introduced.  We report also on the objects
already described in Paper I, since some variations  occurred with the 
improvement of the reduction pipeline and the optimisation of our
procedure.  
 
In the following, all the reported signal to noise ratios (SNR) are
per resolution element. The resolution in the VIS arm is $R
\simeq8800$  and 11000 for the 0.9 and 0.7 arcsec slits, respectively;
while in the NIR arm it is  $R \simeq 5600$ and 8000 for the 0.9 and
0.6 arcsec slits. 
The redshift of absorption systems is reported with the number of
significant digits (in general four or five), in the case of systems with a
complex velocity structure the average redshift is reported with three 
decimal digits.   
%SNR per resolution element
%               4.5-5    5-6
%J1030   80-30    9-26
%J1509   65-30    17-35
%J1306   50-25
%J0818   130-70   26-70
%J0836   80-50    7-37

%--------------------------------------------------
\subsection{SDSS J0818+1722}
%--------------------------------------------------
This object, which has the best SNR in our sample, was already
discussed in Paper I. 

The VIS portion of the spectrum outside the \Lya\ forest is the
richest in absorption lines; having a large SNR ($=120-60$) we could subtract
quite well the telluric features revealing the presence of many
absorption systems. In the NIR portion of the spectrum, the SNR lower
to $\sim 10-15$ in the $z$ band, to increase again to $\sim 50-100$ in the $J$ and $H$
bands. 

\begin{figure*}
\begin{center}
\includegraphics[width=18cm]{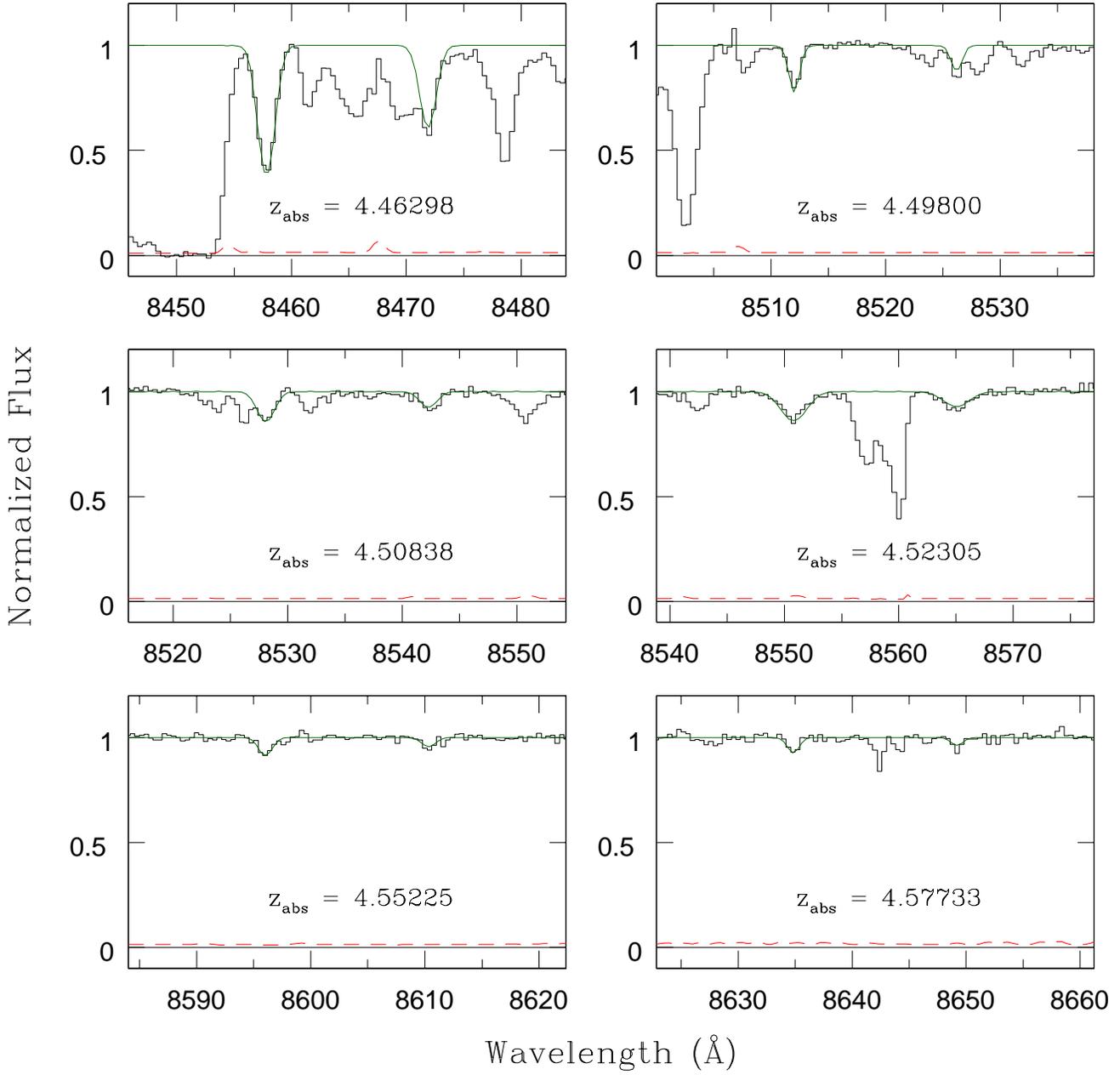}
\caption{Detected \CIV\  absorption systems in the spectrum of SDSS
  J0818+1722. The solid histogram shows the observed spectrum, the 
  continuous green line shows the result of the profile fitting while the
dashed red line denotes the standard deviation of the flux.}   
\label{J0818_sysz4_1}
\end{center}
\end{figure*}

\begin{figure*}
\begin{center}
\includegraphics[width=18cm]{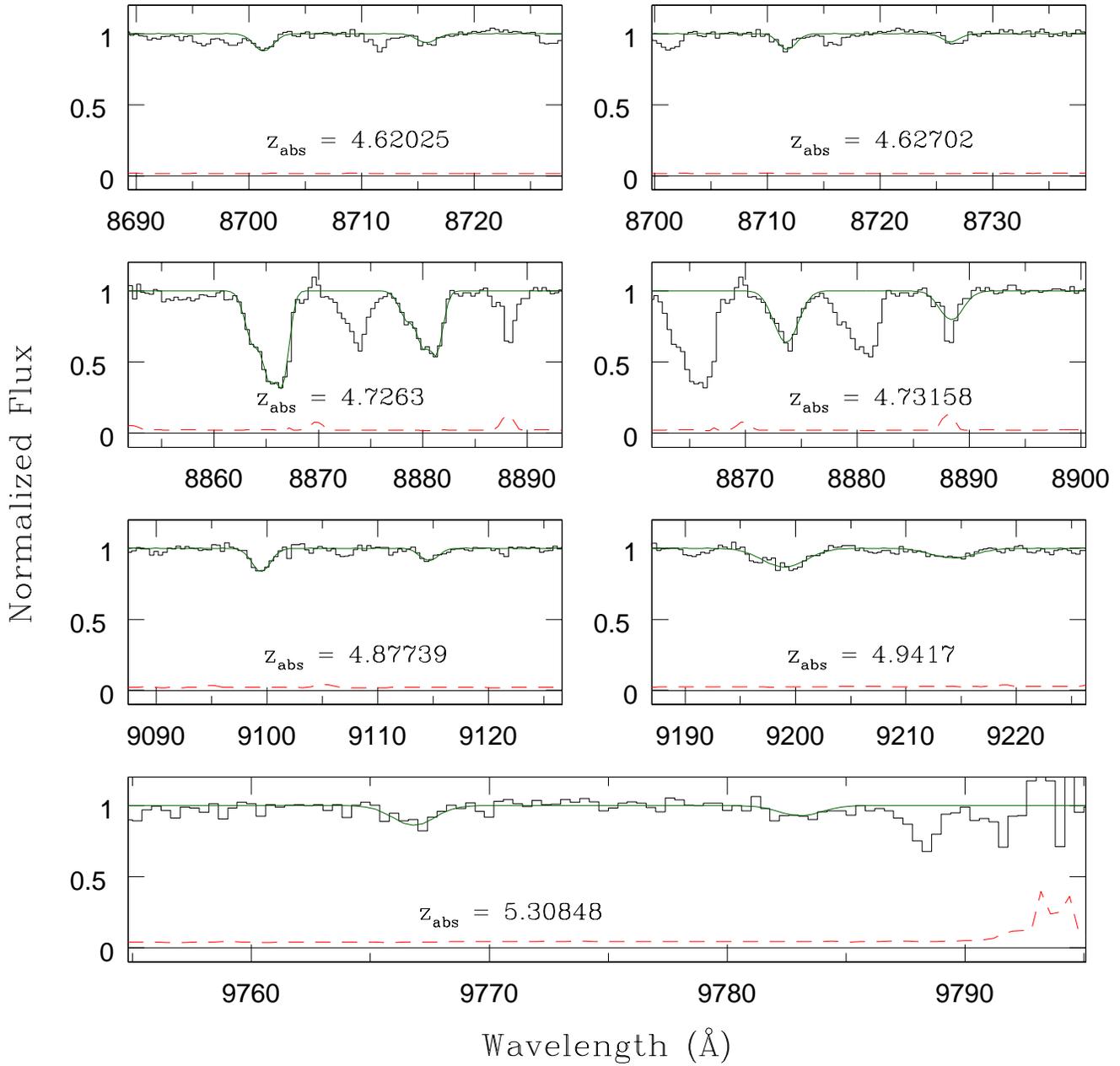}
\caption{Detected \CIV\  absorption systems in the spectrum of SDSS
  J0818+1722 (continuation of Fig.~\ref{J0818_sysz4_1}). }   
\label{J0818_sysz4_2}
\end{center}
\end{figure*}

\FloatBarrier

\begin{figure}
\begin{center}
\includegraphics[width=9cm]{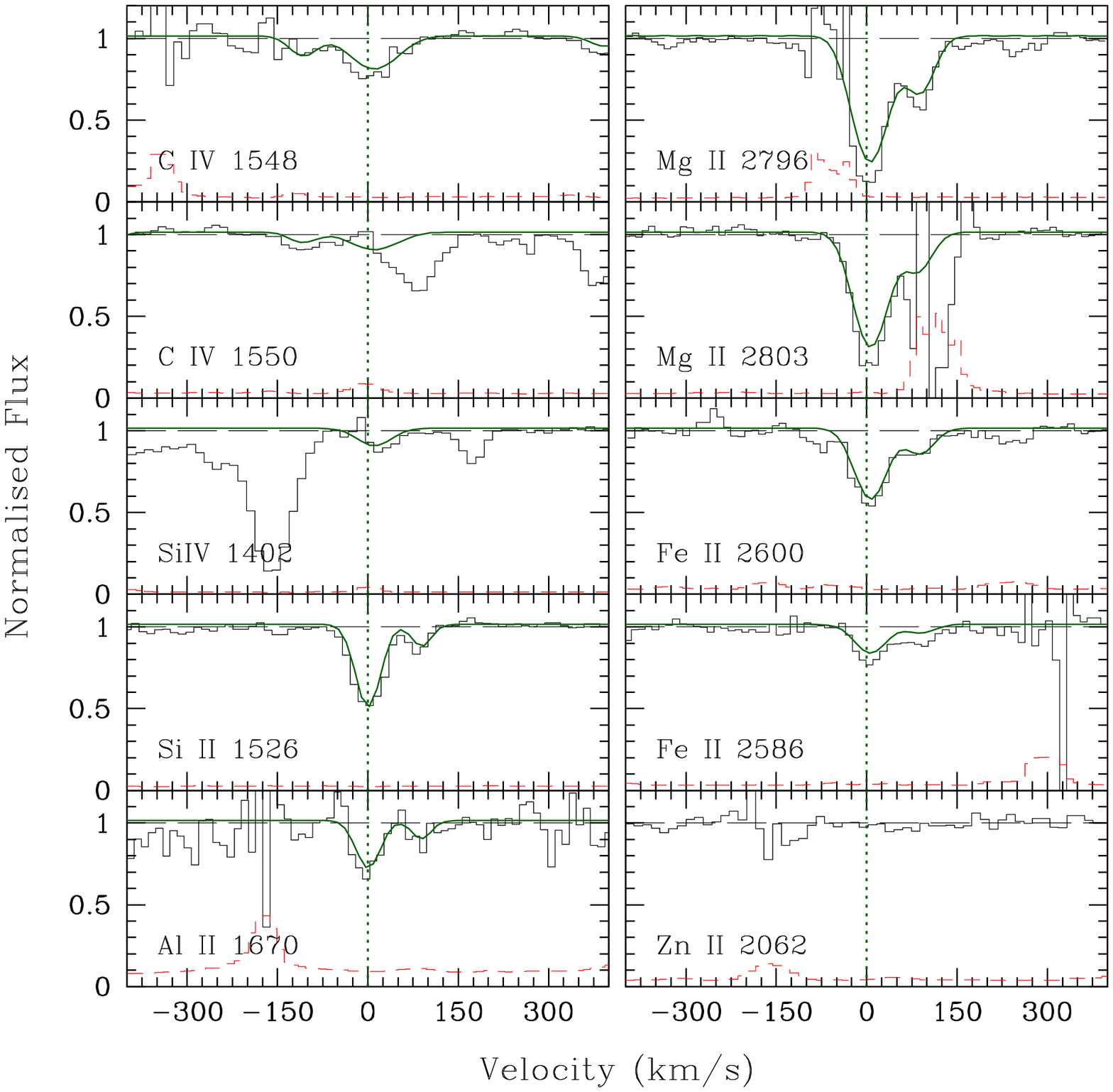}
\caption{Detected transitions in the absorption system at $z_{\rm
    abs}=5.06459$ (\CIV\ is at $\zabs = 5.06467$) in the spectrum of SDSS J0818+1722. }   
\label{J0818_sysz5p06}
\end{center}
\end{figure}

\begin{figure*}
\begin{center}
\includegraphics[width=8cm]{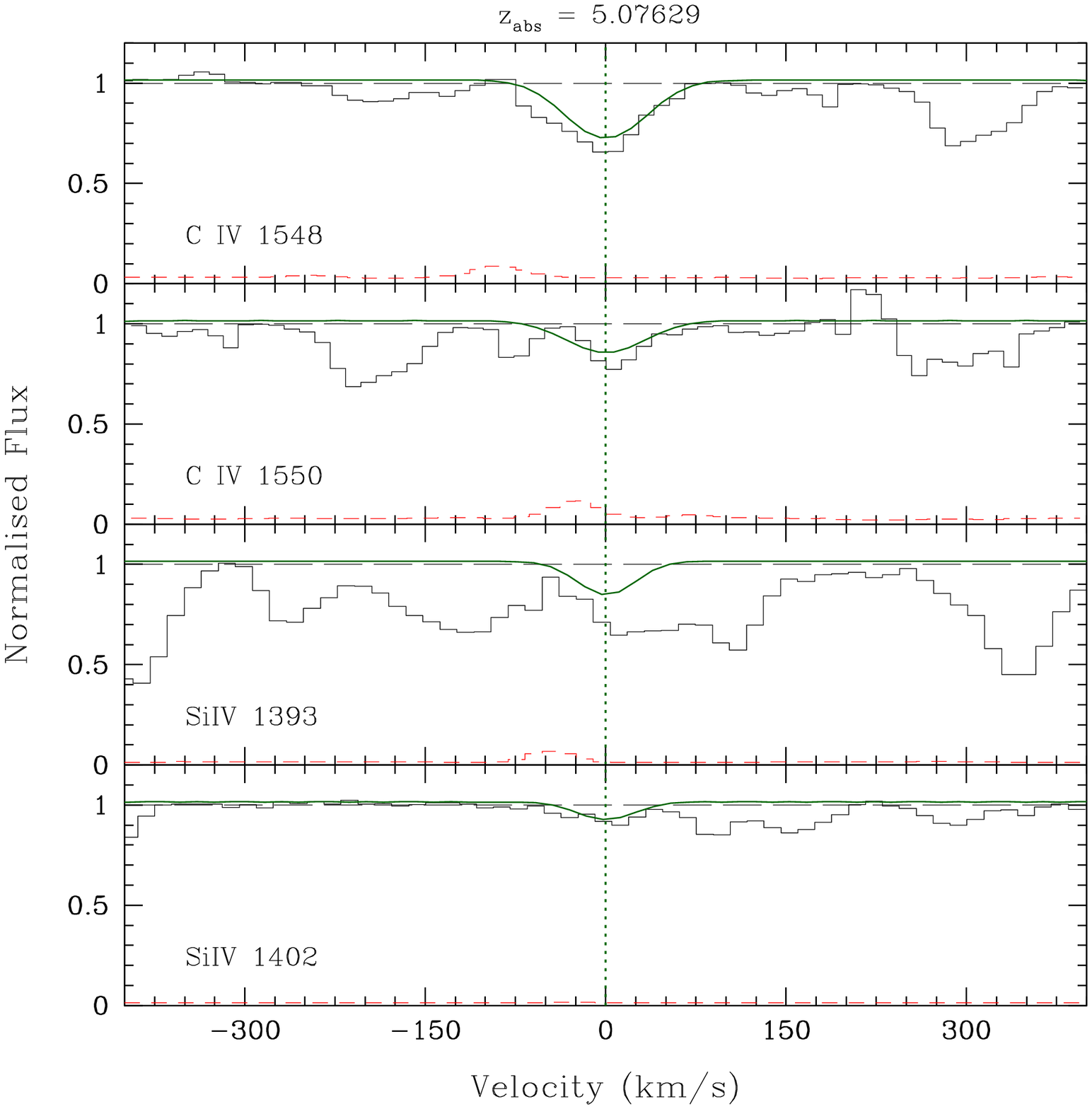}
\includegraphics[width=8cm]{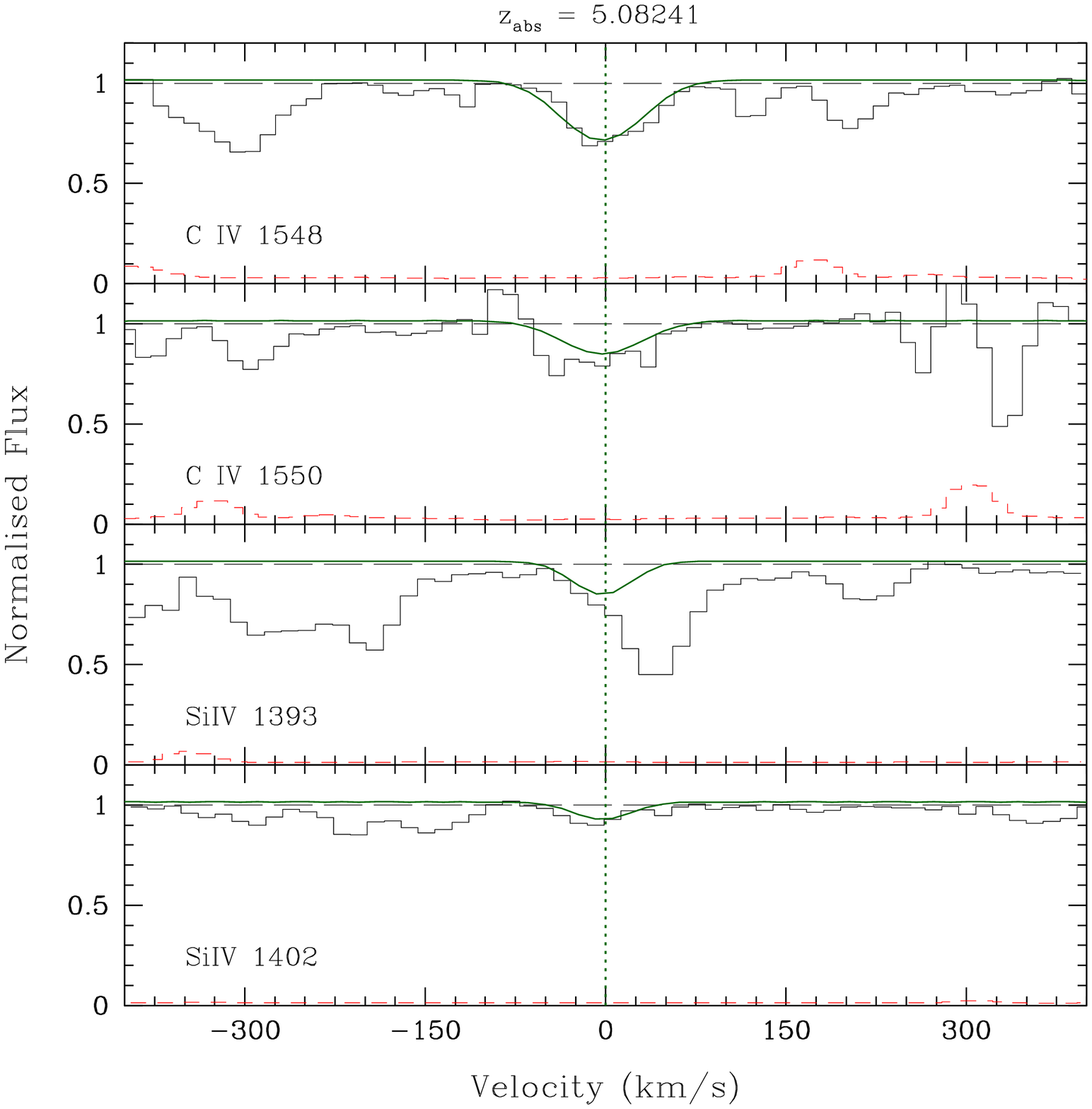}
\includegraphics[width=8cm]{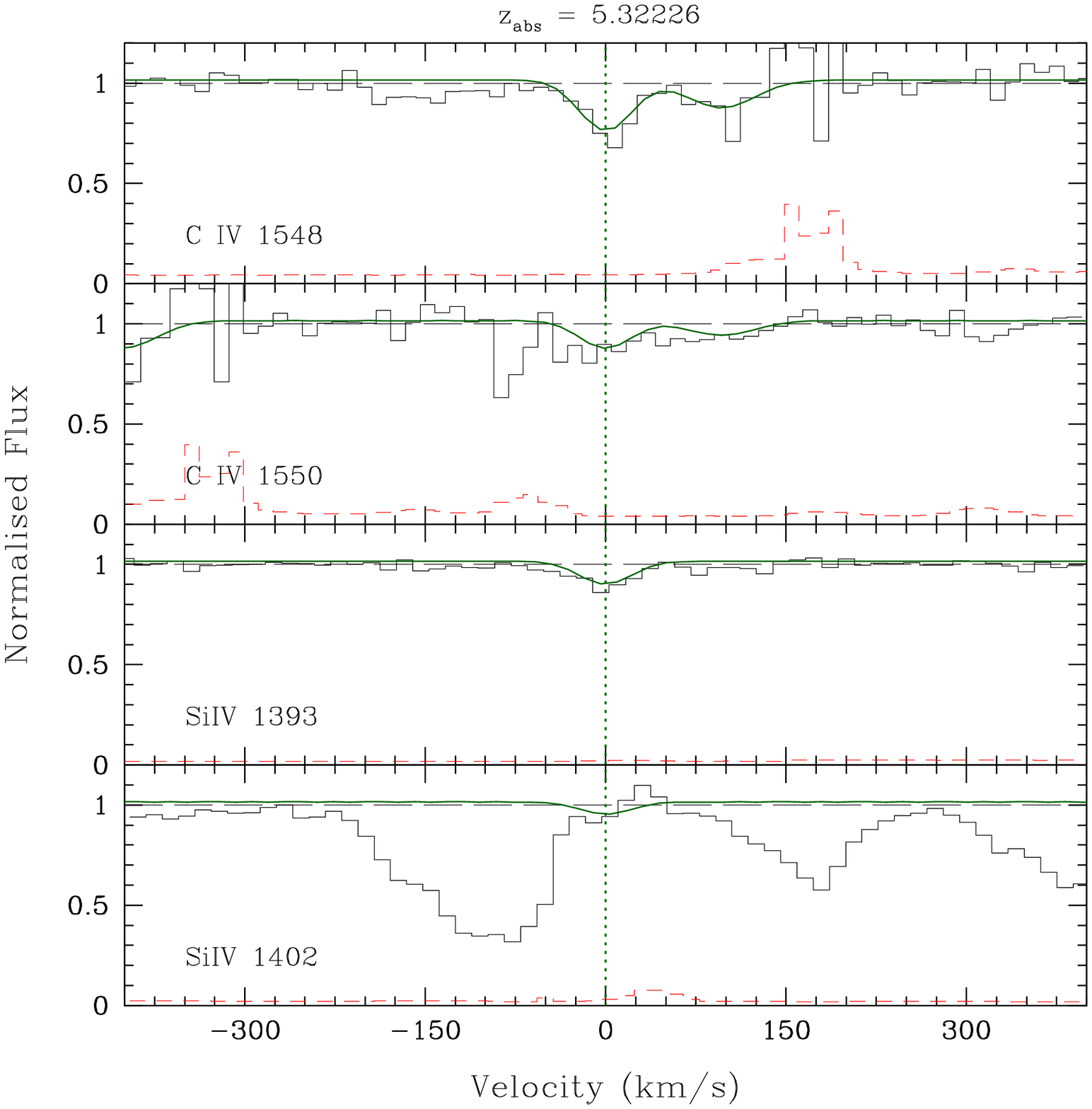}
\includegraphics[width=8cm]{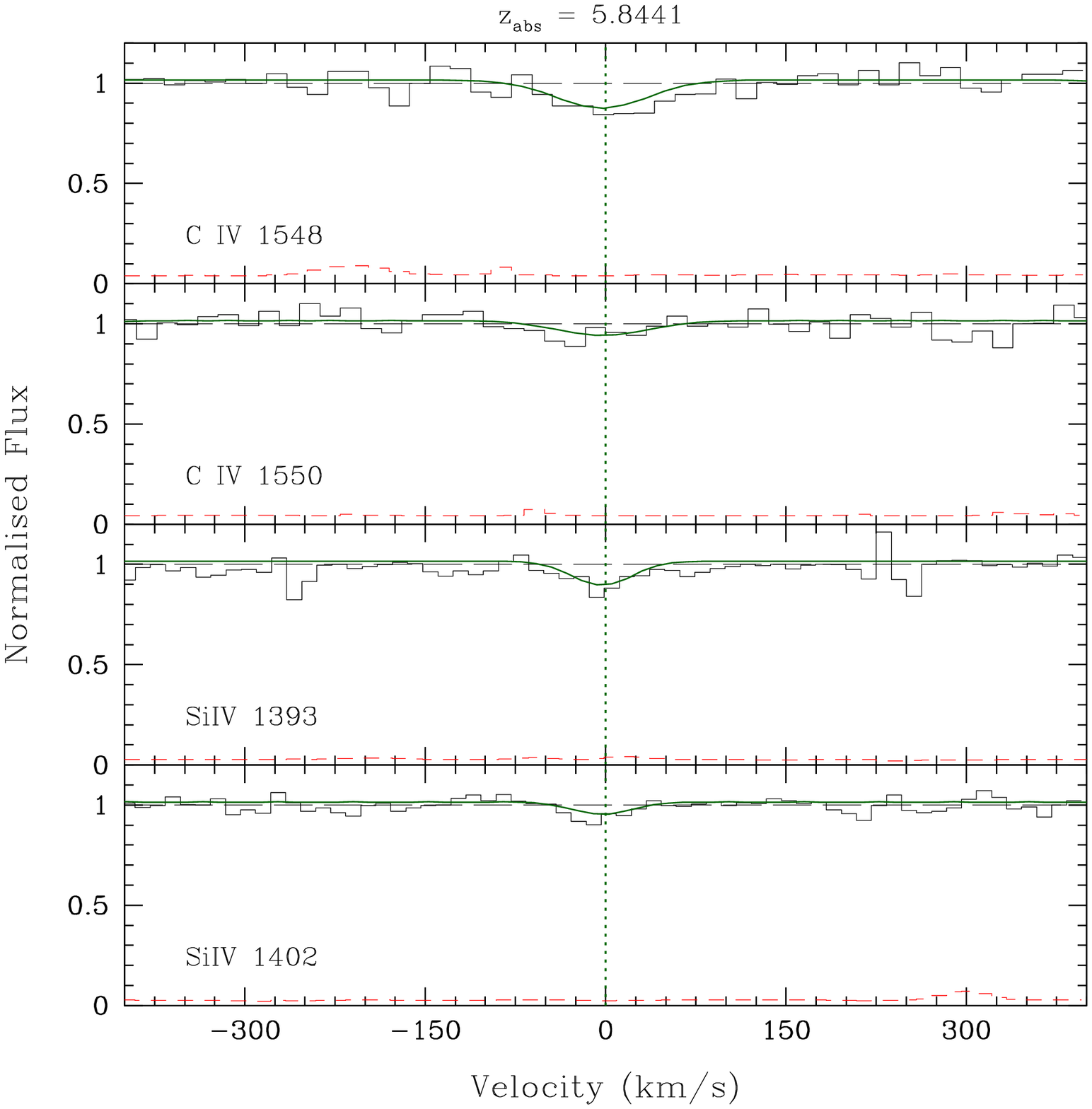}
\caption{Detected \CIV\ absorption systems with associated
  \SiIV\ doublets at $\zabs>5$ in the spectrum of SDSS J0818+1722.  The strong
  absorption lines blended with the \SiIV\ doublets at $\zabs =
  5.07630$, 5.08238 and 5.32226 are telluric lines that we could not
  correct.}    
\label{J0818_sysz5}
\end{center}
\end{figure*}

\FloatBarrier

\begin{figure}
\begin{center}
\includegraphics[width=9cm]{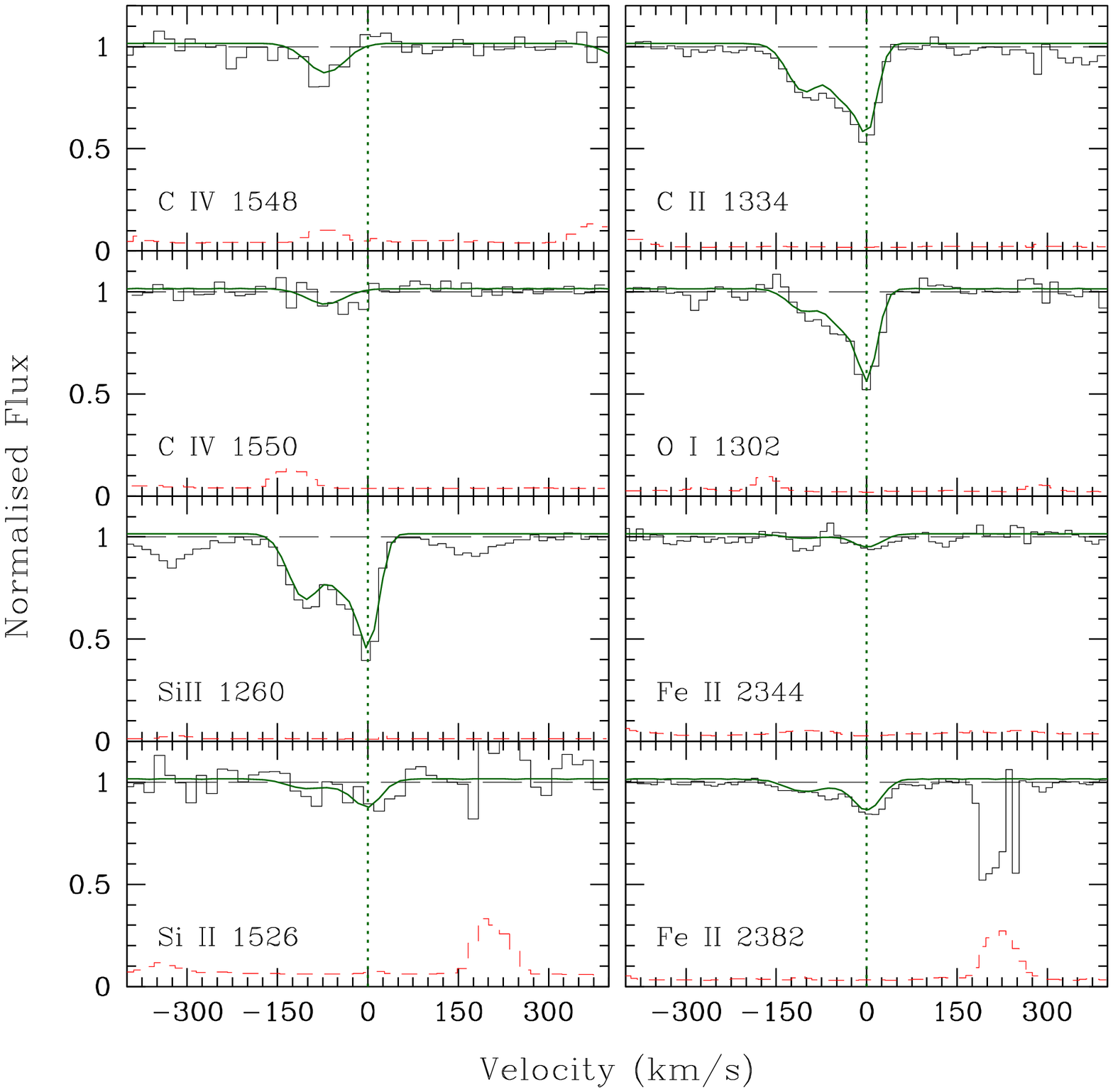}
\caption{Detected transitions in the absorption system at $z_{\rm
    abs}=5.791$ in the spectrum of SDSS J0818+1722. }   
\label{J0818_sysz5p79}
\end{center}
\end{figure}

\begin{figure}
\begin{center}
\includegraphics[width=9cm]{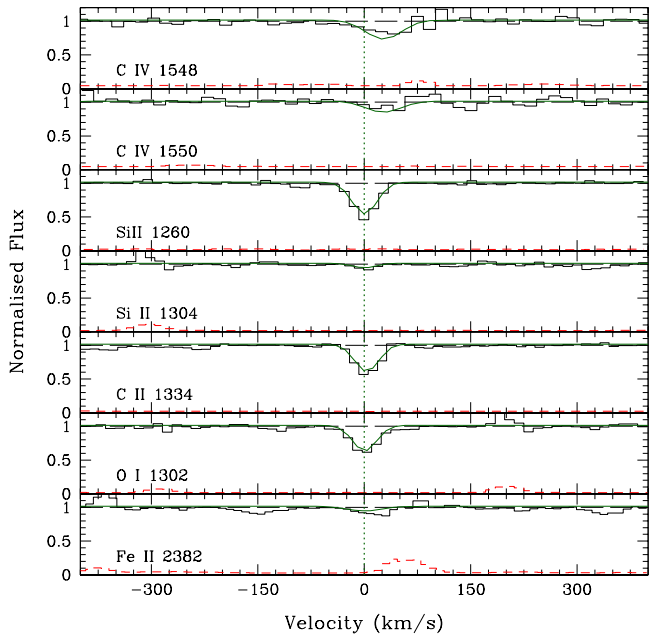}
\caption{Detected transitions in the absorption system at $z_{\rm
    abs}=5.87644$ in the spectrum of SDSS J0818+1722.}   
\label{J0818_sysz5p87}
\end{center}
\end{figure}

\FloatBarrier

We detected \CIV\ doublets at $z_{\rm abs} =4.46298$, 4.49800,
4.50838, 4.52305, 4.62025, 4.62702, 4.7263 (complex, detected also by
Simcoe11), 4.73158, 4.87739, and 4.9417, together with marginal 
detections (below our $3\,\sigma$ limit) at $z_{\rm abs} =4.55225$,
4.57733. Those absorption systems are shown in
Figs~\ref{J0818_sysz4_1} and \ref{J0818_sysz4_2}.   
The $\lambda\,1548$ \AA\ \CIV\ line at $z_{\rm abs} = 4.69148$, reported in
Simcoe11, was identified as the \SiIV\ 1393
\AA\ transition associated with the \CIV\ system at $z_{\rm abs} =5.32226$ (see
below).  

At $z>5$, possible \CIV\ doublets with associated \SiIV\ lines were
detected at $z_{\rm abs} = 5.06467$, 5.07630, 5.08238, 5.32226 and
5.8441, with a marginal detection without \SiIV\ at   $z_{\rm abs}
=5.30848$. Those \CIV\ systems are shown in Figs~\ref{J0818_sysz5p06}
and \ref{J0818_sysz5}. 

The spectrum of J0818+1722 is characterized by the presence of the only
systems at low-ionization above $z\sim5$ detected in our sample, at
$z_{\rm abs} = 5.06459$, 5.791 and 5.87644. For the system at $z_{\rm
  abs} = 5.06459$ (see Fig.~\ref{J0818_sysz5p06}) it was possible to put an upper limit to the column
density of \ZnII\ $\log N($\ZnII$) \lsim 12.5$, implying a ratio
[Zn/Fe]~$\lsim 1.85$. The latter two systems
show also absorption due to \OI\ (see
Figs~\ref{J0818_sysz5p79} and \ref{J0818_sysz5p87}; for all the detected
transitions) they were discussed in the work by \citet{becker11}. 
The presence of the \CIV\ doublet associated with the system at
$z_{\rm abs} = 5.791$ was reported by \citet{ryanweber09}, however both
lines fall on skylines (see Fig.~\ref{J0818_sysz5p79}) and in our
spectrum they are below $3\,\sigma$ detection. Also the \CIV\ at
$z_{\rm abs} = 5.87644$ is very uncertain (see
Fig.~\ref{J0818_sysz5p87}).

Two weak \MgII\ systems were identified at $z_{\rm abs} =2.09059$ and
2.12939, a strong \MgII\ doublet with associated \FeII\ 2344, 2382,
2586, 2600 \AA\ absorptions
was detected at $z_{\rm abs} =3.56285$. We do not confirm the
\MgII\ system at $z_{\rm abs} =2.834$ reported by
\citet{ryanweber09}. The plots for these systems are shown in 
Fig.~\ref{J0818_mgII} (Appendix B).   

%-----------------------------------------------------
\subsection{SDSS J0836+0054}
%-----------------------------------------------------
This quasar, discovered in the SDSS \citep{fan2001}, has the lowest
redshift, $z_{\rm em} = 5.810$, and is the brightest of the sample.

In the past, it was studied with the Keck both at low and high
resolution with ESI, Hires and NIRSPEC
\citep{pettini03,becker06,becker09}, with GNIRS@Gemini \citep{jiang07}
and with ISAAC@VLT \citep{ryanweber09}. 
Due to very good seeing conditions during the observations, we could
observe this object at a higher resolution than the others in the
sample: $R\simeq 11000$ and 8000 in the VIS and NIR arm,
respectively.    
The X-shooter spectrum has an SNR~$\sim 90-50$ in the range
$\lambda\lambda\,8300-9300$ \AA, decreasing to $8-20$ in the $z$
band and increasing again to SNR~$\sim30-37$ in the $J$ and $H$ bands. 
  
We identify a previously unknown \MgII\ system at $\zabs = 2.2990$
with associated \FeII\ 2600 and 2586 \AA, and we confirm the
\MgII\ system reported by \citet{jiang07} at $\zabs = 3.7435$ detecting
also the associated \AlIII\ doublet and \FeII\ 2600 and 
2382 \AA\ features. 
Those absorption systems are plotted in  Fig.~\ref{J0836_mgII} (Appendix B).

\begin{figure*}
\begin{center}
\includegraphics[width=18cm]{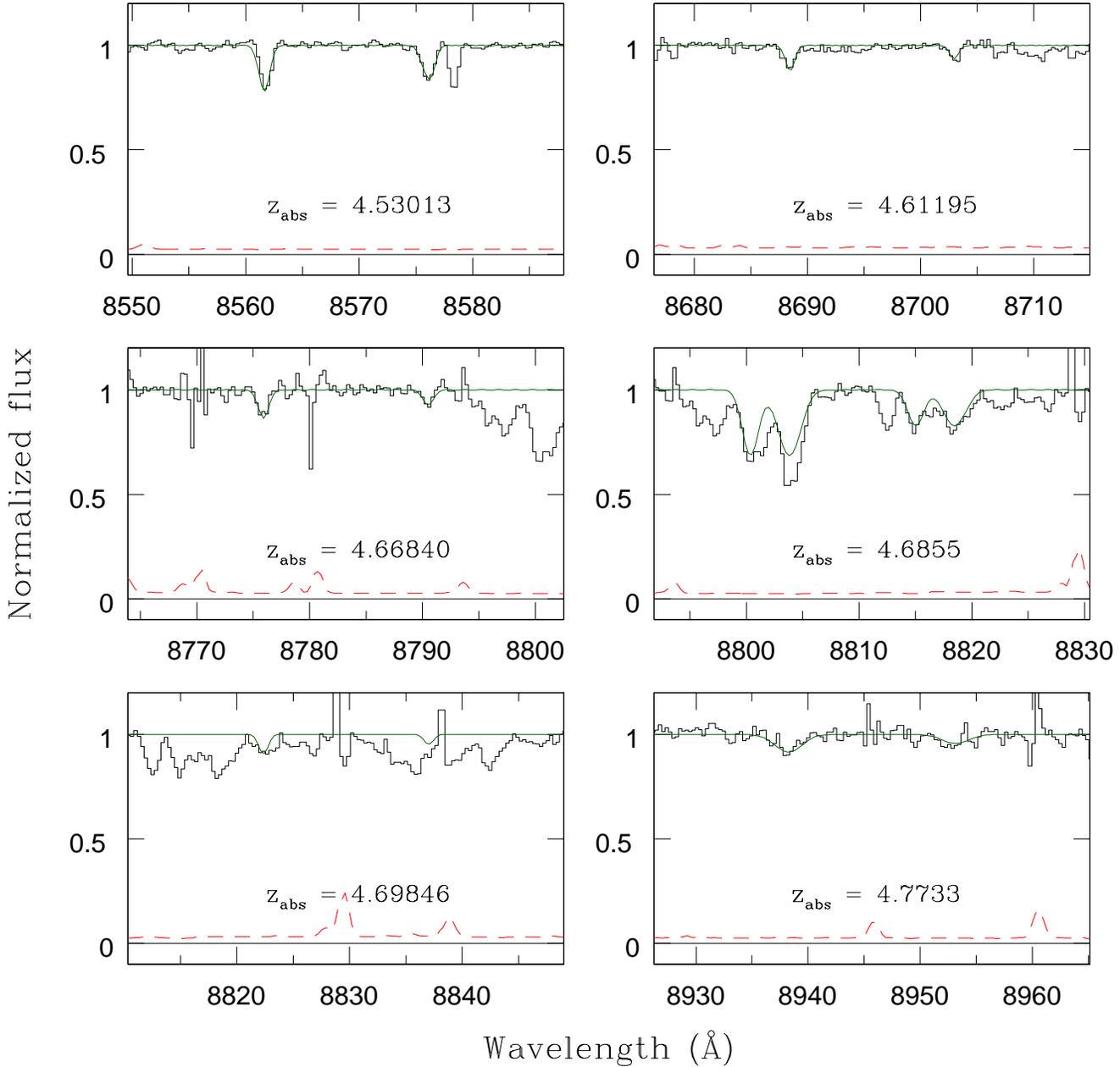}
\caption{\CIV\ absorption systems detected in the spectrum of SDSS
  J0836+0054. } 
\label{J0836_civz4}
\end{center}
\end{figure*}

As for the \CIV\ doublets, a new entry with respect to
\citet[][from now on Pettini03]{pettini03} is found at $z_{\rm abs} =
4.53013$. We also improved the 
fit of the system at $\zabs = 4.68427$ and 4.68651 deblending the
\CIV\ lines from the \AlIII\ doublet at $\zabs = 3.745$.  More \CIV\ 
detections are present at 
$\zabs = 4.61195$, 4.66840, 4.69846, and 4.7733. 
All those \CIV\ systems are shown in Fig.~\ref{J0836_civz4}. 
The system in Pettini03 at $\zabs = 4.5144$ is not confirmed in our
spectrum, the line identified 
as \CIV\ 1548 \AA\ is in fact \SiIV\ 1393 \AA\ at
$\zabs = 5.126$ (see below) and the \CIV\ 1550 \AA\ line is
probably an artefact of a badly subtracted sky line. 

\begin{figure*}
\begin{center}
\includegraphics[width=8cm]{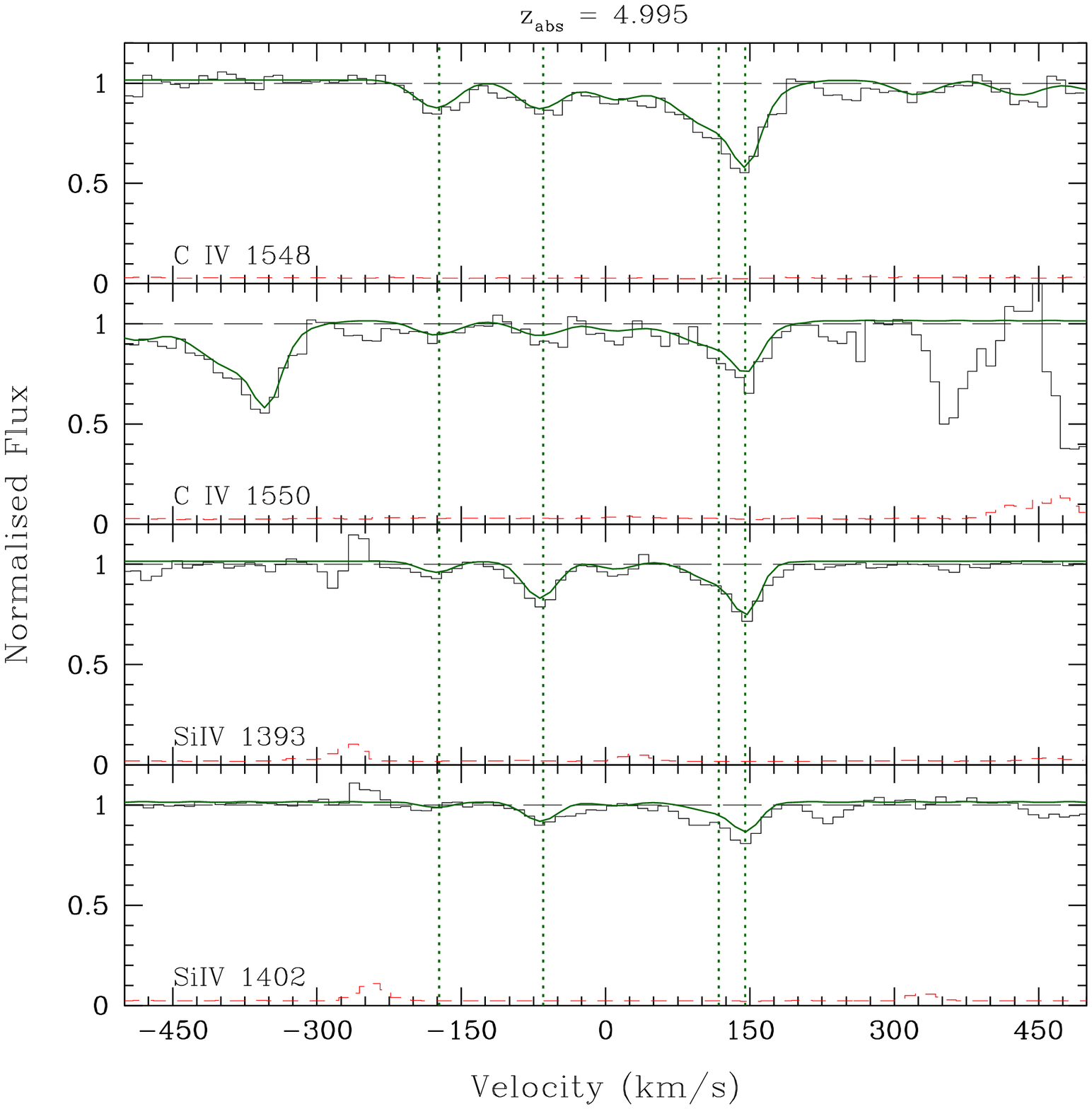}
\includegraphics[width=8cm]{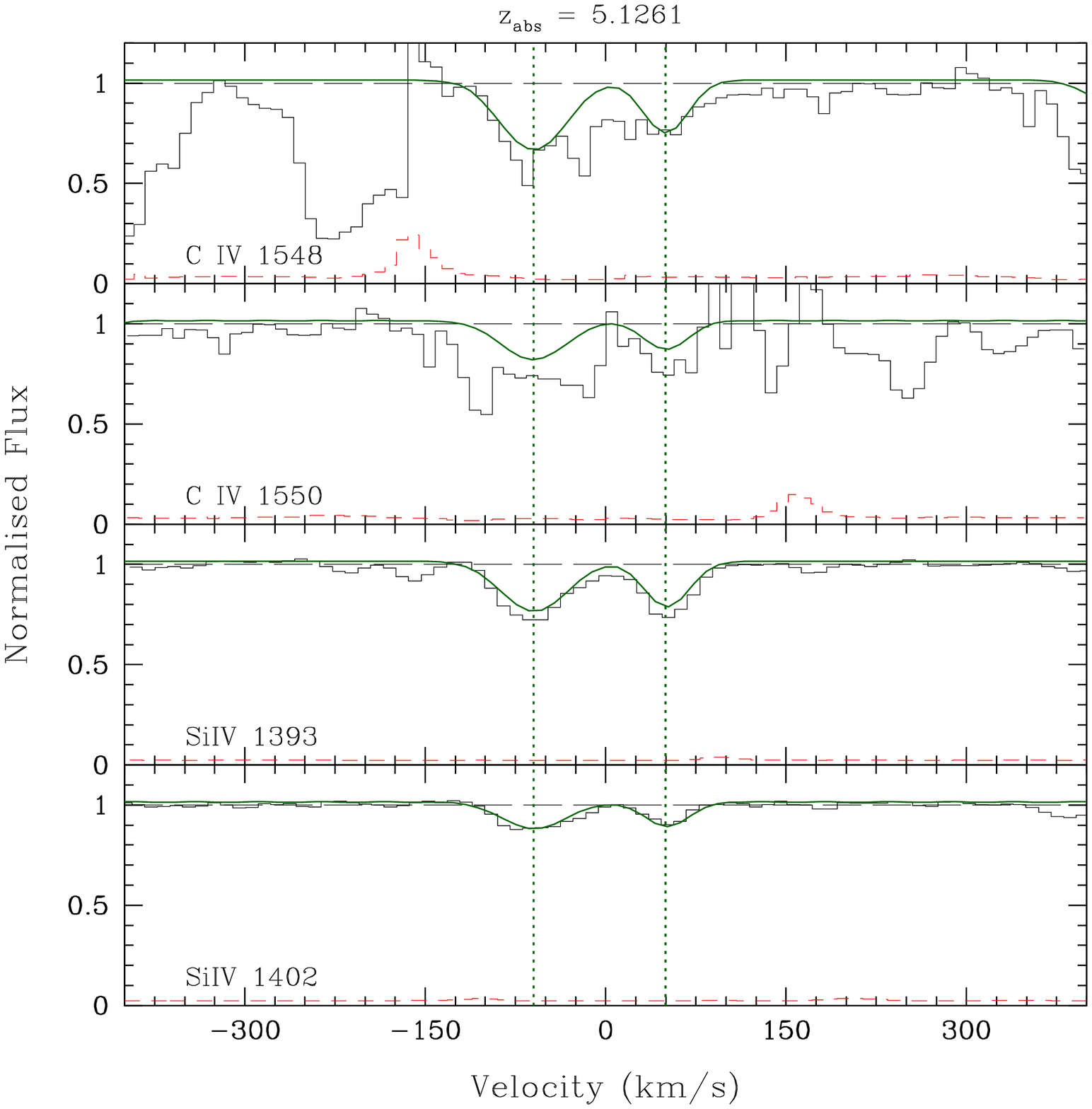}
\includegraphics[width=8cm]{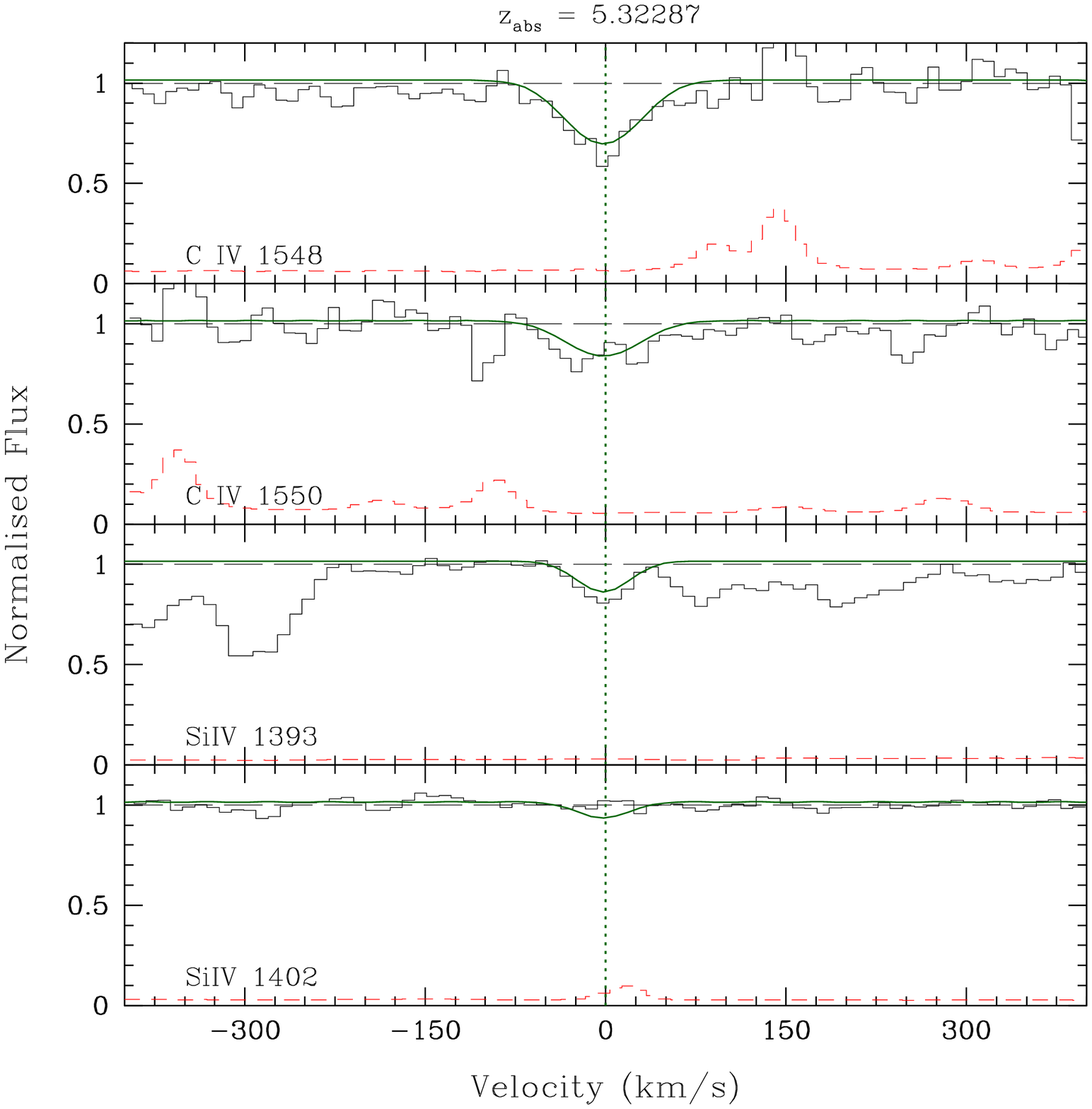}
\caption{\CIV\  absorption systems with associated \SiIV\ doublets in
  the spectrum of SDSS J0836+0054.}  
\label{J0836_civz5}
\end{center}
\end{figure*}

The complex system at $\zabs \simeq 4.994$ detected by  both Pettini03
and Simcoe11 is confirmed by our data. In our spectrum, we detect also
the associated \SiIV\ doublet (see Fig.~\ref{J0836_civz5}), and a
complex velocity structure extending for $\sim 400$ \kms. 

In this work, we claim the detection of two new \CIV\ systems at $z>5$ in
the spectrum of J0836+0054 (see Fig.~\ref{J0836_civz5}). A strong
\SiIV\ at $\zabs \simeq5.126$ with two velocity components is observed
in the VIS region of the spectrum, unfortunately the associated \CIV\ absorption
falls in a region badly contaminated by telluric lines. We partially
corrected for them, but the column density resulting from the fit can
be considered only as an upper limit. A marginal \CIV\ detection
strengthened by the presence of the associated \SiIV\ 1393
\AA\ line is observed at $\zabs = 5.32277$. 

%-----------------------------------------------------
\subsection{SDSS J1030+0524}
%-----------------------------------------------------

This object has been thoroughly studied in the past being the highest
redshift quasar observable from the Southern hemisphere before the
recent discovery of the QSO ULAS J1120+0641 at $z=7.0842$ \citep{mortlock11}. 

Pettini03, by using spectra obtained with ESI at the Keck
telescope, detected a strong \CIV\ system at $\zabs = 4.9482$;
\citet{ryanweber06} identified two \CIV\ doublets at $\zabs = 5.7238$,
and 5.829 in the ISAAC spectrum, which are the main contributors to
their measurement of $\Omega_{\rm CIV}$ \citep[see also][]{ryanweber09}.  

No low ionization systems were detected at $z>5$ both in the low
\citep{jiang07,kurk07} and high \citep{becker06} resolution spectra of
this object.  

The X-shooter spectrum has an SNR~$=140-50$ in the region between the
\Lya\ emission and $\lambda\,9300$ \AA\ decreasing to $\sim 10$ in the
$z$ band and increasing again to $\sim 20-30$ in the $J$ and $H$ bands.  

\begin{figure*}
\begin{center}
\includegraphics[width=18cm]{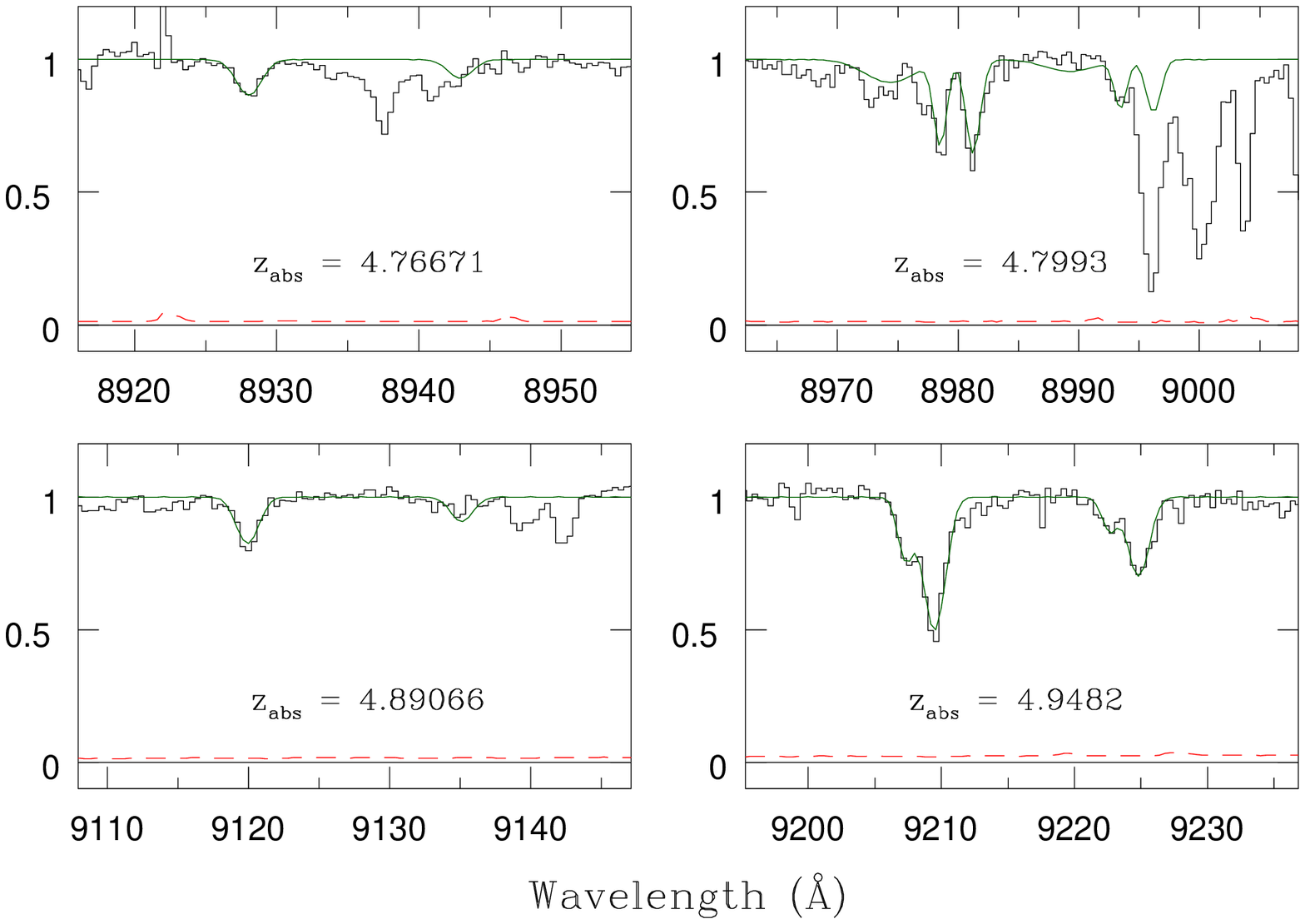}
\caption{Detected \CIV\ absorption systems in the spectrum of SDSS J1030+0524. }   
\label{J1030_civz4}
\end{center}
\end{figure*}

By deblending two complex velocity profiles in the visible region of the
spectrum, we identified two possible new \CIV\ doublets at $\zabs
=4.76671$ (but see Section~4.1) and 4.799 (complex). Another,
\CIV\ doublet was identified at $\zabs =4.89066$ (but see Section~4.1).  
We confirm the presence of the \CIV\ system at $\zabs = 4.9482$
detected by Pettini03, with which we associate also the
\MgII\ doublet in the NIR region of the spectrum. \SiII\ 1526 and
\AlII\ 1670, whose presence is claimed by Simcoe11, are not detected
in our spectrum: in particular, the former if present would be blended
with the \SiIV\ 1393 at  $\zabs =5.5165$. The strong \CIV\ lines at
this redshift are clearly detected in our spectrum. This system shows
associated low ionization lines: \AlII\ 1670 and several
transitions due to \FeII\ ($\lambda\,2344$, 2382, 2586, 2600 \AA) in
the NIR portion of the spectrum.   

\begin{figure*}
\begin{center}
\includegraphics[width=8cm]{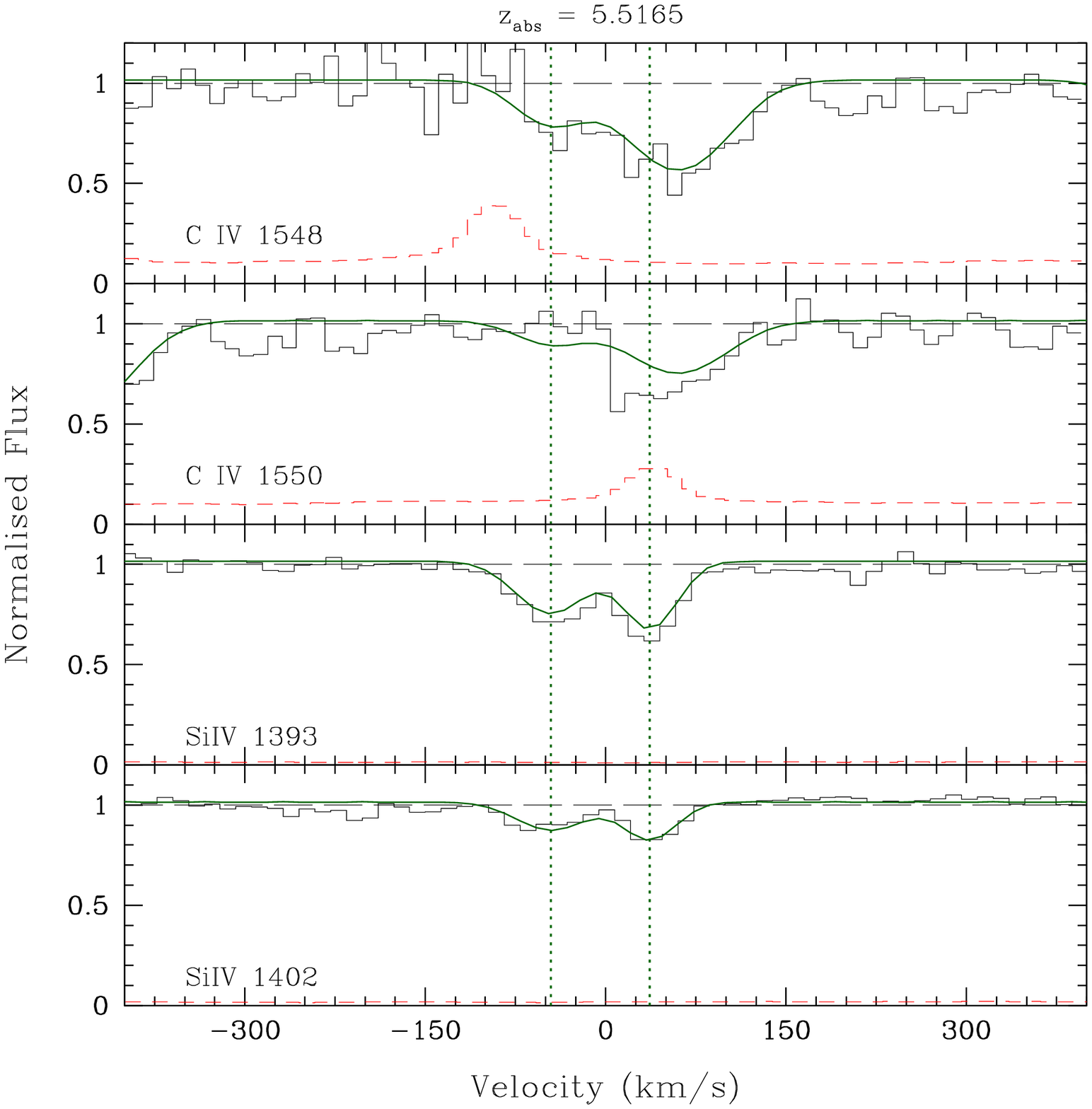}
\includegraphics[width=8cm]{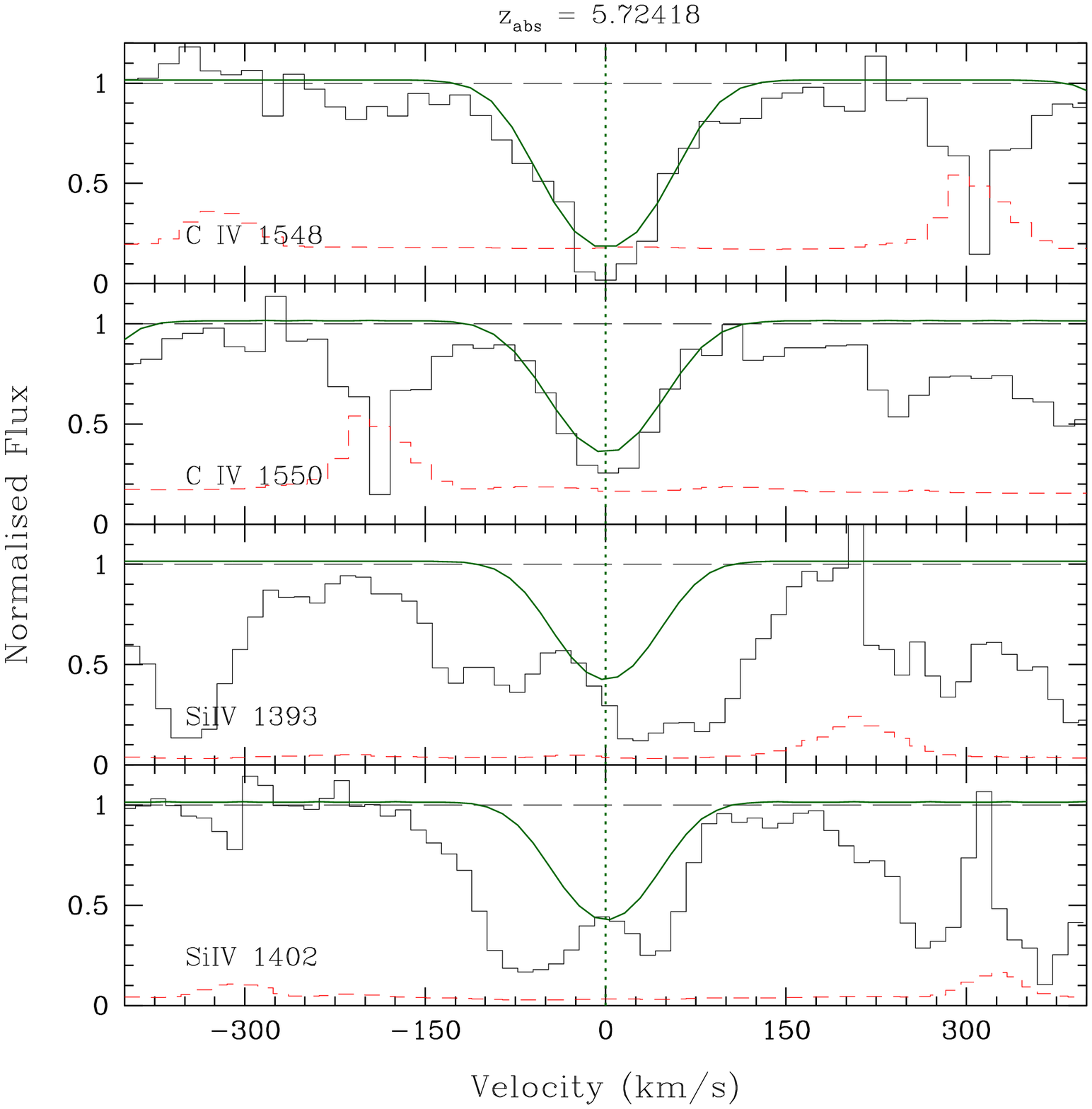}
\includegraphics[width=8cm]{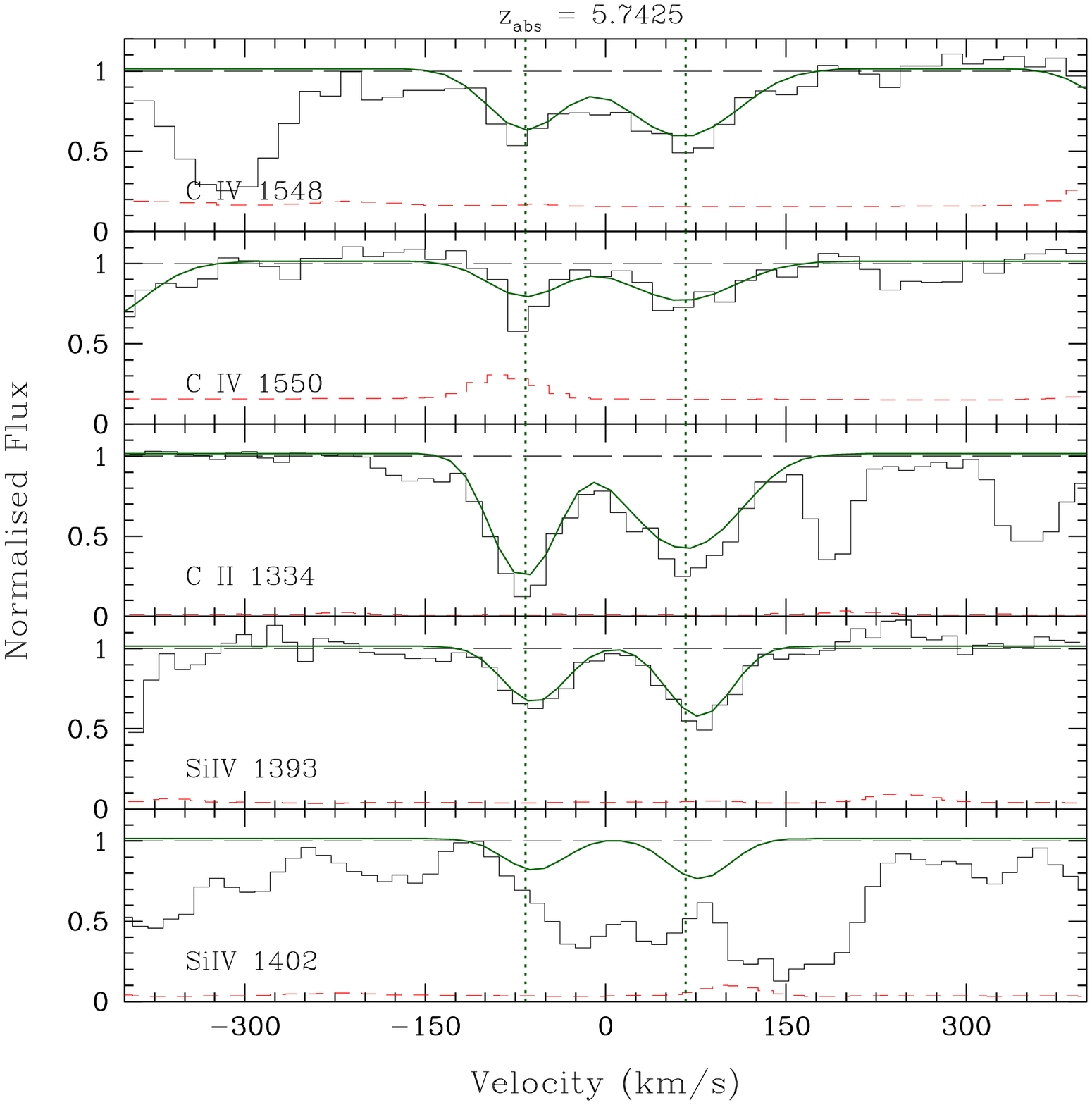}
\includegraphics[width=8cm]{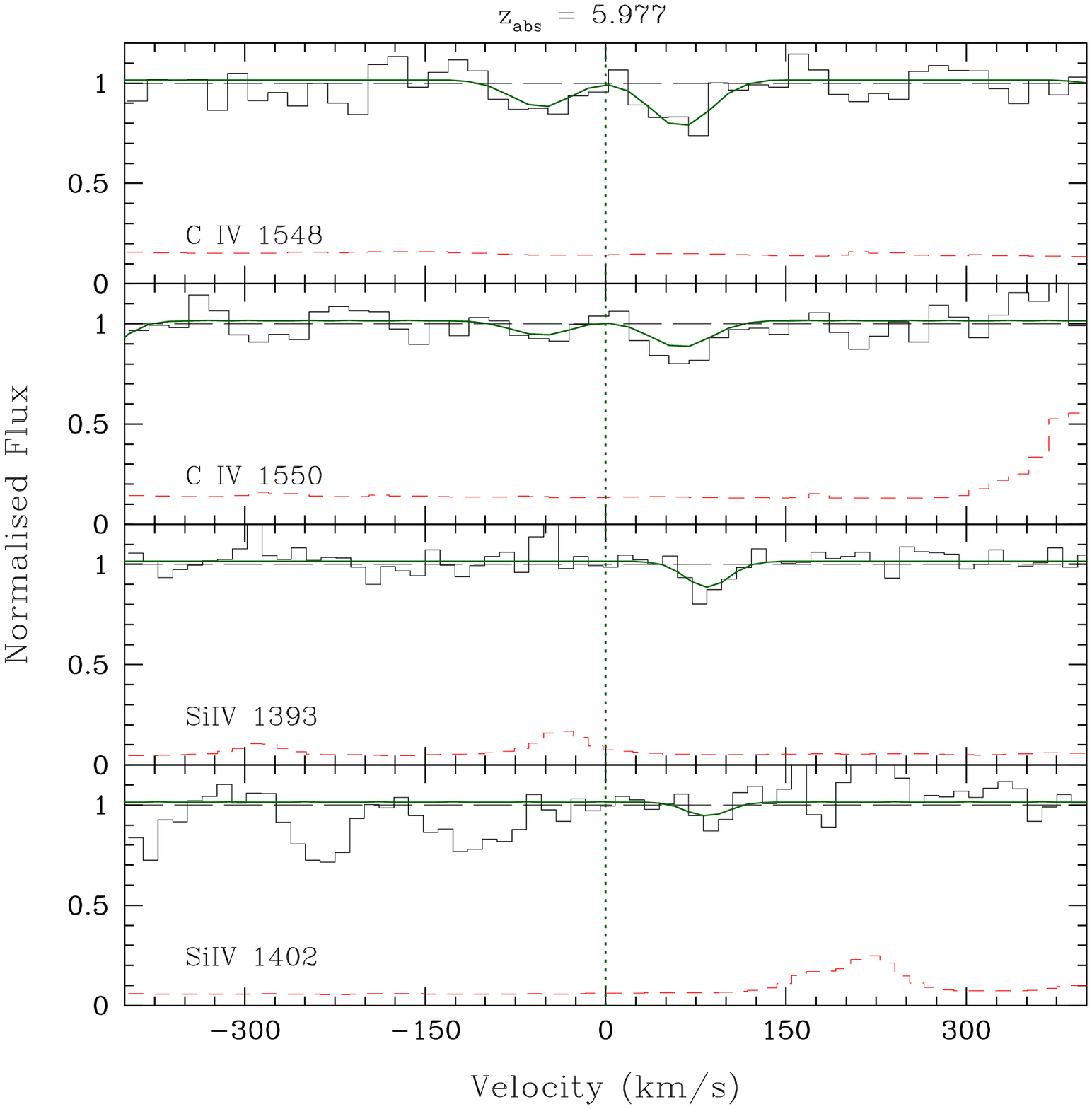}
\caption{Detected \CIV\ absorption systems with the associated
  \SiIV\ doublets at $\zabs>5$ in the spectrum of SDSS J1030+0524. The strong
  absorption lines blended with the \SiIV\ doublets at $\zabs =
  5.72418$ and 5.7425 are telluric lines that we could not correct. }    
\label{J1030_civz5}
\end{center}
\end{figure*}

The previously identified \CIV\ doublet at $\zabs = 5.72418$ is present
also in our spectrum, but no other associated line has been detected
(\SiIV\ is blended with strong telluric lines, but it was detected in
the HIRES spectrum by \citet{becker09}). Recently,
\citet{diaz11} have claimed the association between this
\CIV\ absorption system and a \Lya\ emitter at $z=5.719$, lying at a
projected distance of 79 physical kpc from 
the line of sight. This is the highest redshift galaxy-absorber pair
detected to-date, supporting the idea that galaxy-wide outflows were
already in place at the end of the epoch of re-ionization.

A weak \CIV\ doublet (below $3\,\sigma$) is detected at  $\zabs = 5.7425$, its
nature is confirmed by the presence of the associated transitions due
to \CII\ 1334, \SiIV\ 1393 and \FeII\ 2374, 2382, 2586, 2600 \AA. 
We tentatively identify a \CIV\ doublet at  $\zabs = 5.977$ with
possible associated \SiIV\ doublet (but slightly shifted in
redshift). 
All the \CIV\ doublets in the spectrum of J1030+0524 are shown
in Fig.~\ref{J1030_civz4} and \ref{J1030_civz5}. 

The \CIV\ at $\zabs=5.829$ detected by \citet{ryanweber06} was a
misidentification: it is a strong, complex  \MgII\ doublet at
$\zabs=2.779$ showing also several associated \FeII\ transitions
($\lambda\,2344$, 2382, 2586, 2600 \AA).   
Another strong \MgII\ doublet is present at $\zabs = 4.5836$ with
associated \FeII\ 2344 and 2382 \AA\ lines.  A weak \MgII\ doublet was
detected at  $\zabs = 2.1879$. All \MgII\ systems are plotted in
 Figs~\ref{J1030_mgII_1} and \ref{J1030_mgII_2} (Appendix B).

%\CIV\ at z=5.829 detected by \citet{ryanweber06} is \MgII\ at z=2.78.
 
%-------------------------------------------------------
\subsection{SDSS J1306+0356}
%-------------------------------------------------------

This object was observed during the commissioning of X-shooter. The
VIS portion of the spectrum was discussed in Paper I, while the NIR
portion of the spectrum had an SNR too low to be used. 
 
Here we present new observations taken from the ESO VLT archive which improves
significantly the SNR of the spectrum. Furthermore, these new
observations were obtained with a narrow slit increasing the
resolution to $R=11000$ and 8000 in the VIS and NIR arm,
respectively. 
SNR goes from  $\sim 120$ to 50 in the region between \Lya\ emission
and 9300 \AA, decreasing to $\sim 10$ in the center of the $z$ band to
increase again to $\sim 25$ and 40 in the $J$ and $H$ bands, respectively. 

\begin{figure*}
\begin{center}
\includegraphics[width=18cm]{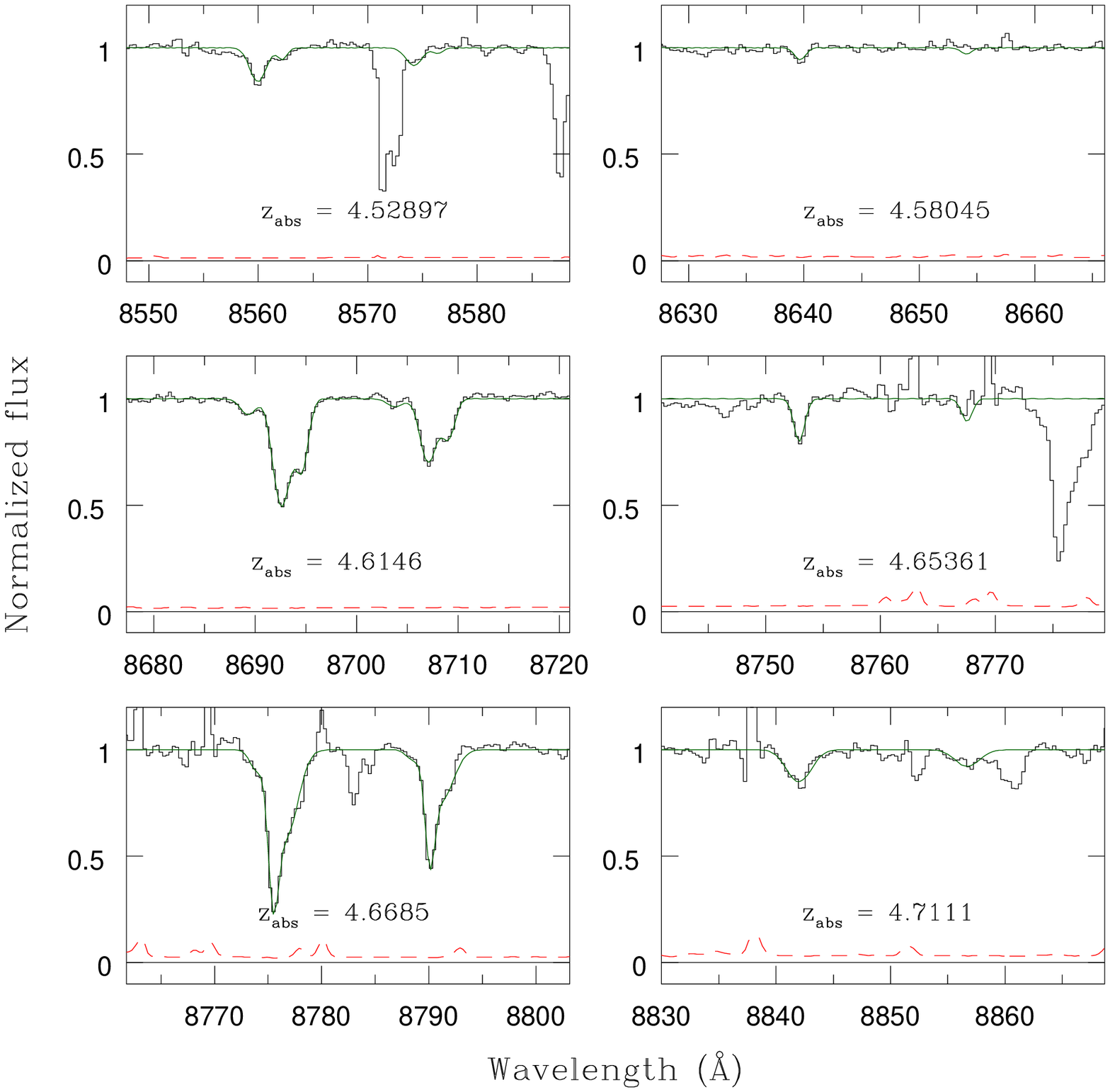}
\caption{Detected \CIV\ absorption systems in the spectrum of SDSS J1306+0356. }   
\label{J1306_sysz4}
\end{center}
\end{figure*}

\begin{figure*}
\begin{center}
\includegraphics[width=18cm]{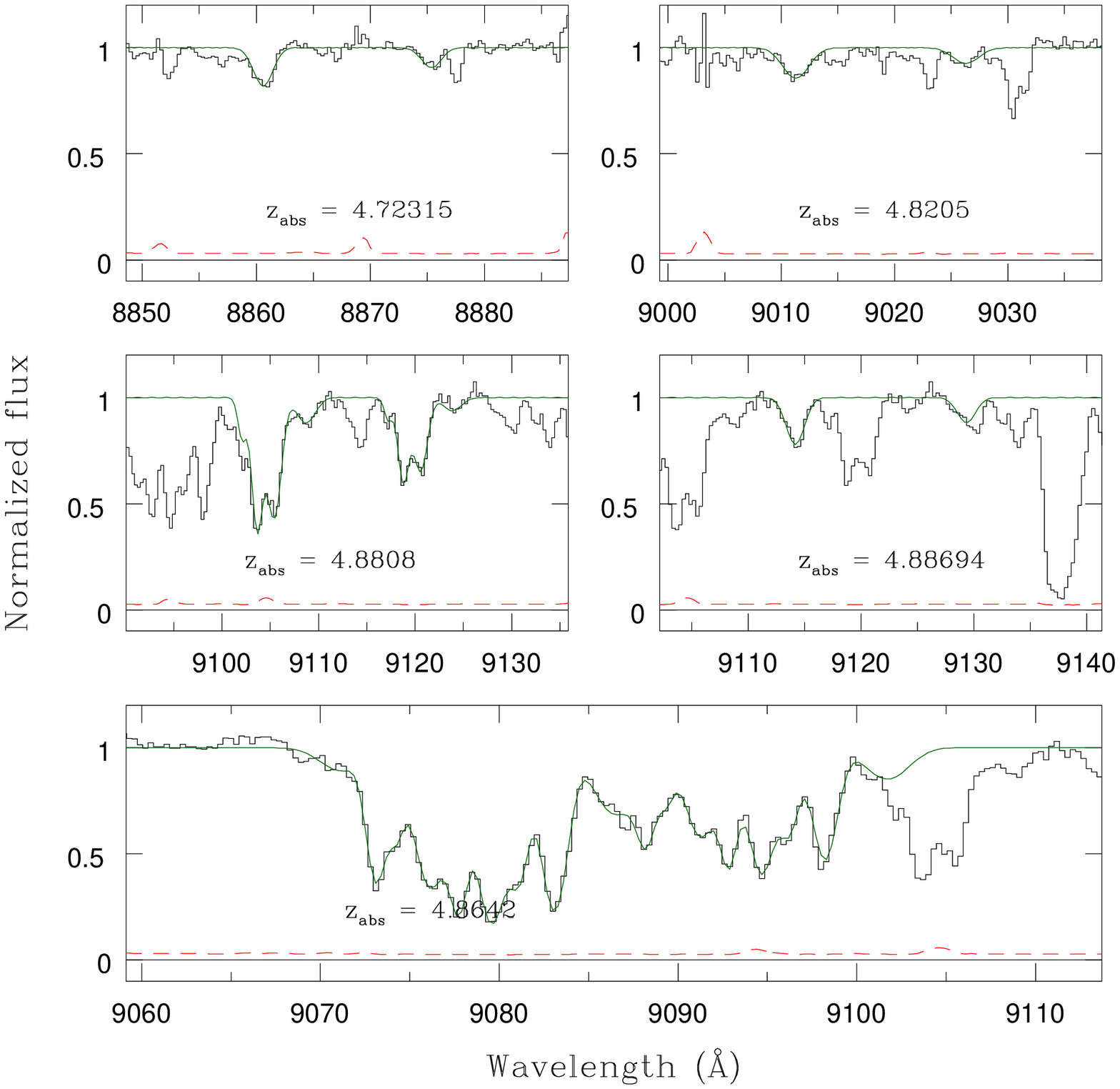}
\caption{Detected \CIV\ absorption systems in the spectrum of SDSS
  J1306+0356 (continuation of Fig.~\ref{J1306_sysz4}). }   
\label{J1306_sysz4_2}
\end{center}
\end{figure*}

The stronger systems at $z_{\rm abs} = 4.6146$,
4.668, 4.8642 (complex system, extending over more than 600 \kms) and
4.8808 were confirmed also by Simcoe11. We report new detections at
$z_{\rm abs} = 4.52897$, 4.65361, 4.7111, 4.72315, 4.82048 (but see Section~4.2), 4.88694 plus
a marginal detection at  $z_{\rm abs} = 4.5804$.  
All the identified \CIV\ systems are shown in Figs~\ref{J1306_sysz4}
and \ref{J1306_sysz4_2}. 

Three new \MgII\ systems were identified at $z_{\rm abs} =  2.3781$
(with associated \FeII\ 2586, 2600 \AA\ lines),
3.4900 and 4.13983 (with associated \FeII\ 2344, 2374, 2382, 2586, 2600
\AA\ and \AlII\ lines), and the system at $z_{\rm abs} 
= 2.5329$ \citep{jiang07,kurk07} is confirmed. Furthermore, we detected the low-ionization lines (\MgII; \FeII\ 1608, 2344, 2374, 2382,
2586, 2600 \AA; \AlII\ and \SiII\ 1526 \AA) associated with the
strong \CIV\ systems at $\zabs = 4.6146$, 4.8642 and 4.8808 (see
Fig.~\ref{J1306_mgII_2} in Appendix B). 

As in previous observations \citep[][; Simcoe11]{ryanweber06,simcoe06}, we
did not identify any reliable metal system at $z \ge 5$  along the
line of sight to J1306+0356.  
%We propose the tentative detection of a
%low ionization system at $z_{\rm abs} =5.4348$ showing
%\CII\ $\lambda\,1334$ \AA and \OI\ in the forest. 

%--------------------------------------------
\subsection{ULAS J1319+0950}
%--------------------------------------------

This QSO at $\zem = 6.13$ was the most recently discovered of the
sample \citep{mortlock09}. We obtained observations from the X-shooter
archive at a resolution of $R=11000$ and 8000 in the VIS and NIR arms,
respectively. This is the only intermediate-resolution spectrum of
this object obtained up to now,together with the FIRE spectrum in
Simcoe11. 

The SNR varies between $\sim 95$ and 65 from the \Lya\ emission to
9300 \AA,
decreasing to $\sim15$ in the centre of the $z$ band. It increases again
to SNR~$\sim30$ and 40 in the $J$ and $H$ band, respectively. 

Four new \MgII\ systems were detected at $\zabs = 2.304$, 2.41
(complex system), 4.21621 (with \AlII\ and \FeII\ 2382 \AA), 4.56837
(with \AlII\ and \FeII\ 2382 \AA) and 4.662 (complex, associated with a
\CIV\ doublet). All \MgII\ systems are shown in  Fig.~\ref{J1319_mgII}
(Appendix B). 

\begin{figure*}
\begin{center}
\includegraphics[width=18cm]{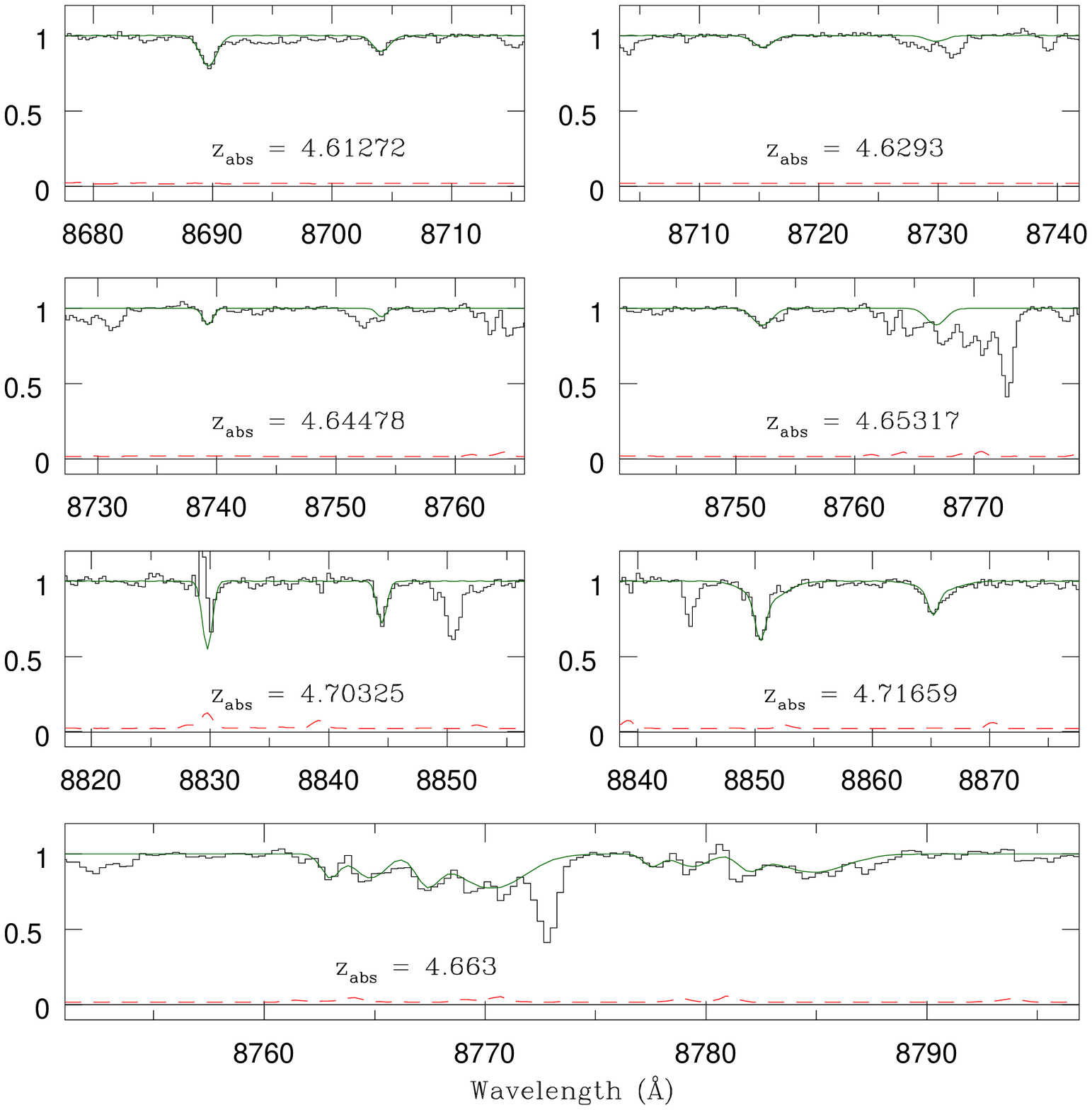}
\caption{Detected \CIV\ absorption systems in the spectrum of SDSS J1319+0950.}   
\label{J1319_sysz4}
\end{center}
\end{figure*}

\begin{figure*}
\begin{center}
\includegraphics[width=8cm]{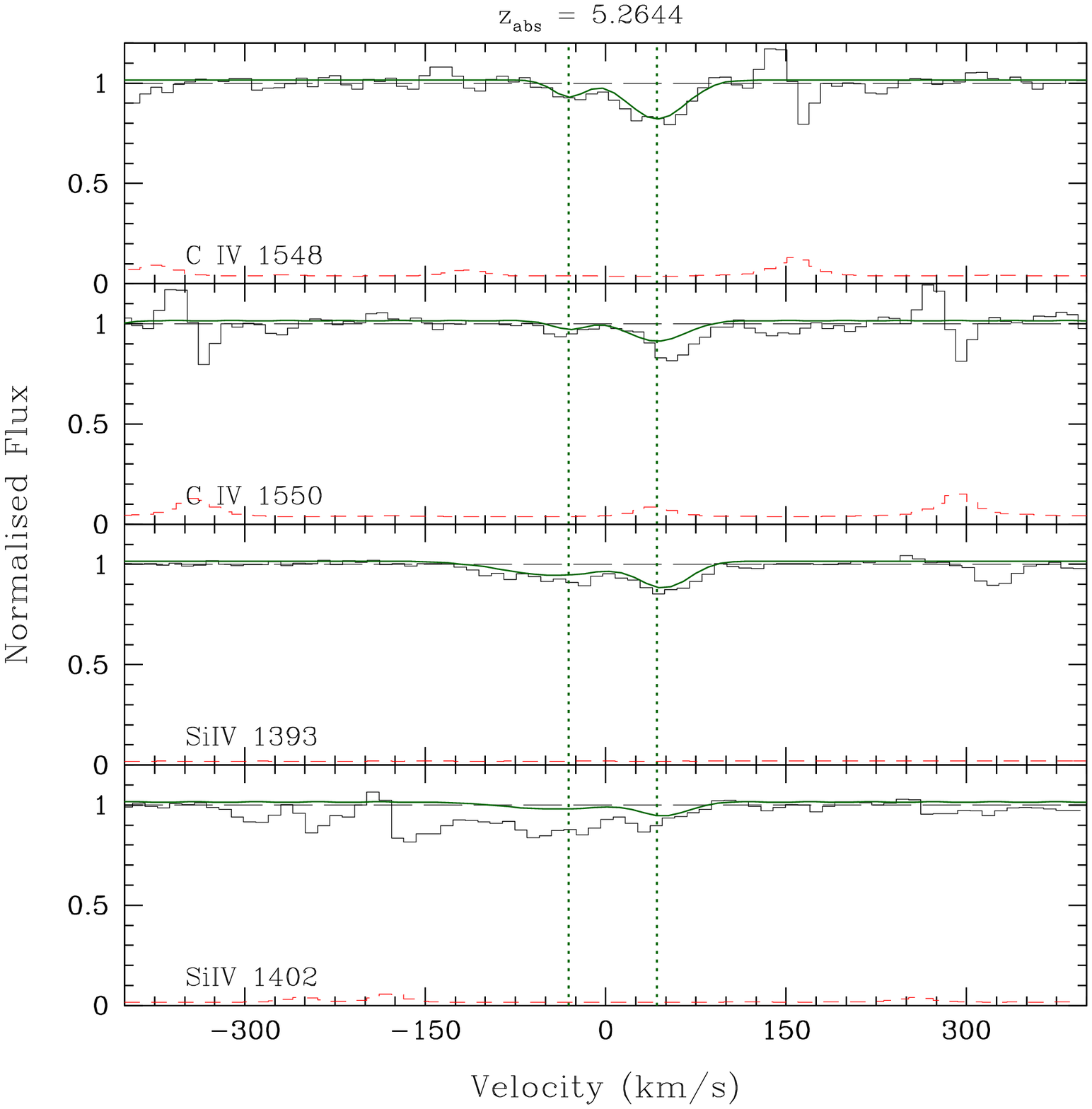}
\includegraphics[width=8cm]{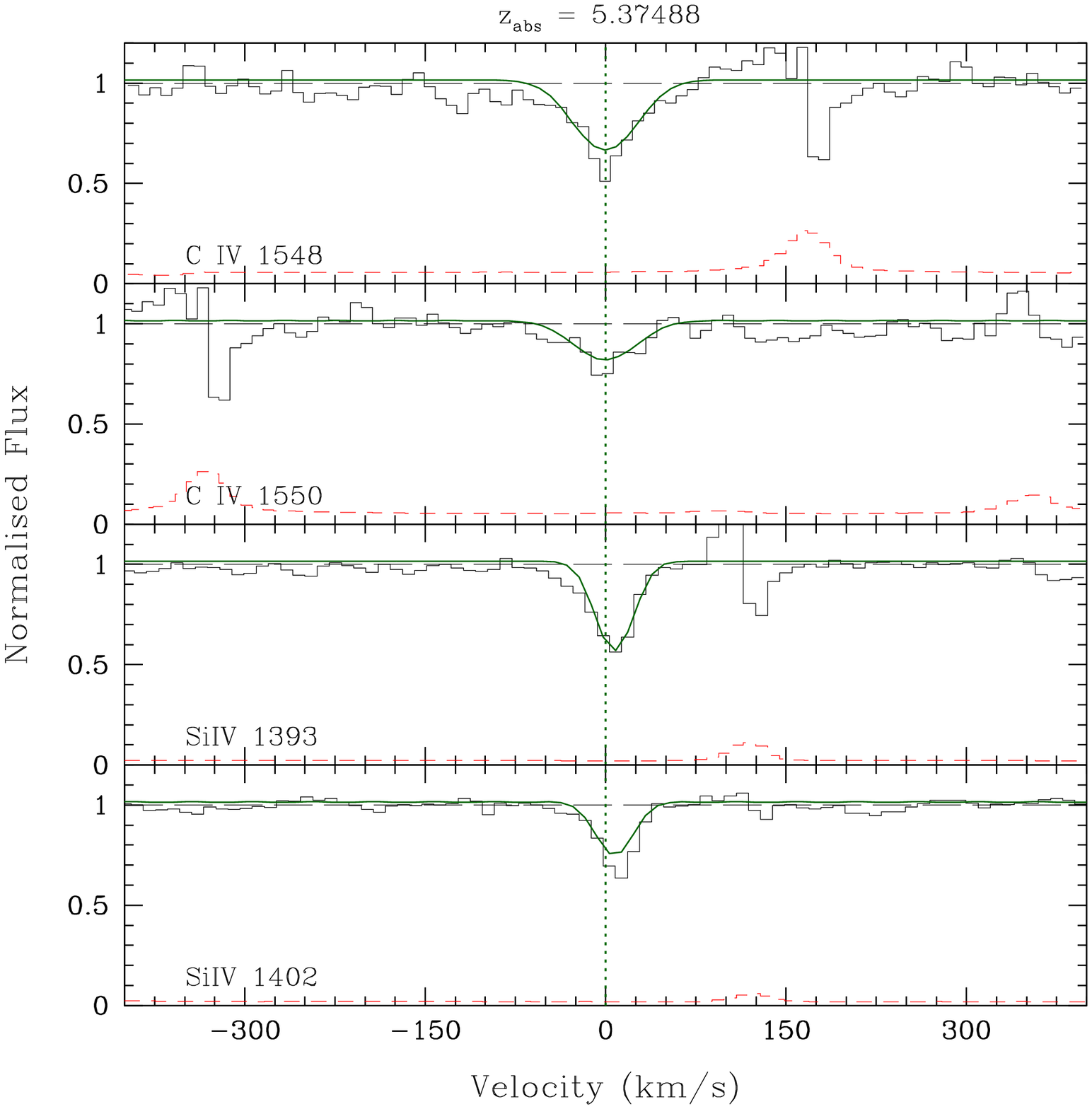}
\includegraphics[width=8cm]{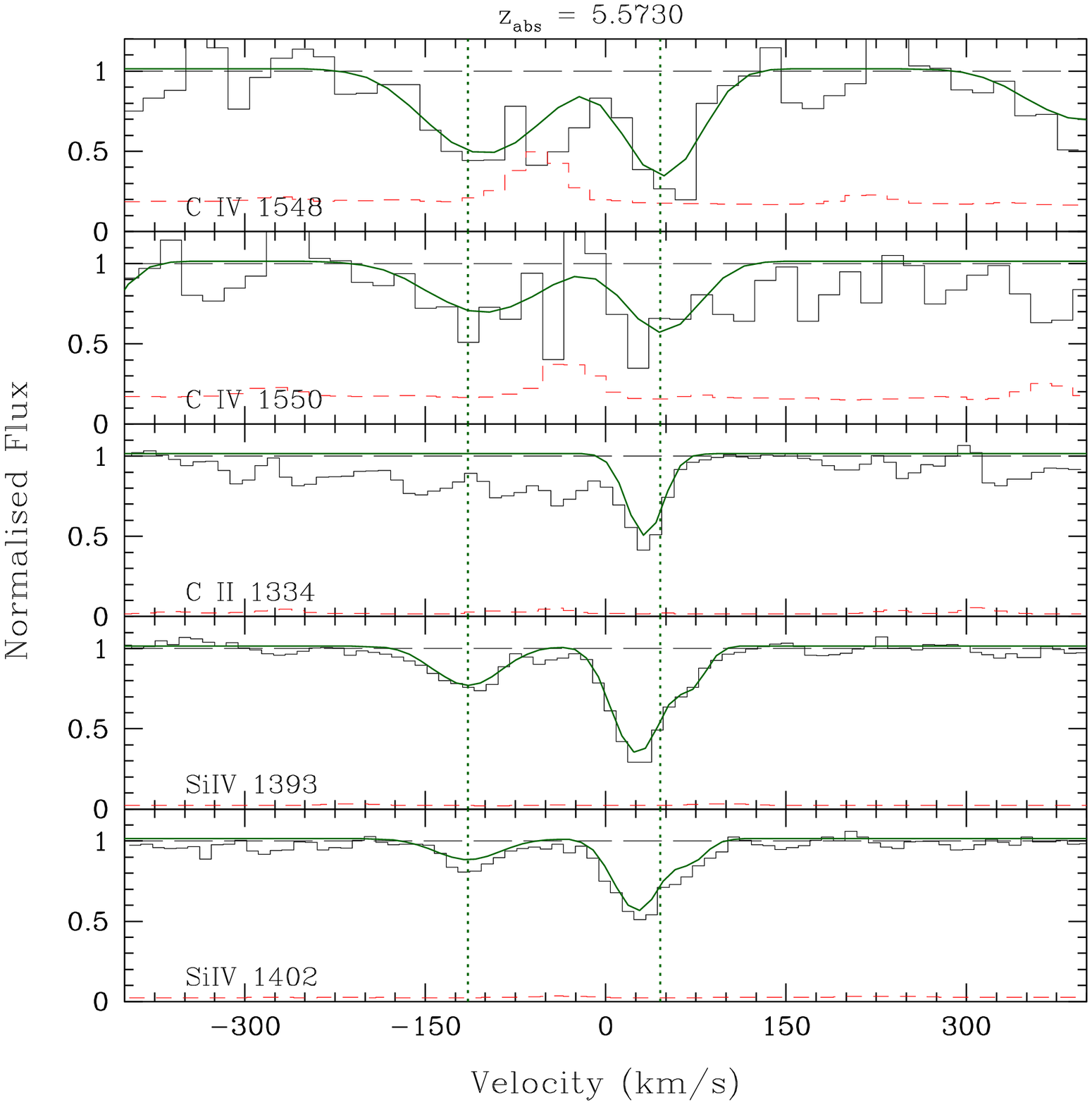}
\caption{Detected \CIV\ absorption systems with associated
  \SiIV\ transitions at $z_{\rm abs} > 5$, in the spectrum of SDSS J1319+0950. }   
\label{J1319_sysz5}
\end{center}
\end{figure*}

Several new \CIV\ doublets were detected in the VIS portion of the
spectrum where the SNR is maximum: at $\zabs = 4.61272$, 4.65317, 4.662
(complex system), 4.70325 and 4.7169, and also marginal detections at
$\zabs = 4.6293$ and 4.64478. Above $z>5$, we report three detections
confirmed by the presence of the associated \SiIV\ doublet:  $\zabs =
5.2644$, 5.37488 and 5.5730 (with the possible detection of
\CII\ 1334 \AA).  Those three systems are shown in
Fig.~\ref{J1319_sysz5}. 
 
%--------------------------------------------
\subsection{CFHQS J1509-1749}
%--------------------------------------------
A preliminary spectrum of this object was already presented in Paper I. 

The VIS spectrum has an SNR~$\sim 65-30$ in the region between the \Lya
emission and $\sim 9600$ \AA, decreasing to $\sim 10$ in the middle of
the $z$ band. The SNR increases again to $\sim 20$ in the $J$ band and to
$25-30$ in the $H$ band. 

We detected  three \MgII\ systems with associated \FeII\ 2344,
2382, 2586, 2600 \AA\ transitions at
$\zabs = 3.1277$, 3.26574 and 4.0108 (in this case \MgII\ is blended with
telluric lines). The saturated \MgII\ doublet at $\zabs
= 3.3915$ presents associated ionic transitions due to \MgI, \mbox{Mn\,{\sc ii}},
\FeII\ 2344, 2382, 2600 \AA\ and \ZnII\ 2026 and
2062 \AA. The plots of these absorption systems are shown in
Fig.~\ref{J1509_mgII} (Appendix B).  

\begin{figure*}
\begin{center}
\includegraphics[width=18cm]{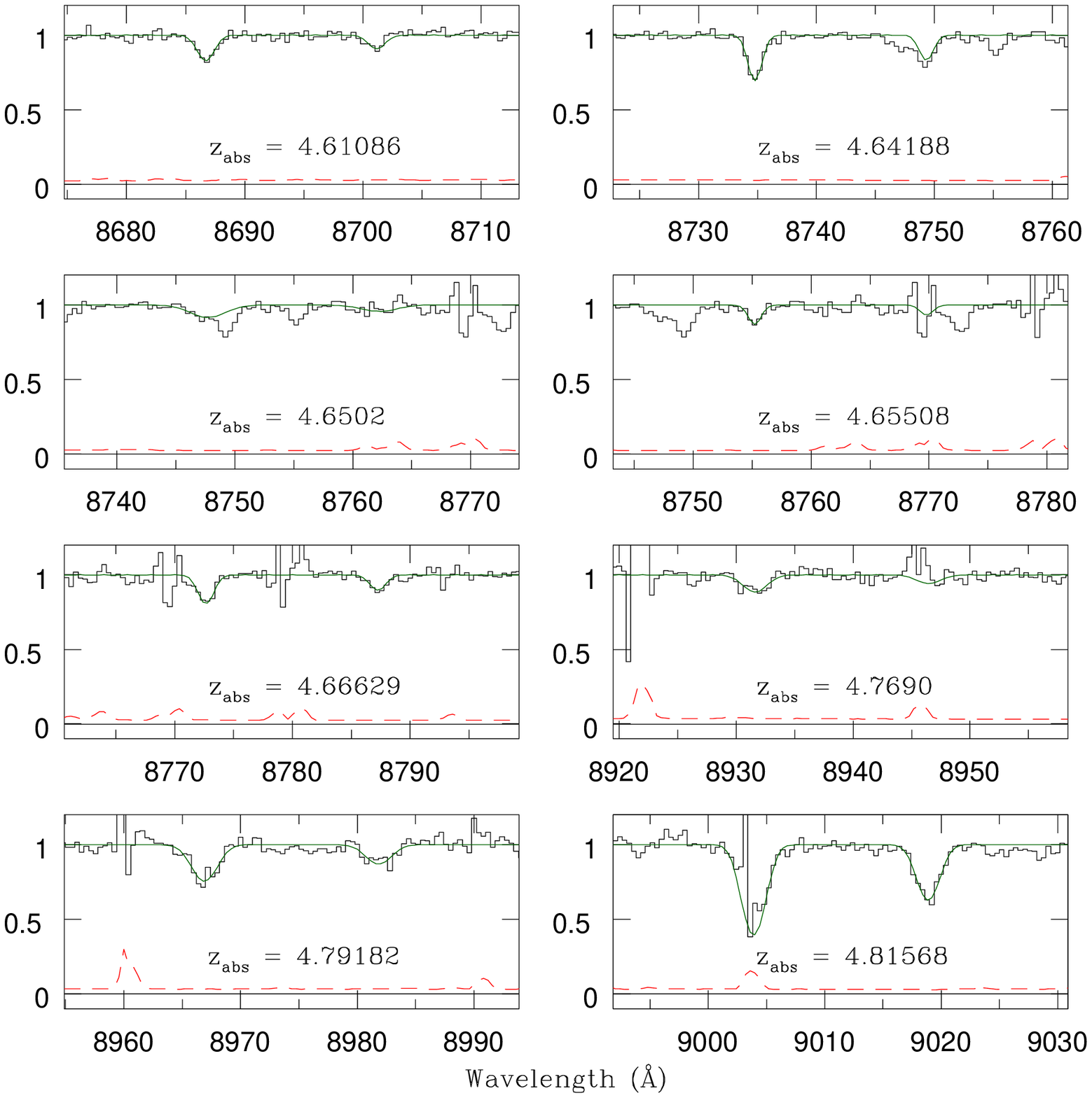}
\caption{\CIV| absorption systems in the spectrum of SDSS J1509-1749. }   
\label{J1509_civz4}
\end{center}
\end{figure*}
Eight \CIV\ doublets have been identified at $z_{\rm abs}=4.61086$, 4.64188, 
4.6501, 4.65508, 4.66629, 4.7690, 4.79182 and 4.81568. 
All \CIV\ doublets are shown in Fig.~\ref{J1509_civz4}. 

\begin{figure}
\begin{center}
\includegraphics[width=9cm]{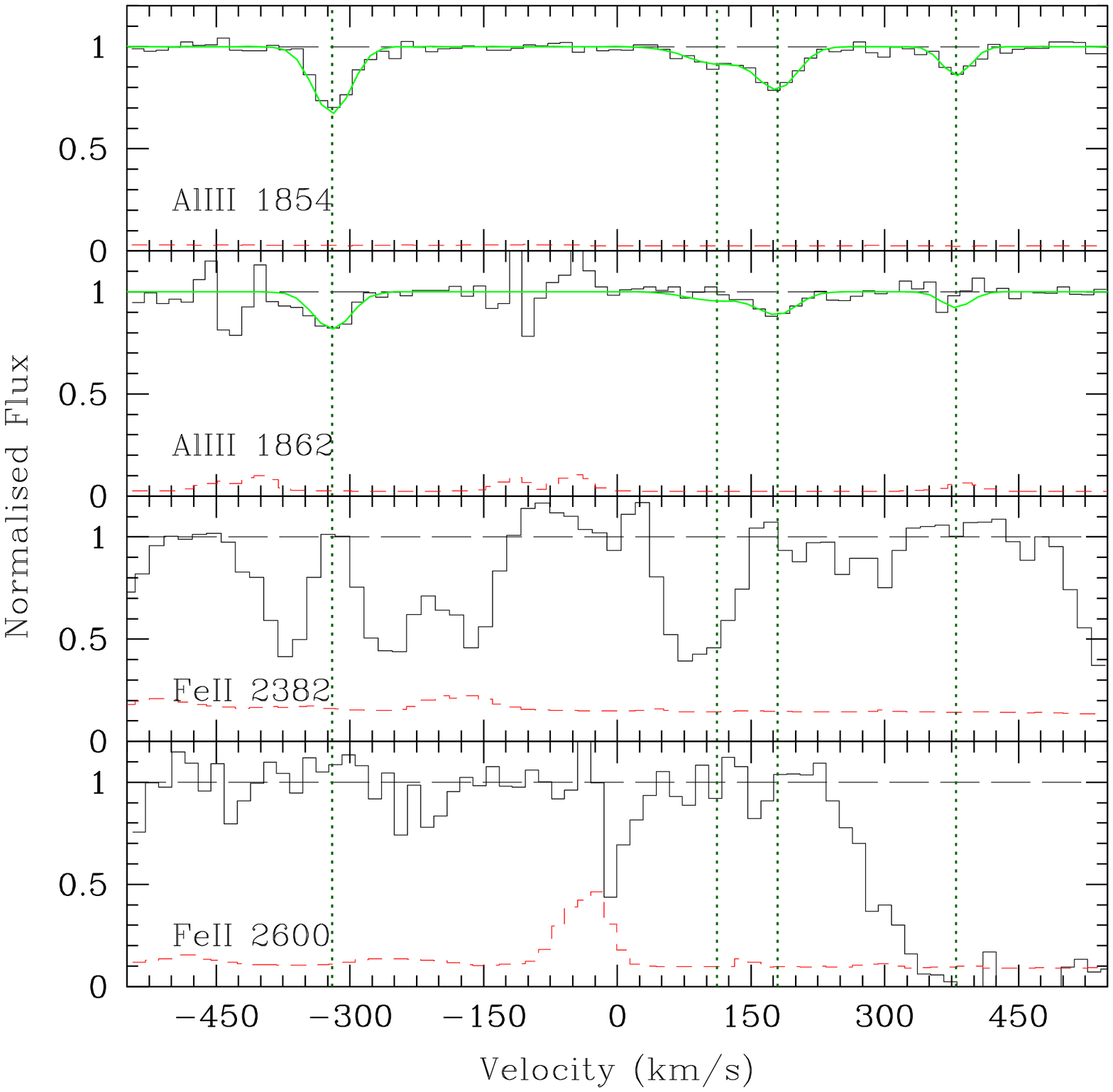}
\caption{Possible complex \AlIII\ doublet at $z_{\rm abs}=3.716$ in the spectrum of SDSS J1509-1749. }   
\label{J1509_alIIIz3p71}
\end{center}
\end{figure}

We propose also an alternative identification for the \CIV\ doublets at $z_{\rm
  abs}=4.64188$,  4.6501 and 4.66629. After the identification of a
weak \AlIII\ doublet at $z_{\rm abs} = 3.7205$, we realized that the
lines identified as \CIV\ could also be part of a complex \AlIII\ system at
$z_{\rm abs} \sim 3.716$,  
where the first two velocity components happen to have the separation
and the column density ratio of a \CIV\ doublet. The system is plotted
in Fig.~\ref{J1509_alIIIz3p71}. The absence of \FeII\ transitions
could be an argument in favour of the \CIV\ nature of the lines;
however there are no other pieces of strong evidence to make us prefer one or
the other identification.  This alternative identification of the
three \CIV\ doublets does not affect the computation of $\Omega_{\rm
  CIV}$ since the lines all have $\log N($\CIV$) < 13.4$.

\begin{figure}
\begin{center}
\includegraphics[width=9cm]{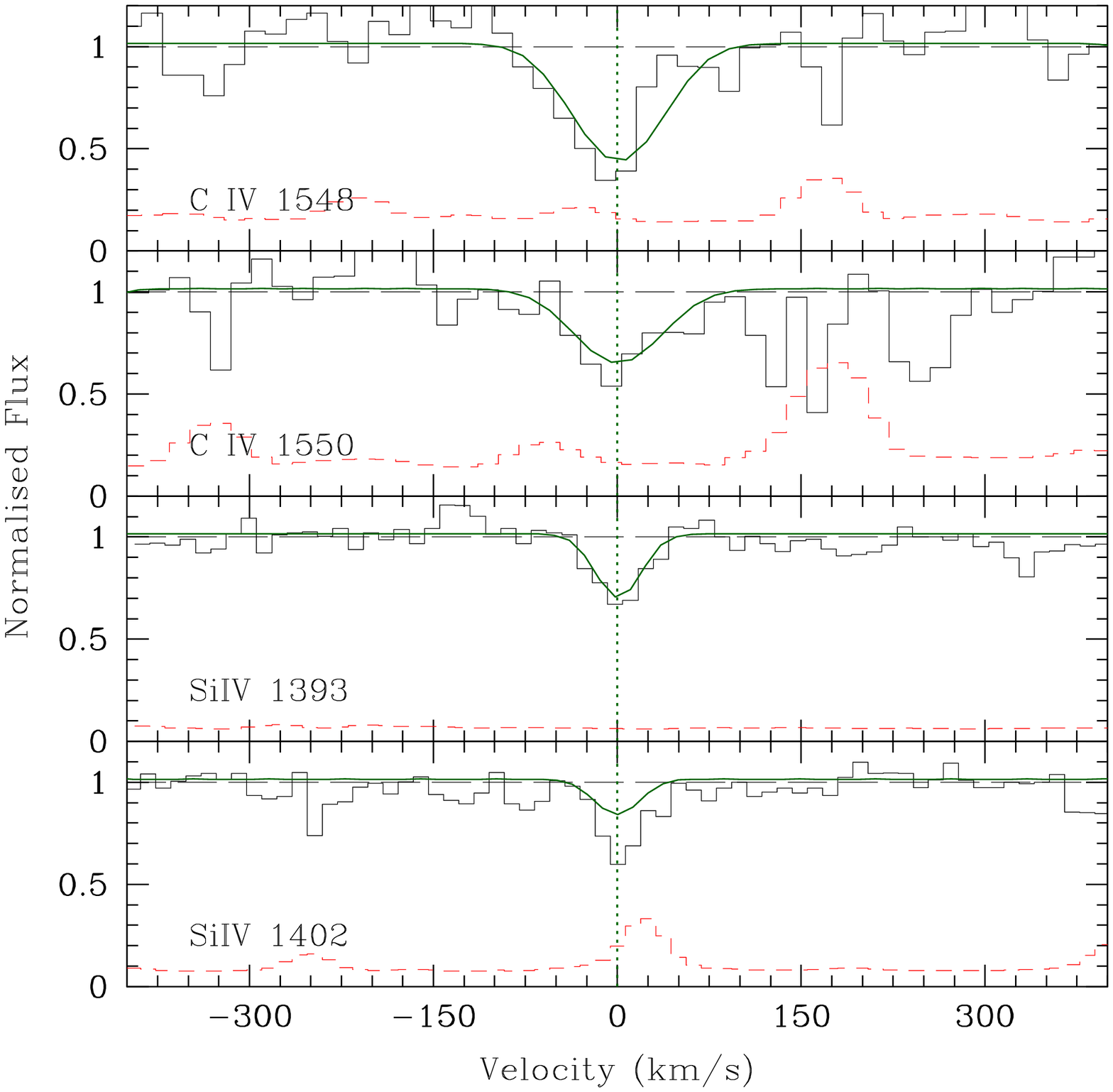}
\caption{Absorption system at $z_{\rm abs} = 5.91572$ in the spectrum of SDSS J1509-1749. }   
\label{J1509_civz5p91}
\end{center}
\end{figure}

At $z>5$, the presence of the \CIV\ system at $z_{\rm abs} = 5.91572$ claimed by
Simcoe11 is confirmed in our NIR spectrum and is
reinforced by the detection of the associated \SiIV\ doublet. The
system is shown in Fig.~\ref{J1509_civz5p91}.

%%%%%%%%%%%%%%%%%%%%%%%%%%%%%%%%%%%%%%%%%%%%%%%%%%%%%%%%%%%%%%%%%%%%%%%%%
%Da discutere: sistemi in J0818+1722
% Sistemi con AlIII a z=3.7 e rapporto con FeII (cambio della shape
% dello ionizing flux?)
% 
%\section{Low-ionization systems at $z>5$}

%%%%%%%%%%%%%%%%%%%%%%%%%%%%%%%%%%%%%%%%%%%%%%%%%%%%%%%%%%%%%%%%%%%%%%%%%

%\begin{table}
%\begin{center}
%\caption{\OI\ Absorption systems in J0818+17}
%\begin{minipage}{80mm}
%\label{tab_obs}
%\begin{tabular}{@{} l c c c}
%\hline  
%Ion & $z$ & $b$ & $\log N$   \\
%& & (km s$^{-1}$) & (cm$^{-2}$) \\
%\hline
%\SiII\ 1260 & & & \\
%\OI\ 1302 & & & \\
%\CII\ 1334 & & & \\
%
%\hline
%\end{tabular}
%
%\end{minipage}
%\end{center}
%\end{table}
%%%%%%%%%%%%%%%%%%%%%%%%%%%%%%%%%%%%%%%%%%%%%%%%%%%%%%%%%%%%%%%%%%%%%%%%%

\section{UVES spectra}

%%%%%%%%%%%%%%%%%%%%%%%%%%%%%%%%%%%%%%%%%%%%%%%%%%%%%%%%%%%%%%%%%%%%%%%%%
In this section, we briefly report on the analysis of the high resolution spectra of SDSS
J1030+0524, SDSS J1306+0356 and SDSS J1044-0125 obtained with the UV and
Visual Echelle Spectrograph \citep[UVES,][]{dekker} at the ESO VLT
telescope. They were all observed by our group but the data have not
been published before.  The journal of observations is reported in
Table~\ref{uves_obs}. 

The  individual spectra were reduced and wavelength calibrated by
using the UVES pipeline \citep{UVESpipe}. Wavelengths were then corrected to the vacuum
heliocentric reference system.     
Fast rotators stars observed the same nights as the QSO were also
reduced in order to identify telluric absorptions in QSO spectra.
The final spectrum is computed as a  weighted mean of the single
frames. The SNR is in general quite low for all spectra,
however the higher resolution allowed us to exclude the presence of
some of the weakest systems detected in the X-shooter spectra. 

\begin{table}
%\begin{center}
\caption{UVES observations.  }
%\begin{minipage}{80mm}
\label{uves_obs}
\begin{tabular}{@{} l l l r r c}
\hline  
QSO & $z_{\rm em}$ & $J_{\rm mag}$ & $T_{\rm exp}$ & $\lambda_{\rm cen}$  & $\Delta\,z_{\rm CIV}$\\
&& & (s) & (\AA) &\\
\hline
J1030+0524 & 6.308 & 18.87 & 15300 & 7940 & $4.757-5.394$ \\
             & & & 22600 & 814 &\\
J1306+0356 & 6.016 & 18.77 & 12219 & 7940 & $4.527-5.459$\\
        & & & 6300 & 804 &\\
         & & & 4277 & 814 &\\
         & &  &9900 & 826 &\\
%         & & & 30994 & 860 (2x3)\\
         & & & 28500 & 860 &\\
J1044-0125 & 5.7824 & 18.31 & 5400 & 7940 & $4.343-5.365$\\
        & & & 30270 & 860 &\\
\hline
\end{tabular}
%\end{minipage}
%\end{center}
\end{table}

%-------------------------------------------------------
\subsection{SDSS J1030+0524}
%------------------------------------------------------- 
We used all the seven spectra existing in the UVES archive, obtained
in six consecutive nights at  the end of April and beginning of 2002 May. 
Only the red arm was used for these observations at central
wavelengths $\lambda\lambda7940$ and 8140 \AA,  with slit width of 1 arcsec, and
binning of $2\times2$. 
%The spectra were recorded  with slit width of 1 arcsec, and binning 2x2  for a total exposure time of 4.25h and 6.25h in the 794 nm and 814 nm settings respectively.   
 
We looked for metals in the region $8760 < \lambda < 9900$ \AA, outside
the \Lya\ forest and
common to the two settings. 
%We only used the spectra registered with the up CCD   with coverages
%8000-9900 \AA and  8196-9921 \AA for the 794 nm and 814 nm settings respectively, since the flux is observed at  lambda  $>$ 8760 \AA
%We then end up with a final spectrum obtained by  averaging  the
%common spectral regions of both settings.  
The SNR varies from 20 to 5 in the reddest part of the spectrum. 

Inspection of the high-resolution spectrum of J1030+0524 did not
confirm the presence of two systems detected in the X-shooter spectrum: the
\CIV\ doublets at $\zabs = 4.76671$ and $4.89066$.

%-------------------------------------------------------
\subsection{SDSS J1306+0356}
%-------------------------------------------------------

We used two sets of UVES spectra available in the archive, obtained in
2002 and 2007. All the spectra were recorded with binning of $2\times2$ and slit
width of 1 arcsec.  Six spectra are from 2002 with a total exposure
time of 8.6h, and central wavelengths $\lambda\lambda7940$, 8140, 8260, 8040 \AA. 
Ten spectra are from 2007 with a total exposure time of 7.9 h recorded
at central wavelength $\lambda$ 8600 \AA. 
%For each set of data 
%the  individual spectra were reduced and wavelength calibrated by using the UVES pipeline.
%Then the spectra were corrected to the heliocentric and vacuum scale,
%and finally   averaged   to obtain a weighted mean final spectrum for
%each set of data. 
In the final spectrum, the region outside the \Lya\ forest ranges
between 8430 and 10000 \AA. 
For the analysis, we used the average spectra of both sets of data for
$\lambda <  9550$ \AA, with SNR varying from $\sim 24$ to 10 in the
interval $\lambda$$\lambda8725-9550$ \AA, respectively. 
For $\lambda$ $> $ 9550 \AA, only the average of the 2002
observations was used with SNR varying between 8 and 5 for the reddest
wavelengths, because the spectra taken in 2007 have 
very poor quality with SNR~$\sim2$. 

The presence of the \CIV\ doublet at $\zabs = 4.82048$, detected in the X-shooter
spectrum, was not confirmed in this spectrum.  

%-------------------------------------------------------
\subsection{SDSS J1044-0125}
%-------------------------------------------------------
This quasar was excluded from the sample collected with the X-shooter
spectrograph because it is a BAL object. 

In the UVES archive there are six spectra obtained in March and April
of 2002 with the red arm at central wavelength $\lambda8600$ \AA, with slit width
of 1 arcsec and binning of $2\times2$.

The final spectrum covers the ranges $\lambda\lambda6710-8520$ and $8669-9855$
\AA, even though the flux is not zero at $\lambda > 8150$ \AA. The quality
of the final spectrum is quite poor with an SNR of the order of 10.  
  
A single feature is observed which was identified as \CII\ 1334
\AA\ at $\zabs \simeq 5.2847$. No other associated transitions  
were detected in the UVES spectrum. 

\citet{song_cowie02} claimed the detection of the \FeII\ 1608
\AA\ line at the same redshift in the ESI spectrum and derived 
an estimate of the metallicity of [Fe/H] $= -2.65$. 
On the other hand, no absorption lines were detected in the higher
resolution spectrum obtained with MIKE@Keck by \citet{becker11}.

%%%%%%%%%%%%%%%%%%%%%%%%%%%%%%%%%%%%%%%%%%%%%%%%%%%%%%%%%%%%%%%%%%%%%%%%%

\section{\CIV\ line statistics}

%%%%%%%%%%%%%%%%%%%%%%%%%%%%%%%%%%%%%%%%%%%%%%%%%%%%%%%%%%%%%%%%%%%%%%%%%
% Compute CDDF 
\subsection{ The \CIV\ CDDF}

The CDDF, $f(N)$, is defined as
the number of lines per unit column density and per unit redshift
absorption path, d$X$ \citep{tytler87}. The CDDF is a fundamental
statistics for absorption lines, similar for many aspects to the
luminosity function for stars and galaxies.   

The redshift absorption path is used to remove the redshift dependence
in the sample and put everything on a comoving coordinate scale. In the assumed
cosmology, it is defined as

\begin{equation}
{\rm d}X \equiv (1+z)^2 [ \Omega_{\rm m}(1+z)^3 + \Omega_{\Lambda}]^{-1/2}
{\rm d}z.
\end{equation}  

With the adopted definition, $f(N)$ does not evolve at any redshifts
for a population whose physical size and comoving space density are
constant. 

\begin{figure}
\begin{center}
\includegraphics[width=9cm]{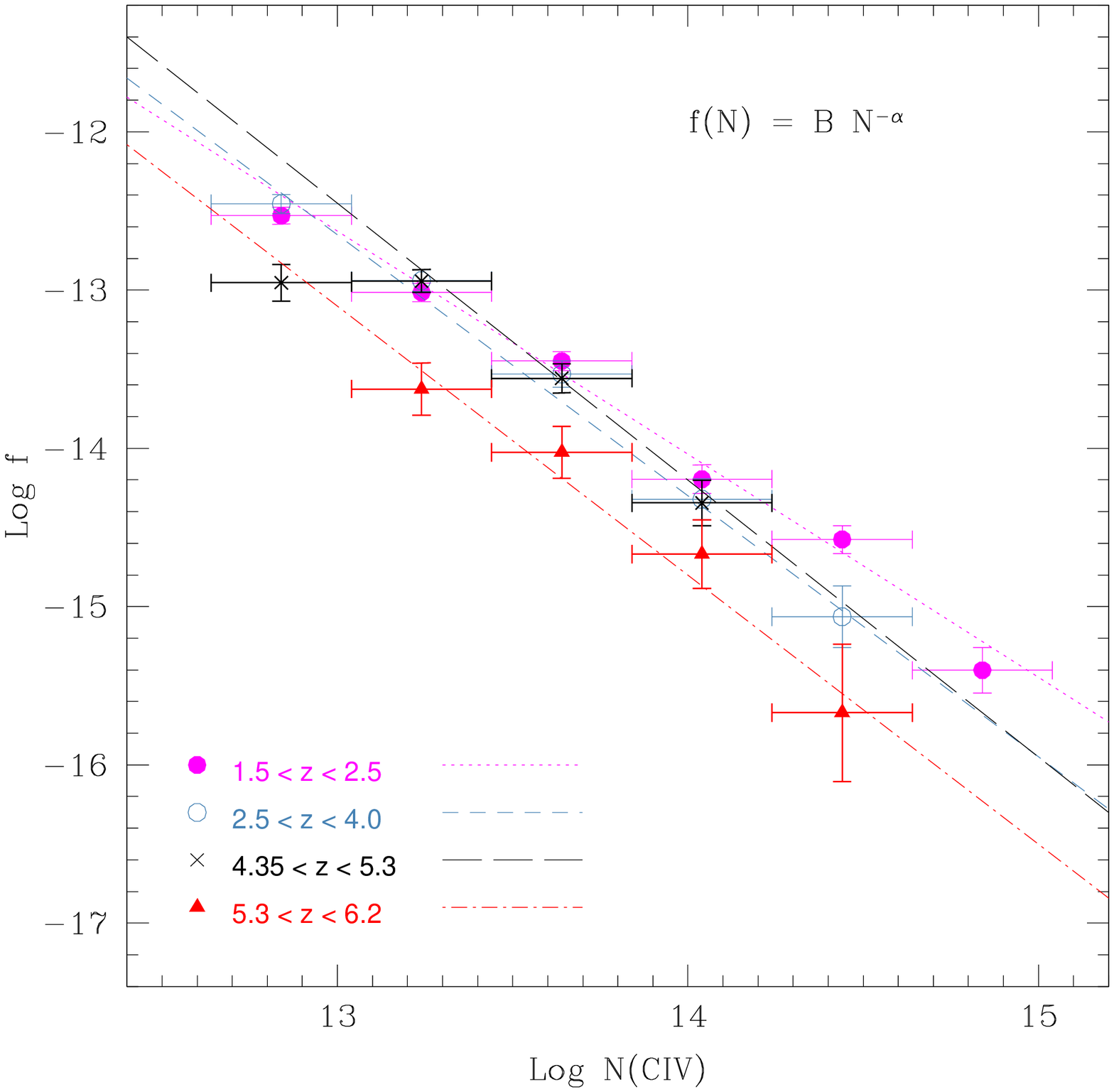}
\caption{Comparison of the CDDF of
  \CIV\ systems with $12.6 < \log N($\CIV$) < 15$ in our sample ($4.35
  < z < 6.2$) and in the lower redshift sample by \citet{dodorico10}
  ($1.5 < z < 4$) where \CIV\ lines closer than $50$ \kms\ have been
  merged. Both redshift ranges were split into two smaller intervals:
  see the legend for the correspondence between the symbol and redshift
  interval. 
% {\sl Black open dots}: \CIV\ systems at 
%  $1.5 < z < 2.5$; {\sl Red solid dots}: \CIV\ systems at
%  $2.5 < z < 4$; {\sl Magenta crosses}: \CIV\ lines at $4.35 < z < 5.3$;
%        {\sl Cyan triangles}: \CIV\ lines at $5.3 < z < 6.2$
  The bin size in $\log N($\CIV$)$ cm$^{-2}$ is 0.4 and the error bars
  are $\pm1\,\sigma$ based on the number of points in each bin. 
  The lines are power laws of the form $f(N) = BN^{-\alpha}$ fitting
  the four samples with indices: $\alpha=1.41\pm0.07$, $1.65\pm0.09$,
  $1.75\pm0.1$ and $1.7\pm0.2$ in order of
  increasing redshift. See the legend for the correspondence between
  the line type and redshift interval.}   
\label{comp_cddf}
\end{center}
\end{figure}

\begin{figure}
\begin{center}
\includegraphics[width=9cm]{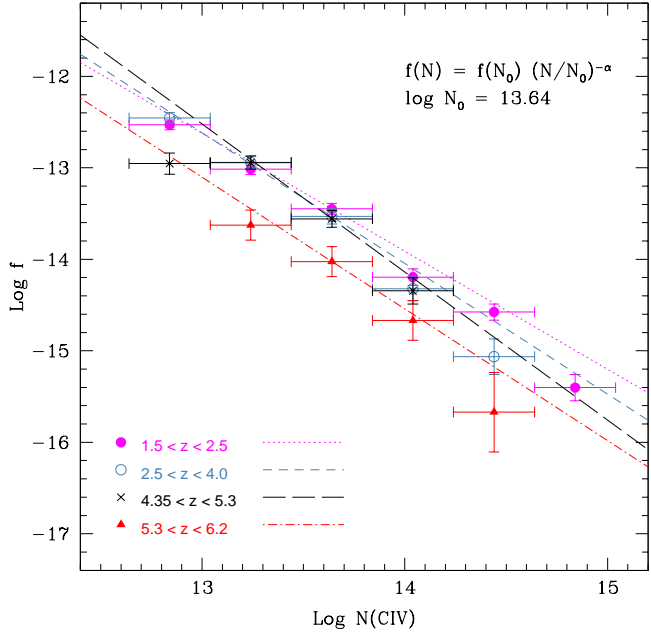}
\caption{Same as Fig.~\ref{comp_cddf} where the different redshift
  samples have been fitted with a power law of the form $f(N) =
  f(N_0)(N/N_0)^{-\alpha}$ and $N_0 = 10^{13.64}$. 
  In this case the exponential indices are $\alpha=1.29\pm0.05$,
  $1.43\pm0.06$, $1.62\pm0.2$ and $1.44\pm0.3$  in order of
  increasing redshift. See the legend for the correspondence between
  the symbol type, line type and redshift interval.}   
\label{comp_cddf_pivot}
\end{center}
\end{figure}

\FloatBarrier

We have computed $f(N)$ splitting our data into two redshift bins in
order to have comparable $\Delta X_{\rm tot}$ in both of them:
$4.35 < z < 5.3$ ($\Delta X_{\rm tot}=20.893$) and $5.3 < z < 6.2$ ($\Delta
X_{\rm tot}=19.526$), considering column densities $12.6 \le \log N($\CIV$) \le 15$. 
The possible evolution with redshift of the CDDF has been verified by
comparing $f(N)$ for the present data with $f(N)$ for the lower
redshift sample in \citet{dodorico10}. Due to the higher resolution of
the spectra in the latter work, \CIV\ lines with velocity separation
smaller than $50$ \kms\ have been merged into \CIV\ systems. 
 
The result is shown in Fig.~\ref{comp_cddf}, where $f(N)$ is plotted
for four redshift intervals: $1.5 < z < 2.5$,  $2.5 < z < 4$, $4.35 < z <
5.3$ and $5.3 < z < 6.2$ binned at $10^{0.4}\, N($\CIV$)$ cm$^{-2}$.  
The two CDDFs at higher redshift are complete starting from larger column
densities, and a constant decrease with increasing
redshift is observed in the bin corresponding to $14.2 < \log
N($\CIV$) < 14.6$. 

To quantify the redshift evolution of the CDDF, we have computed a
least-squares fit of the four samples with a power law of the form $f(N)
= BN^{-\alpha}$ (see Fig.~\ref{comp_cddf}). From
the fit, we observe that in the highest redshift bin the CDDF is lower
by a factor of $\sim 3-4$ with respect to the CDDF at $2.5 < z < 4$ and $4.35 < z <
5.3$. This is in good agreement with the result by
\citet{becker09} which was based on NIRSPEC spectra at $R\approx
13000$ of four $z\sim6$ QSOs, two of which are also in our sample. 
The CDDF in the lowest redshift bin has a different
slope due to the presence of a larger number of absorption systems in
the column density range $\log N($\CIV$) > 14.2$. At $\log N($\CIV$) =
14.4$ there is a factor of $7.5$ difference between the CDDF at $1.5 < z
< 2.5$ and that at $5.3 < z < 6.2$. 
Due to the small number of points, the power law fit with two
variables gives very large errors, in particular for the power-law
normalization (highly correlated with the index errors). To obtain
more solid results, we have tried also a power-law fit with a pivot column
density, $f(N) = f(N_0) (N/N_0)^{-\alpha}$, where we chose $\log
N_0=13.64$. The results of this fit are shown in
Fig.~\ref{comp_cddf_pivot}. The power-law indices are lower than in
the previous fit, although always consistent within $3\,\sigma$. The
net decrease of the highest redshift CDDF is less evident but
still present: there is a factor of $\sim 2-3$ difference between the
CDDF in the redshift bins  $4.35 < z < 5.3$ and $5.3 < z < 6.2$.

%The statistics in the highest
%redshift bin, $5.3 < z < 6.2$, is still quite poor (19 lines spread in
%4 bins), however the fact that all four points are low adds
%significance to the result.   

%*****************************************************************
\subsection{The redshift evolution of the cosmic mass density of \CIV}
%*****************************************************************

\begin{table}
\begin{center}
\caption{Values of $\Omega_{\rm CIV}$ for the X-shooter sample in two
  column density ranges. }
\label{tab_omega}
\begin{tabular}{c r r c c c }
\hline  
$z$ range & $z_{\rm med}$ & $\Delta X$ & Lines & $\Omega_{\rm CIV}$  & $\delta
\Omega$  \\ 
 & & & & ($\times 10^{-8}$) & ($\times 10^{-8}$) \\
\hline
\multicolumn{3}{c}{ $13.4 \leq \log N($\CIV$) \leq 15$} & & & \\
$4.35-5.30$ & 4.818 & 19.21&  30 & 1.4 & 0.3 \\
$5.30-6.20$ & 5.634 & 19.53 & 12 & 0.84 & 0.33  \\
&&&&& \\
\multicolumn{3}{c}{ $13.8 \leq \log N($\CIV$) \leq 15$} & & & \\
$4.35-5.30$ & 4.802 & 19.21 &   9 & 0.65 & 0.22 \\
$5.30-6.20$ & 5.706 & 19.53 &   5 & 0.61 & 0.32 \\
\hline
\end{tabular}
\end{center}
\end{table}

The CDDF can be integrated in order to
obtain the cosmological mass density of \CIV\ in QSO absorption
systems as a fraction of the critical density today:

\begin{equation}
\label{omega}
\Omega_{\rm CIV} = \frac{H_0\, m_{\rm CIV}}{c\, \rho_{\rm crit}} \int N
f(N) {\rm d}N, 
\end{equation}
where $H_0 = 100\, h$ \kms Mpc$^{-1}$ is the Hubble constant, $ m_{\rm
  CIV}$ is the mass of a \CIV\ ion, $c$ is the speed of light,
$\rho_{\rm crit} = 1.88 \times 10^{-29} h^2$ g cm$^{-3}$ and $f(N)$ is
the CDDF.  
The above integral can be approximated by the sum: 

\begin{equation}
\label{omega_approx}
\Omega_{\rm CIV} = \frac{H_0\, m_{\rm CIV}}{c\, \rho_{\rm crit}}
\frac{\sum_i  N_i (\mbox{\CIV})}{\Delta X},
\end{equation}
with an associated fractional variance:

\begin{equation}
\label{omega_err}
\left( \frac{\delta \Omega_{\rm CIV}}{\Omega_{\rm CIV}} \right)^2 = \frac{\sum_i  [N_i (\mbox{\CIV})]^2}{\left[\sum_i  N_i (\mbox{\CIV})\right]^2}
\end{equation}
as proposed by \citet{storrie96}. Note that the errors determined with this
  formula could be underestimated, in particular in the case of
  small line samples. In \citet{dodorico10}, we found that errors
  on  $\Omega_{\rm CIV}$ computed with a bootstrap technique were, at maximum, a
  factor of $\sim 1.5$ larger than those estimated with equation
  (\ref{omega_err}). For a fair comparison with previous results,
  however, we report in Table~\ref{tab_omega} the errors computed 
  with equation (\ref{omega_err}).
 
Great care has to be taken when comparing results obtained from
different samples, since the values of $\Omega_{\rm CIV}$
significantly depend on the column density range over which the sum or
the integration is carried out, and as a consequence on the
resolution and SNR of the available spectra. 
To take this aspect into account, we have computed two sets
of values to be compared consistently with different data in the
literature: the first with $13.4 \le \log N($CIV$) \le 15$ and the
second with $13.8 \le \log N($CIV$) \le 15$.  
The final results (reported in Table~\ref{tab_omega}) are not
corrected for completeness. We have computed the correction to
$\Omega_{\rm CIV}$ due to the undetected \CIV\ lines with column
densities $\log N($CIV$) \sim 13.4-13.5$ in the redshift bin $z =
5.3-6.2$, which results to be negligible, of the order of $\sim 6 \times 10^{-10}$. 

\begin{figure*}
\begin{center}
\includegraphics[width=8.5cm]{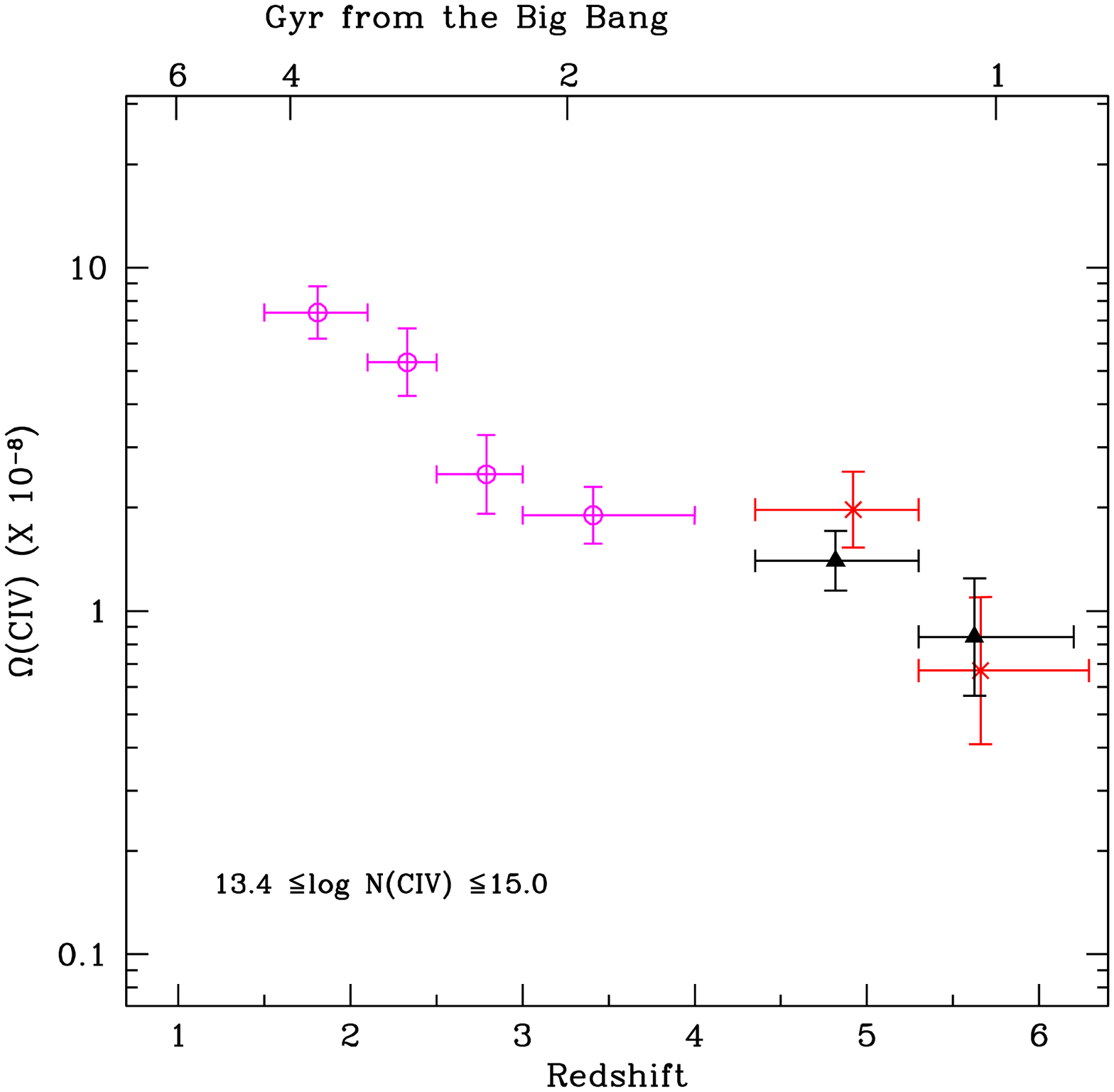}
\includegraphics[width=8.5cm]{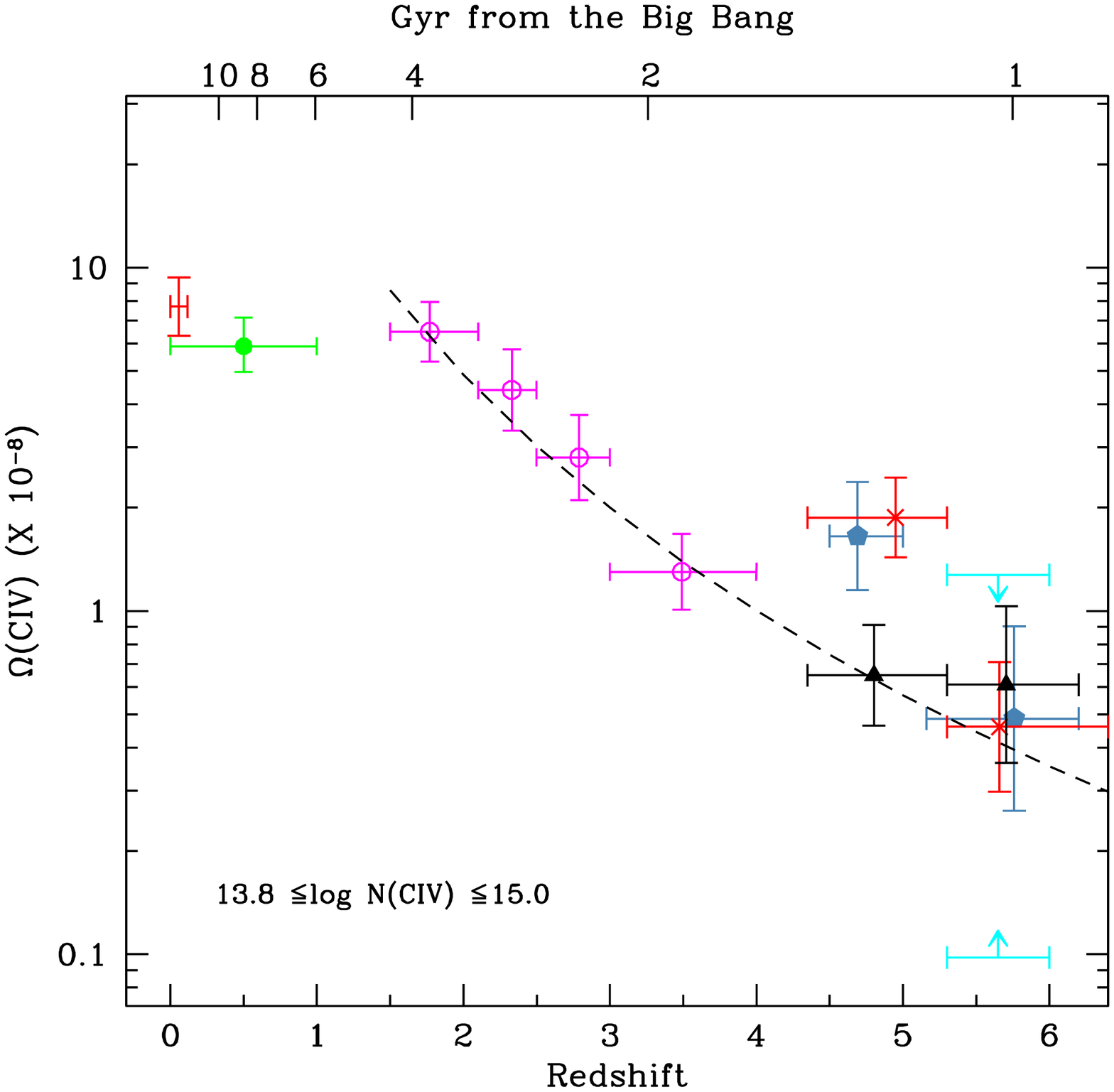}
\caption{Evolution with redshift of the \CIV\ cosmic mass density,
  $\Omega_{\rm CIV}$, computed for two column density ranges as
  specified in the figures. Left-hand panel: data points refer to this work 
  (black triangles), Simcoe11 (red crosses) and
  \citet[][magenta open dots]{dodorico10} for the \CIV\ lines in the
  desired column density range.    
   Right-hand panel: data points refer to this work (black
   triangles),  Simcoe11 (red crosses)
   Pettini03 and \citet[][blue pentagons]{ryanweber09},
   \citet[][cyan upper and lower limit 95 per cent confidence interval]{becker09}
   \citet[][magenta open dots]{dodorico10}, \citet[][green
     cross]{cooksey} and \citet[][red plus]{danforth:shull}. The
   dashed line denotes a fit to points with $z>1.5$ (considering only our
   measurements in the range $z > 4.0$) of the form $\Omega_{\rm CIV} = (2\pm1)\times
   10^{-8} [(1+z)/4]^{-3.1\pm0.1}$. The error
   bars are $1\,\sigma$.}    
\label{omegaciv}
\end{center}
\end{figure*}

%\begin{figure}
%\begin{center}
%
%\caption{Estimates of $\Omega_{\rm CIV}$ in the redshift range
%  $z\simeq 0-6.2$ for \CIV\ systems with $13.8 \le \log N($\CIV$) \le
%  15$: {\it red plus} \citet{danforth:shull}; {\it green cross}
%  \citet{cooksey}; {\it magenta open dots} \citet{dodorico10};{\it
%    black solid triangles} this work; {\it
%    blue solid penthagons} \citet{pettini03} and
%  \citet{ryanweber09};  {\it 95 \% confidence interval} \citet{becker09}}.
%\label{omegaciv_13p8}
%\end{center}
%\end{figure}

The evolution with redshift of $\Omega_{\rm CIV}$ is shown in the
  two panels of Fig.~\ref{omegaciv}. The determination based on our
  X-shooter sample has been split into two redshift bins $4.35 \le z
  \le 5.3$ and $5.3 \le z \le 6.2$, to make a reliable comparison with
  the other most recent determination of $\Omega_{\rm
  CIV}$ at high redshift by Simcoe11. 
The behaviour of $\Omega_{\rm CIV}$ in the redshift range $1.5
\le z \le 4$ was determined with the high-resolution sample by
\citet{dodorico10}. 
In the left plot, where $13.4 \le \log N($CIV$) \le 15$, our measurement is compared with the result obtained
by Simcoe11 using only the FIRE spectra, which have the same degree of
completeness as our observations. We have not used their enlarged
sample for which the completeness column density limit is not well
defined. 
The two estimates are in very good agreement (within one sigma) as was expected since the two samples are almost identical.  
In the whole inspected redshift range, $\Omega_{\rm CIV}$ slowly
increases from $z \sim6.2$ to lower redshifts, and then it increases more
steeply with a factor of $\sim 4$ difference between $z\sim3$ and $z\sim1.5$.

The determination of $\Omega_{\rm CIV}$ was extended
also to very low redshift \citep[$z<1$,][]{danforth:shull,cooksey}
using UV spectra at lower resolution and lower SNR limiting
the detectability of \CIV\ lines to larger column densities.
In order to study the evolution of  $\Omega_{\rm CIV}$ in the whole
redshift range between 
$z\sim0$ and 6.2, we have carried out a second computation of the values
of  $\Omega_{\rm CIV}$ for \CIV\ systems in the column density range $13.8 \leq \log
N($\CIV$) \leq 15$. The results  are shown in the right plot of
Fig.~\ref{omegaciv} where we  
have added also the results by \citet{dodorico10} and the older
determinations at high redshift by  Pettini03 (corrected for the
considered column density range), \citet{ryanweber09} and
\citet{becker09}. In this case, we compare our estimate of
$\Omega_{\rm CIV}$ with the result obtained by Simcoe11 for their
enlarged sample. 
Our point in the highest redshift bin is in agreement with all previous
determinations. The value of $\Omega_{\rm CIV}$ in the redshift range
$4.35 \le z \le 5.3$ is slightly lower than the estimate by
Simcoe11, resulting in a smoother increase of the mass density towards
lower redshifts. The discrepancy between our measurements and the
result by Simcoe11 is due, in our opinion, to the fact that in
Simcoe11 the two estimates at $4.35 \le z \le 5.3$ and at $5.3 \le z
\le 6.2$ have different completeness limits. In the lower redshift
range the FIRE spectra are complete down to $N($\CIV$) \sim 13.4$,
while the enlarged sample in the higher redshift bin is likely complete to
higher column densities. 

In summary, we observe a slow increase of the \CIV\ content from
$z\sim6$ to $z\sim1.5$ and then a flattening towards $z\sim0$. Our data at high redshift together with the data points of \citet{dodorico10} were fitted with the function: $\Omega_{\rm CIV} = (2\pm1)\times
   10^{-8} [(1+z)/4]^{-3.1\pm0.1}$ (see the right-hand panel of Fig.~\ref{omegaciv}). Since we
expect that these strong lines arises in the halo of star forming
galaxies \citep[e.g.][]{adelberger}, what we observe could be the
effect of the progressive enrichment due to in situ star formation as
the star formation rate density increases with time and then decreases
below redshift $\sim1$.   

%%%%%%%%%%%%%%%%%%%%%%%%%%%%%%%%%%%%%%%%%%%%%%%%%%%%%%%%%%%%%%%%%%%%%%%

\section{Redshift evolution of ionic ratios}

%%%%%%%%%%%%%%%%%%%%%%%%%%%%%%%%%%%%%%%%%%%%%%%%%%%%%%%%%%%%%%%%%%%%%%%%%%

%The strength and spectrum of the metagalactic ionizing radiation
%background at high redshift and the nature of the sources which
%ionized the IGM are outstanding issues in cosmology. 
It is generally thought that the IGM is kept ionized by the integrated
UV emission from active nuclei and star-forming galaxies, but the relative
contributions of these sources as a function of epoch are poorly
known. 
%Because of the high ionization threshold (54.4 eV) and
%small photoionization cross section of He ii, and of the rapid
%recombination rate of He iii, the double ionization of helium is
%expected to be completed by hard UV-emitting quasars around
%the peak of their activity at  (e.g., Madau & Meiksin
%1994; Sokasian et al. 2002; McQuinn et al. 2009), much later
%than the reionization of H i and He i. 
At $z > 3$, the declining population of bright quasars appears to make
an increasingly small contribution to the 1 Ryd radiation background
\citep[e.g.][]{bianchi+01}, and it is believed that massive stars in
galactic and sub-galactic systems may provide the additional ionizing
flux needed at early times (e.g., Madau et al. 1999; Gnedin 2000;
Haehnelt et al. 2001; Wyithe \& Loeb 2003; Meiksin 2005;
Faucher-Gigu\`ere et al. 2008; Robertson et al. 2010). This idea may
be supported by the detection of 
escaping ionizing radiation from individual Lyman break galaxies at
$z\sim 3$ \citep[but see][]{vanzella12a,vanzella12b}. 

The spectral shape of the ultraviolet background (UVB) radiation should be
reflected into the ionization pattern of QSO metal system absorbers. 
In particular, the ionic ratios \SiIV/\CIV\ and \CII/\CIV\ are
sensitive to the shape of the high-energy end of the 
ionizing spectrum where the \HeII\ break is expected
\citep{giroux_shull,savaglio97,agafonova07,bolton_viel11}.

\begin{figure}
\begin{center}
\includegraphics[width=9cm]{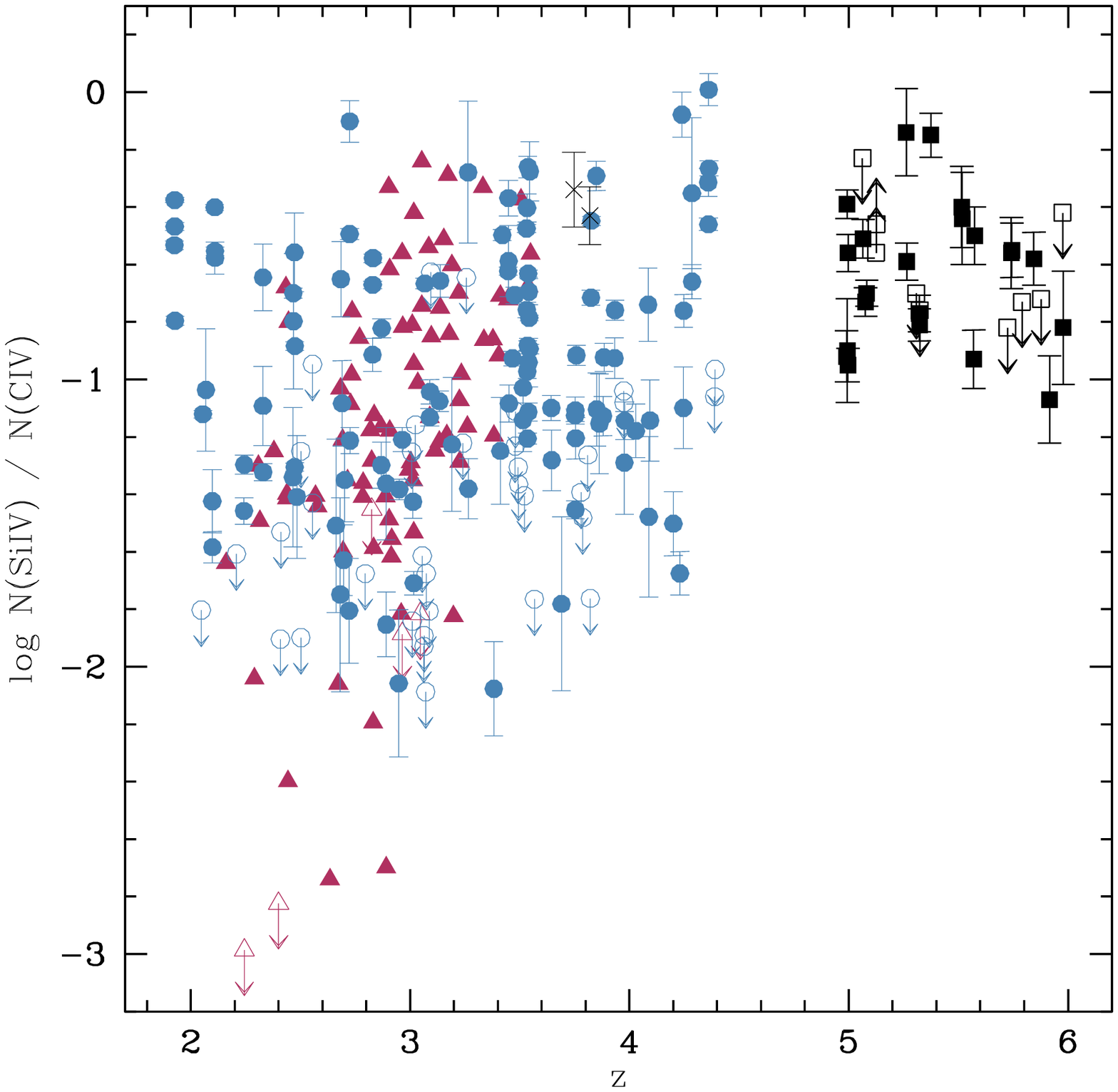}
\caption{Evolution of the \SiIV/\CIV\ column density ratio as a function of
  redshift for the selected systems in our sample (black
    squares). The other data points are from: \citet[][triangles]{songaila98}, \citet[][crosses]{savaglio97} and \citet[][dots]{BSR03}. The open
  symbols with arrows indicate upper or lower limits.} 
\label{si4c4_z}
\end{center}
\end{figure}

In this work, we have detected for the first time a significant sample of \SiIV\ doublets
at $z\gsim 5$. This fact allows us to study the behaviour of the
$N($\SiIV$)/N($\CIV$)$ ratio for all those absorption systems in our
sample for which both \SiIV\ and \CIV\ are outside the 
\Lya\ forest. The result is shown in Fig.~\ref{si4c4_z} where this
ionic ratio is plotted as a function of redshift, together with data
from works at lower redshift \citep{savaglio97,songaila98,BSR03}. In
order to compare the data by \citet{BSR03}, obtained from the fit of high resolution
HIRES@Keck spectra, with the other samples, we merged velocity components
closer than 50 \kms\ before computing the \SiIV/\CIV\ ratios. 
Qualitatively, our X-shooter sample is characterized by large values
and a small dispersion.  
\citet{songaila98}
claimed to observe a jump in the median value of the \SiIV/\CIV\  ratio at
redshift $\sim 3$ that she explained with a variation of the shape of
the UVB spectrum due to the end of the \HeII\  re-ionization
process. This result was not confirmed in the work by
\citet{kim02}. The authors of this work selected \CIV\ systems in
the range $1.6<z<3.6$ on the base of the associated \HI\ absorber in
order to consider only metals associated with \Lya\ forest lines. Also \citet{BSR03}
found that the median values of \SiIV/\CIV\ obtained from summed column
densities in systems are consistent with being constant over the whole
observed range $1.9<z<4.4$.
%In Fig.~\ref{si4c4_z} we report the median values computed for systems
%with $z<3$ and $z>3$ in \citet{songaila98},  for our data. 

\begin{figure}
\begin{center}
\includegraphics[width=9cm]{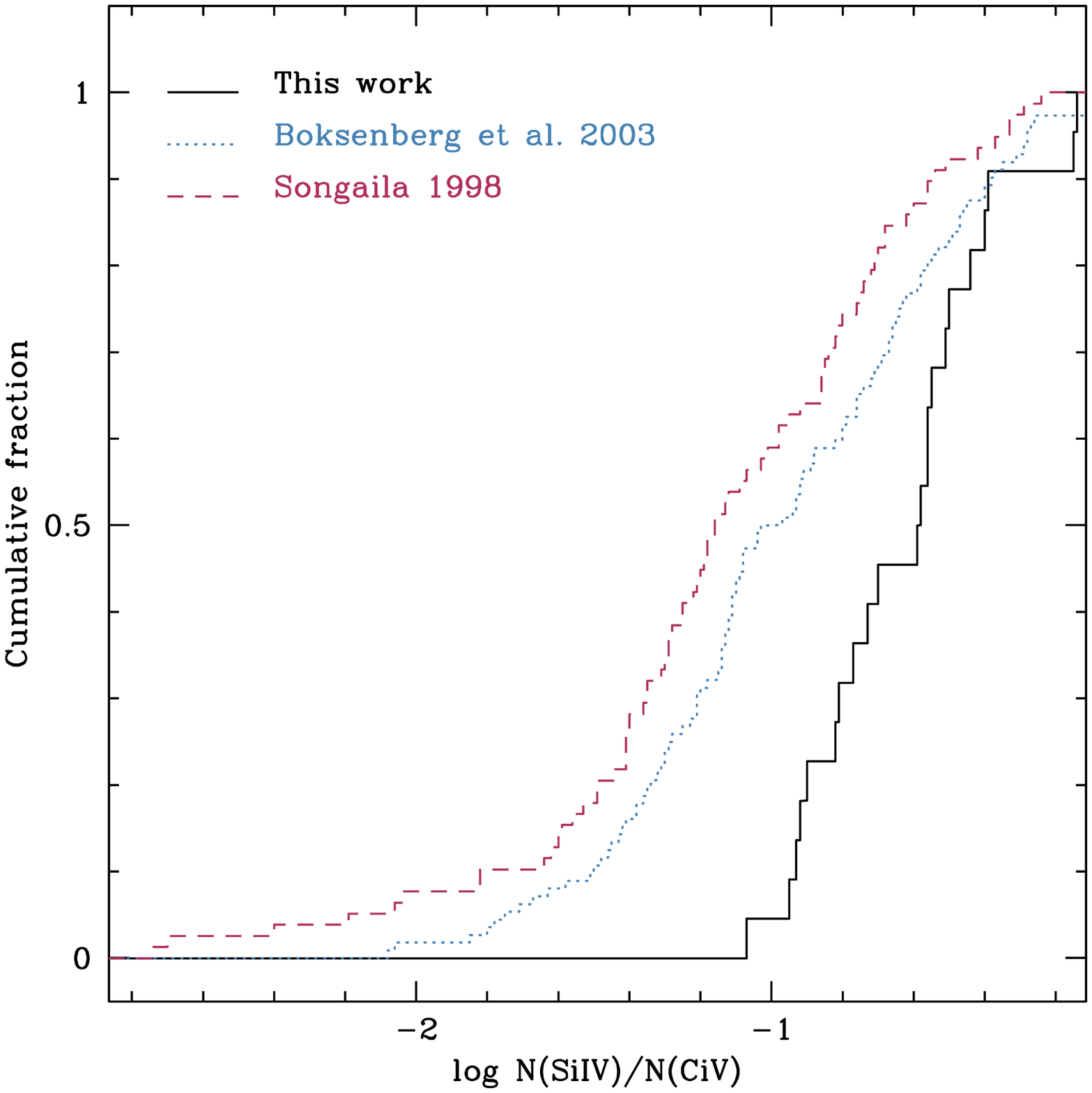}
\caption{Cumulative distributions for the \SiIV/\CIV\  ratios of the
  three main samples used in the analysis: the X-shooter 
  sample at $z\gsim5$ (solid line) and the lower redshift
  samples of \citet[][dashed line]{songaila98} and \citet[][dotted line]{BSR03}.}
\label{kstest}
\end{center}
\end{figure}

A quantitative comparison of our sample with the lower redshift ones
using the Kolmogorov-Smirnov test indicates that the X-shooter high
redshift sample is likely drawn from a different parent distribution. The cumulative
distributions for the \SiIV/\CIV\  ratios of the three main samples are shown in
Fig.~\ref{kstest}. The probability that the X-shooter sample is drawn
from the same distribution of the Songaila et al. sample or of the
Boksenberg et al. one is tiny: $P = 1.3 \times 10^{-5}$ and $3\times
10^{-4}$, respectively. The probability that the two lower redshift
samples are drawn from the same distribution is significantly larger: $P = 6 \times
10^{-2}$.   
%The median ratio for the high redshift systems in our sample is
%consistent at 1\,$\sigma$ with the median value for the systems at
%$z>3$ of \citet{songaila98} while it represents a slight increase with
%respect to the median value measured by \citet{BSR03}.   

\subsection{Photoionization models}

\begin{figure*}
\begin{center}
\includegraphics[width=13cm]{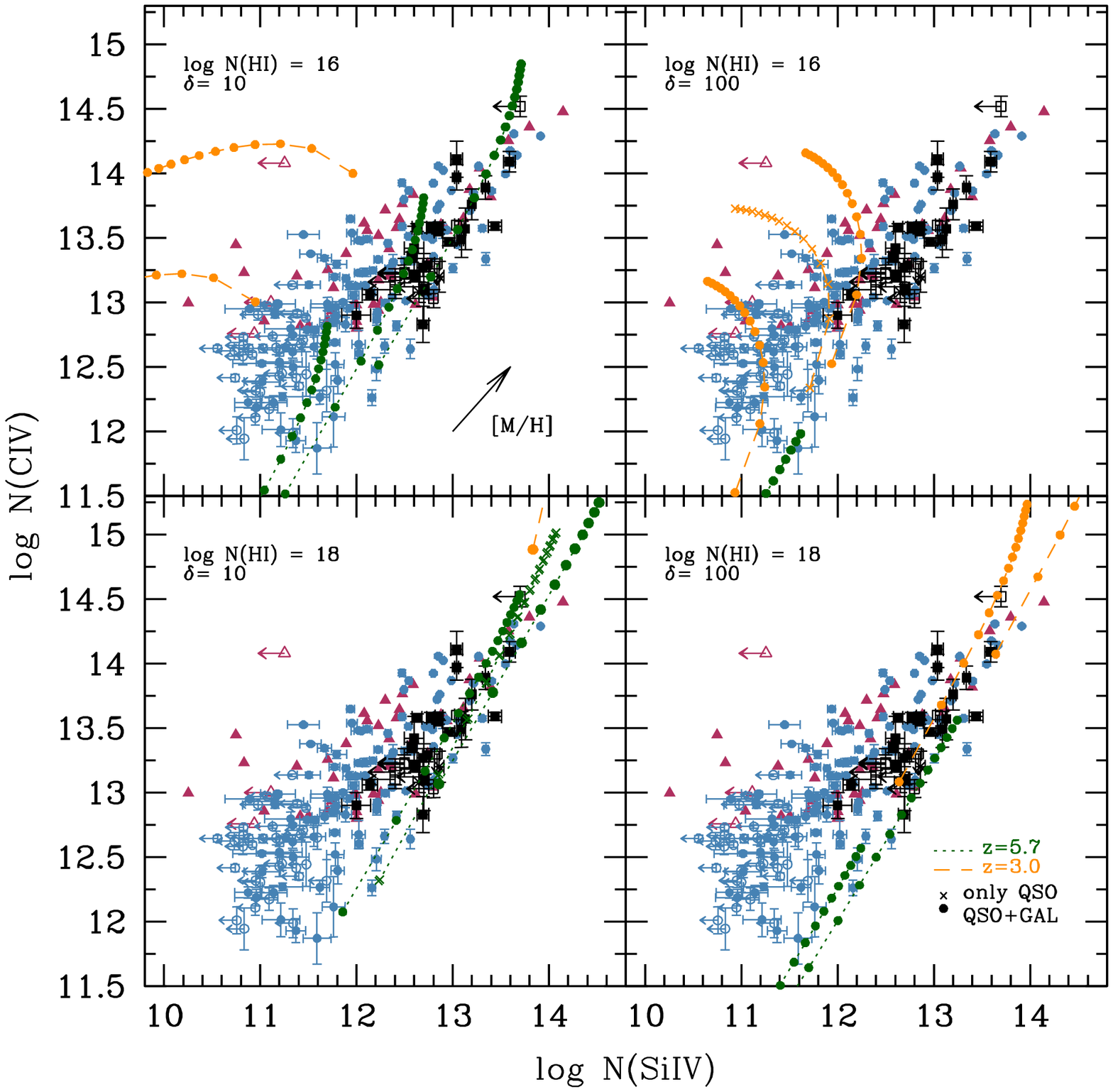}
\caption{\CIV\ column density as a function of the corresponding
  \SiIV\ column density for all the systems of our sample (black
    squares). The other data points are from  \citet[][crosses]{savaglio97}, \citet[][triangles]{songaila98} and \citet[][dots]{BSR03}. The open
  symbols are limits. Superimposed are the {\small CLOUDY} models at
  $z=5.7$ (green dotted lines) and $z=3$ (orange dashed 
    lines) with metallicities 
  [$M$/H]~$=-3$ and $-2$. In the upper-left panel, we show also the
  predictions at $z=5.7$ for a metallicity [$M$/H]~$=-1$. The arrow
  indicates approximately the direction of increasing metallicity. The
  predictions for a QSO only background and [$M$/H]~$=-2$ are shown in
  the lower-left panel (at $z=5.7$) and in the upper-right panel (at
  $z=3$). See the legend and the text for more details.} 
\label{si4_c4}
\end{center}
\end{figure*}

\begin{figure*}
\begin{center}
\includegraphics[width=13cm]{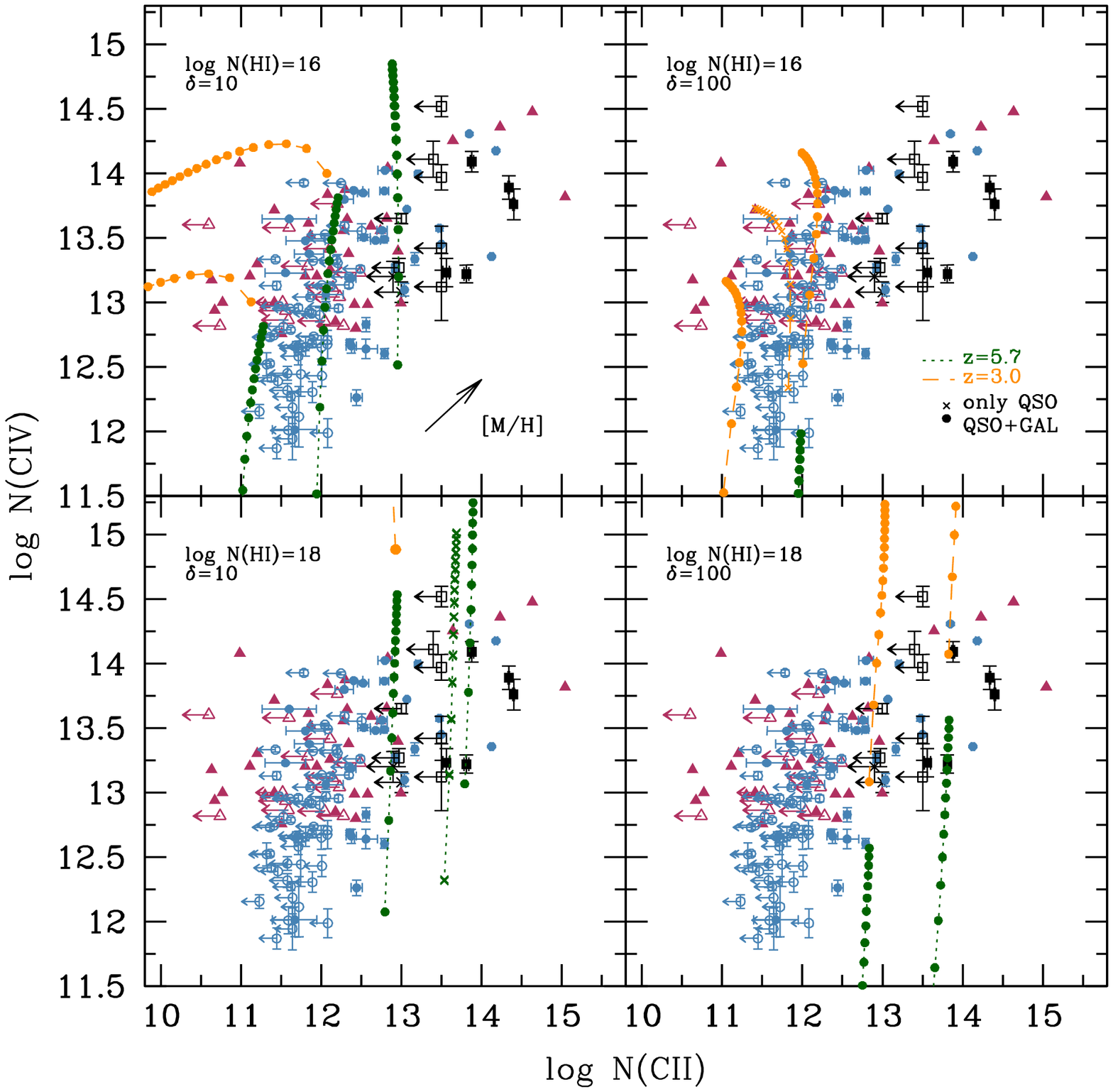}
\caption{\CIV\ column density as a function of the corresponding
  \CII\ column density for the 12 systems in our sample with both 
  transitions outside the \Lya\ forest (black and open squares).  The
  other data points are from  \citet[][crosses]{savaglio97}, 
  \citet[][triangles]{songaila98} and \citet[][ dots]{BSR03}. The open
  symbols are limits. Superimposed are the {\small
    CLOUDY} models. In the upper-left panel, we show also the
  predictions at $z=5.7$ for a metallicity [$M$/H]~$-1$. The arrow
  indicates approximately the direction of increasing metallicity. The
  predictions for a QSO only background and [$M$/H]~$=-2$ are shown in
  the lower-left panel (at $z=5.7$) and in the upper-right panel (at
  $z=3$). See the legend and the text for more details.}
\label{c2_c4}
\end{center}
\end{figure*}

\begin{figure*}
\begin{center}
\includegraphics[width=13cm]{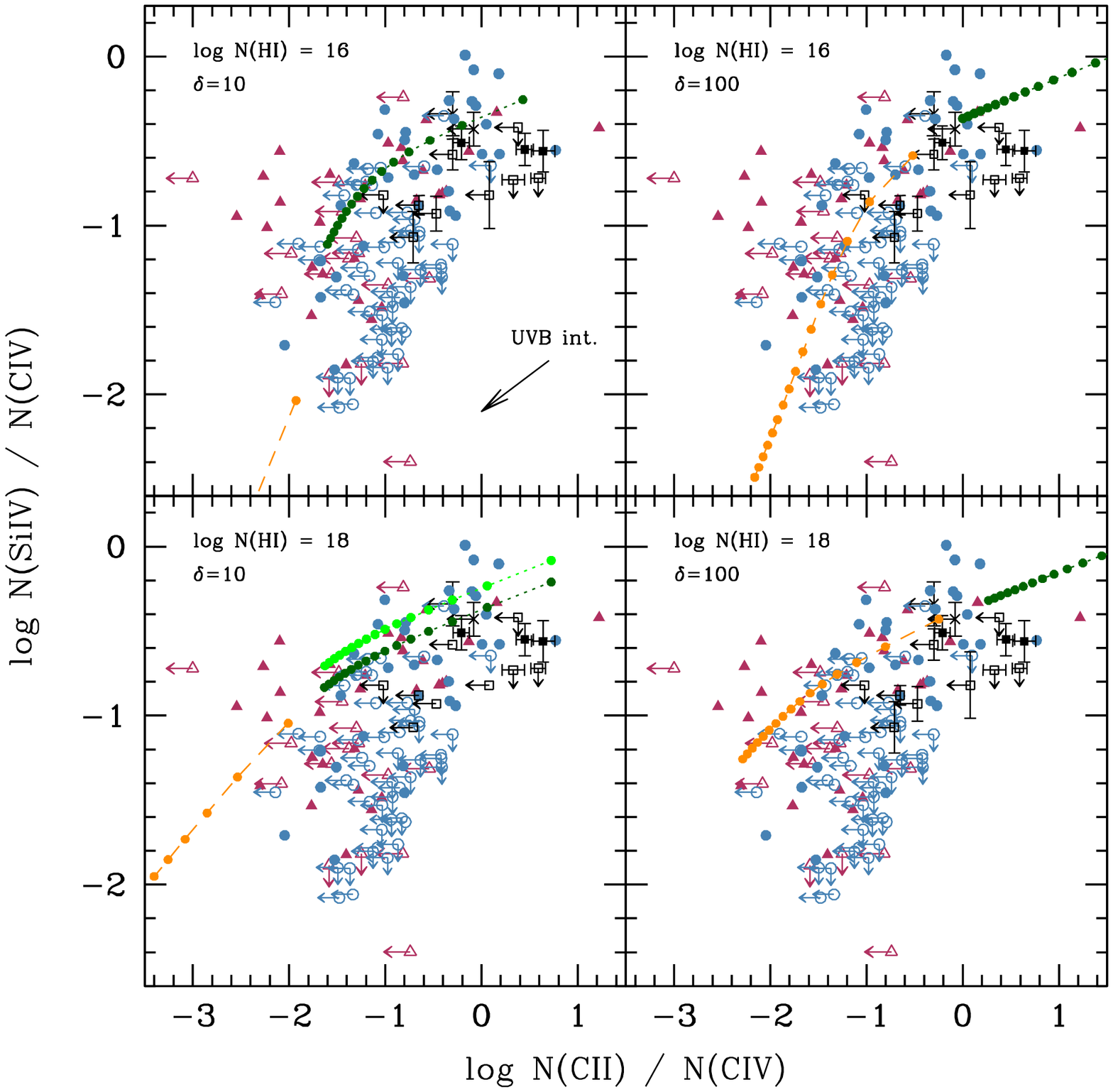}
\caption{\SiIV/\CIV\ ionic ratio as a function of the \CII/\CIV\ ratio
  for the 12 systems in our sample for which all three transitions are 
  outside the \Lya\ forest ( black and open squares).  The other data
  points are from  \citet[][ crosses]{savaglio97}, 
  \citet[][ triangles]{songaila98} and  \citet[][ dots]{BSR03}. The open
  symbols are limits. Superimposed are the {\small
    CLOUDY} models for a metallicity [$M$/H]~$=-2$. The model in light green in the lower-left panel has a relative abundance
  [C/Si]~$=-0.2$ for the gas at $z=5.7$. The arrow indicates approximately the direction of
  increasing UVB intensity. See Fig~\ref{c2_c4} for the legend of the models
  and the text for more details.} 
\label{si4c4_c2c4}
\end{center}
\end{figure*}

We have run a set of ionization models with version 10.00 of the
  {\small CLOUDY} code, last described by \citet[][]{ferland}. It is
  not conceivable to constrain the whole set of parameters of a
  photoionization model with just a few ionic transitions as in our
  case, due to the high level of degeneracy between e.g. total metallicity
  and chemical composition, or density and intensity of the ionizing
  flux. For this reason, we decided to run a limited set of models with
  parameters varying in reasonable ranges with the main aim of verifying that their 
  predictions were compatible with the observed ionic column
  densities. We chose the HM05 option in
{\small CLOUDY}, which consists of a UVB made by quasars and
galaxies with a 10 per cent photon escape fraction reprocessed by the
IGM \citep{hm01}. Two sets of models were run: one at $z=5.7$ to represent our data
and one at $z=3$ to represent the lower redshift sample. At each
redshift, we considered two over-densities, $\delta \equiv \rho/\rho_0
-1 = 10$ and 100, where $\rho$ is the IGM baryonic density and $\rho_0$ is the
mean density at the considered redshift;  two metallicities,
[$M$/H]~$=-2$ and $-3$, although in some cases also the predictions for
[$M$/H]~$=-1$ are plotted, and two \HI\ column 
densities, $\log N($\HI$) = 16$ and 18. In all cases, solar
relative abundances were adopted as a ground base. We ran also models
with a relative abundance [C/Si]~$=-0.2$ as measured recently in metal-poor damped Lyman$\alpha$ systems \citep{cooke11} and in $z>5$
low-ionization absorption systems \citep{becker12}.
The column densities of the interesting ions were changed by
multiplying the intensity of the UVB spectrum by a factor ranging from
0.25 to 4 in steps of 0.25. 

Even assuming that C and Si are evolving in locksteps, their ratio
could be affected not only by the ionization condition of the gas but
also by the possible variation in the relative abundances of
elements with respect to solar.  
%the variation of the
%\SiIV/\CIV\ ionic ratio as a function of redshift is 
%interesting as an indicator of the change in the ionization state of
%absorption systems at different epochs, but it is not very efficient to
%constrain the shape of the ionizing flux. 
In order to isolate ionization effects, we have looked for the \CII\ 1334 \AA\ lines
associated with our \CIV\ systems, which fall outside the 
\Lya\ forest.  Only 12 systems in our sample satisfy 
this requirement, of which 7 are upper limits. The \CII/\CIV\ ratio is
not affected by possible variation in the relative abundances of
elements with respect to solar; on the other hand, the transition
\CII\ 1334 \AA\ is saturated in most cases and the measured column
density could be underestimated. Assuming that this is not the case for the
five detections in our sample, we see from Fig.~\ref{c2_c4} that the
measured column densities are, in general, larger than those measured
at lower redshift for the same range of \CIV\ column densities. This
is evident also in Fig.~\ref{si4c4_c2c4}, where most of the
low-redshift  data have a lower \CII/\CIV\ ratio with respect to our points.   

The comparison between observed data and {\small CLOUDY} models is
shown in Figs~\ref{si4_c4}-\ref{si4c4_c2c4}. Each one of the four boxes
present in every figure, shows a different combination of $\delta$ and
$\log N($HI$)$ for the {\small CLOUDY} runs with the considered
metallicities. In Fig.~\ref{si4c4_c2c4}, we showed only the model results for
[$M$/H]~$=-2$, because the different metallicities give very similar
results. When the model track is not shown in the figure it is because
it falls outside the plotted range of values.
In the upper-right and bottom-left plots of Figs~\ref{si4_c4} and
\ref{c2_c4}, we show also the result of a model with a UVB due only to
quasars and metallicity [$M$/H]~$=-2$. At a fixed metallicity and UV
intensity, the QSO-only background produces (a factor of $\sim2-4$)
less \CII, \CIV\ and \SiIV\ with respect to the QSO-galaxy background,
this effect is more evident at $z=5.7$ where the contribution of QSOs
to the UVB is less important.  
   
Some qualitative considerations can be drawn as follows: the high-redshift absorbers
are better explained by gas with an over density of 10 and an indicative total
\HI\ column density of $\log N($\HI$) \sim 18$, while the lower redshift
absorbers trace gas with $\delta \sim 100$ and preferentially have
lower \HI\ column density ($\log N($\HI$) \sim 16$).
This is in agreement with what was found in simulations
\citep[e.g.][]{opp_dave06,cen_chisari11}, where it is shown that \CIV\ is a good
tracer of the metallicity in the low-density gas at high redshift while becoming less and
less  representative going towards lower redshift where it traces gas
at larger over densities ($\sim 100$). 
It has to be noted that the indication on the column density of
absorbers could be an observational bias, since in our sample we are
observing only the strongest systems due to the resolution and SNR of
our spectra.       

Models with a relative abundance [C/Si]~$=-0.2$ do not reproduce
  the high-redshift data better than solar abundance model (see
  Fig.~\ref{si4c4_c2c4}). This is somewhat puzzling since
  \citet{becker12} found this abundance result for a sample of
  absorption systems at $z>5$ among which there are also the two
  absorption systems at $\zabs \sim5.7899$ and 5.8770 along the line
  of sight to J0818+1722. A possible explanation is that since those
  systems are likely damped or sub-damped Lyman$\alpha$ systems, they are
  subject to a local ionizing flux whose shape is different from the
  adopted one.   
%Regarding the source of ionization and the shape of the ionizing
%spectrum: it is clear from Fig.~\ref{si4c4_c2c4} (top right panel)
%that the UVB at $z\sim3$ is dominated by the QSO contribution (the
%``QSO only'' track is superimposed to the ``QSO+GAL'' one), while in
%the bottom left panel one can see that at high redshift there is a
%significant contribution of galaxies. Indeed, in Fig.~\ref{c2_c4} and
%\ref{si4_c4} the tracks with a UVB due only to QSOs are not explaining
%the observed high redshift ionic column densities. This result
%suggests that the metals observed at high redshift are ionized by
%galactic sources and confirm the hypothesis that they could originate
%from in-situ star-formation.         

A more detailed analysis of the properties of the high-$z$ absorbers will be carried out in a forthcoming paper with a set of cosmological hydro-simulations coupled with cloudy models. 
%%%%%%%%%%%%%%%%%%%%%%%%%%%%%%%%%%%%%%%%%%%%%%%%%%%%%%%%%%%%%%%%%%%%%%%%

\section{Conclusions}

%%%%%%%%%%%%%%%%%%%%%%%%%%%%%%%%%%%%%%%%%%%%%%%%%%%%%%%%%%%%%%%%%%%%%%%%%%
The aim of this study was to investigate the metal content of QSO 
absorbers at very high redshift ($z>4.35$) in the framework of the
general picture of the enrichment history of the Universe.    

The reported results are based on a programme devoted to the
observations of QSOs at $z\sim6$ with 
the X-shooter spectrograph at the unit 2 of the ESO VLT telescope. 
The final sample consists of the spectra of six objects, of which four were observed in our
programme and two were downloaded from the X-shooter archive. 
Each QSO spectrum was inspected to look for metal absorption lines
outside the \Lya\ forest. Then, we focused our attention on the
properties of the \CIV\ sample  looking, in particular, for the
presence of the associated \SiIV\ doublet and \CII\ 1334
\AA\ line. A total of 102 \CIV\ lines were detected in the redshift
interval $4.35 < z < 6.2$, of which 25 with $z\gsim5$ have
associated \SiIV, and 5 show the \CII\ line. This is the first time
that a significant sample of $z\gsim5$ \SiIV\ doublets is reported.                      

We have reduced and analysed also the UVES spectra of $z\sim6$ QSOs
that were available in the archive. Unfortunately, those observations
have too low signal-to-noise ratio to add significant information to
the X-shooter observations. In a few cases, thanks to the higher resolution,
it was possible to discard the weak metal systems detected with
X-shooter.     

The main results are described in the following. 

\begin{description}
\item{-}\CIV\ {\it CDDF} the data
  sample has been divided into two redshift bins. The CDDF for the bin
  at $4.35 < z < 5.3$ is in good agreement with the CDDF of the low
  redshift systems ($1.5 < z <4$). On the other hand, the distribution
  function of the lines in the high redshift
  bin ($5.3 < z <6.2$) shows systematically lower values in all four
  column density bins, by a factor of $\sim 2-4$ depending on the
  considered fitting function. This indicates that the properties
  (e.g. the number density or the physical size) of the
  \CIV\ absorber population are varying at those redshifts. The
  observed decrease at $z>5.3$ is in agreement with what was found by
  \citet{becker09} with a sample of four QSOs observed at slightly higher
  resolution.

\item{-} \CIV\ {\it cosmic mass density:} the contribution of
  \CIV\ lines to the cosmic mass density has been computed for our
  line sample and compared with previous results at lower and similar
  redshift. We considered two column density regimes which give
  slightly different results (see Table~\ref{tab_omega}). 
  $\Omega_{\rm CIV}$ computed for the absorption lines with $13.4
    \le \log N($\CIV$) \le 15$ is in very good agreement with the
    recent determination by Simcoe11. We observe a slow increase from 
  the range $5.3 < z < 6.2$ to $4.35 < z < 5.3$ and then to $3<z<4$,
  then the mass density evolves more steeply with an increase of a   
 factor of $\sim4$ to $z\simeq 1.5$. Considering the evolution in time,
 we see approximately an increase of a factor of $\sim2$ in
 $\Omega_{\rm CIV}$ every Gyr. In the case of the
  stronger \CIV\ lines ($13.8 \le \log N($\CIV$) \le 15$) a gentle rise of
  a factor of $\sim 10$ is observed between $z\simeq 6.2$ and $z\simeq
  1.5$ with a possible flattening towards $z\sim 0$. For the strong lines, we do not observe
  a drop in   $\Omega_{\rm CIV}$ between $\langle z \rangle = 4.802$
  and $5.725$ as claimed by other authors
  \citep[e.g.][Simcoe11]{ryanweber09}. The increase between
  $z\sim6.2$ and 1.5 is well fitted by a power law: $\Omega_{\rm CIV}
  = (2\pm1)\times10^{-8} [(1+z)/4]^{-3.1\pm0.1}$. 
The observed behaviour is suggestive of a
  progressive accumulation of the metals produced by stars inside
  galaxies in the circumgalactic medium and IGM. When the
  star formation rate density is  observed to decrease, below
  $z\sim1$, the increase in  $\Omega_{\rm CIV}$ slows down. 
  It is important to remember that the abundance of our
  observable, the triply ionized carbon, is determined by the total
  amount of carbon but also by the ionization conditions in the gas:
  cosmological simulations could help to fully interpret our results
  and determine the true evolution of the metal content in the IGM. 
  
%\Omega_CIV(z)=\Omega_CIV(z=3)*((1+z)/4)^slope

\item{-} \CIV, \SiIV\ {\it and} \CII\ {\it combined constraints on the
  nature of absorbers:} starting from our \CIV\ sample, we built a
  subsample of lines with associated \SiIV\ doublets and a further
  subsample of lines for which it was possible to detect both the
  \SiIV\ and the \CII\ 1334 \AA\ lines, outside the \Lya\ 
  forest. The \SiIV/\CIV\ column density ratios for our sample have a
  small dispersion around a logarithmic median value of $-0.58$. This
  is somewhat higher than the median value of $\sim -1.2$ at $z \sim 3$ found
  by \citet{BSR03}. A Kolmogorov-Smirnov test run on our sample
  in comparison with lower redshift samples ($z\sim3$) indicates that
  they are likely not drawn from the same parent distribution,
  suggesting a change in the ionization
  conditions of the gas with increasing redshift. A qualitative test 
  was carried out by comparing the observed ionic column densities
  with the results of a (limited) set of {\small CLOUDY} ionization
  models. We adopt a Haardt-Madau UVB, solar abundances
  and metallicities [$M$/H]~$=-2$ and $-3$. Indicatively, the
  observed \CIV\ absorbers at high redshift are better explained by
  gas with over density of $\sim10$, while at $z\sim3$ better
  agreement is obtained with gas with $\delta \simeq 100$. This
  is in agreement with the predictions of metal enrichment simulations
  \citep[e.g.][]{opp_dave06,cen_chisari11}. {\small CLOUDY} models
  with [C/Si]~$=-0.2$ give a worse representation of the high-redshift
  data although this relative abundance was obtained for $z>5$
  absorption systems. This discrepancy suggests that probably some of
  these absorbers trace very dense and neutral environments ionized
  by local sources. 
\end{description}

The present study shows how metal enriched ionized gas is present even
at very high redshift possibly in the form of dense systems affected
by the presence of local ionizing stellar sources. 
Unfortunately, the modest number of  lines of sight and the
relatively low SNR in the NIR region are still preventing us to reach
solid results on the statistics of \CIV\ absorbers at $z>5$. Only a
major observational effort could improve significantly this situation
due to the faintness of the targets and their paucity. This will be an
extremely interesting and driving science case for the next
generation of visual and NIR high-resolution spectrographs at the
ELTs. 

\section*{Acknowledgements}

This work is supported by PRIN-INAF 2010. MV is supported by
PRIN-MIUR, INFN/PD51 and the ERC Starting Grant CosmoIGM. We would like to thank the
anonymous referee for the very careful reading of the paper and the
helpful suggestions.

\appendix

\section[]{\CIV\ and \SiIV\ absorptions in the X-shooter spectra of the QSOs at
  $\zem \sim 6$}

\begin{table*}
\begin{center}
\caption{\CIV\ absorption systems in the spectrum of SDSS
  J0818+1722. In the first column, the letters denote the single
  velocity components which was not possible to deblend in the
  computation of the equivalent width.}
\begin{minipage}{120mm}
\label{tab_CIV_0818}
\begin{tabular}{@{} l l c c  c c}
%\hline  
%\multicolumn{4}{c}{SDSS J0818+1722} \\
\hline
System & Ion & $z$ & $W_0$ & $b$ & $\log N$   \\
& & & (\AA) & (km s$^{-1}$) & (cm$^{-2}$) \\
\hline
1 & \CIV\ 1548 & $4.46298\pm0.00001$ & $0.258\pm0.004$ & $27.3\pm0.7$ & $13.90\pm0.01$ \\
& \CIV\ 1550 & & $0.115\pm0.002$ & & \\
2 & \CIV\ 1548 & $4.49800\pm0.00001$ & $0.054\pm0.003$ & $11\pm1$ & $13.16\pm0.02$ \\
& \CIV\ 1550b &  & $<0.059\pm0.003$ & & \\
3 & \CIV\ 1548b & $4.50838\pm0.00004$ & $>0.045\pm0.003$ & $31\pm4$ & $13.12\pm0.03$ \\
& \CIV\ 1550 &  & $0.039\pm0.004$ & & \\
4 & \CIV\ 1548 & $4.52305\pm0.00004$ & $0.080\pm0.006$ & $51\pm4$ & $13.28\pm0.02$ \\
& \CIV\ 1550 & & $0.055\pm0.004$ & & \\
%\CIV$^{\rm a}$ & $4.55225\pm0.00005$ & $21\pm8$ & $12.78\pm0.05$ \\
%\CIV$^{\rm a,b}$ & $4.57733\pm0.00006$ & 15 & $12.65\pm0.07$ \\
5 & \CIV\ 1548 & $4.55225\pm0.00005$ & $0.030\pm0.003$ & $21\pm8$ & $12.78\pm0.05$ \\
& \CIV\ 1550 & & $0.020\pm0.003$ & & \\
6 & \CIV\ 1548 & $4.57733\pm0.00006$ & $0.016\pm0.005$ & 15 & $12.65\pm0.07$ \\
& \CIV\ 1550 & & $0.013\pm0.003$ & & \\
7 & \CIV\ 1548b & $4.62025\pm0.00005$ & $>0.050 \pm0.004$ & $36\pm5$ & $13.09\pm0.04$ \\
& \CIV\ 1550 & & $0.034\pm0.004$ & & \\
8 & \CIV\ 1548 & $4.62702\pm0.00005$ & $0.050\pm0.005$ & $30\pm6$ & $13.00\pm0.04$ \\
& \CIV\ 1550 & & $0.030\pm0.004$ & & \\
%\hline
9 & \CIV\ 1548m & $4.7263$ & $0.539\pm0.008$ & & \\
& \CIV\ 1550m &  & $0.316\pm0.007$ & & \\
9a& \CIV & $4.7252\pm0.0001$   & & $29\pm6$ & $13.59\pm0.08$ \\
9b& \CIV & $4.72615\pm0.00005$ & & $15\pm4$ & $13.73\pm0.07$ \\
9c& \CIV & $4.72699\pm0.00005$ & & $18\pm2$ & $13.89\pm0.03$ \\
%\hline
10 & \CIV\ 1548 & $4.73158\pm0.00003$ & $0.225\pm0.008$ & $45\pm2$ & $13.70\pm0.01$ \\
& \CIV\ 1550s & & $0.12\pm0.02$ & & \\
11 & \CIV\ 1548 & $4.87739\pm0.00004$ & $0.070\pm0.006$ & $32\pm4$ & $13.20\pm0.03$ \\
& \CIV\ 1550 & & $0.037\pm0.005$ & & \\
12 & \CIV\ 1548 & $4.9417\pm0.0001$ & $0.136\pm0.009$ & $87\pm7$ & $13.46\pm0.03$ \\
& \CIV\ 1550 & & $0.06\pm0.01$ & & \\
%\CIV$^{\rm a}$ & $5.0623\pm0.0001$  & $26\pm7$ & $13.03\pm0.07$ \\
13 & \CIV\ 1548 & $5.0623\pm0.0001$  & $0.045\pm0.009$ & $26\pm7$ & $13.03\pm0.07$ \\
& \CIV\ 1550s & & $0.040\pm0.007$ & & \\
& \SiIV & & & & $< 12.8$ \\
14 & \CIV\ 1548 & $5.06467\pm0.00005$ & $0.127\pm0.008$ & $45\pm4$ & $13.47\pm0.03$ \\
& \CIV\ 1550sb & & $0.14\pm0.01$ & & \\
& \SiIV\ 1402 & & $0.038\pm0.006$ &  & $12.96\pm0.06$ \\
15 & \CIV\ 1548 & $5.07629\pm0.00004$ & $0.18\pm0.01$ & $38\pm4$ & $13.59\pm0.03$ \\
& \CIV\ 1550 & & $0.11\pm0.02$ & & \\
& \SiIV\ 1402 & & $0.033\pm0.003$ &  & $12.86\pm0.04$ \\
16 & \CIV\ 1548 & $5.08241\pm0.00004$ & $0.147\pm0.009$ & $37\pm3$ & $13.55\pm0.02$ \\
& \CIV\ 1550 & & $0.126\pm0.007$ & & \\
& \SiIV\ 1402 & & $0.036\pm0.003$ &  & $12.85\pm0.04$ \\
%\CIV$^{\rm a}$ & $5.3085\pm0.0001$  & $31\pm8$ & $13.10\pm0.07$ \\
%\CIV$^{\rm a}$ & $5.32226\pm0.00005$  & $22\pm4$ & $13.35\pm0.04$ \\
17 & \CIV\ 1548 & $5.3085\pm0.0001$  & $0.052\pm0.009$ & $31\pm8$ & $13.10\pm0.07$ \\
& \CIV\ 1550 & & $0.05\pm0.01$ & & \\
& \SiIV & & & & $< 12.4$ \\
18 & \CIV\ 1548 & $5.32226\pm0.00005$  & $0.09\pm0.01$ & $22\pm4$ & $13.35\pm0.04$ \\
& \CIV\ 1550 & & $0.07\pm0.01$ & & \\
& \SiIV\ 1393 & & $0.037\pm0.004$ & & $12.58\pm0.05$ \\
& \SiIV\ 1402 & & $0.011\pm0.003$ & & \\
%\CIV$^{\rm a}$ & $5.3243\pm0.0002$ &  $33\pm11$ & $13.2\pm0.1$ \\
%\CIV$^{\rm a}$ & $5.7899\pm0.0002$ &  $30\pm14$ & $13.2\pm0.1$ \\
%\CIV$^{\rm a}$ & $5.8441\pm0.0001$ &  $35\pm8$ & $13.27\pm0.07$ \\
19 & \CIV\ 1548s & $5.3243\pm0.0002$ & $0.06\pm0.02$ & $33\pm11$ & $13.2\pm0.1$ \\
& \CIV\ 1550 & & $0.036 \pm 0.009$ & & \\
& \SiIV & & & & $< 12.4$ \\
20 & \CIV\ 1548s & $5.7899\pm0.0002$ & $0.08\pm0.02$ & $30\pm14$ & $13.2\pm0.1$ \\
& \CIV\ 1550s & & $0.04\pm0.01$ & & \\
& \SiIV & & & & $< 12.5$ \\
& \CII\ 1334 & $5.78909\pm0.00004$ & & $26\pm2$ & $13.56\pm0.04$ \\
21 & \CIV\ 1548 & $5.8441\pm0.0001$ & $0.11\pm0.02$ & $35\pm8$ & $13.27\pm0.07$ \\
& \CIV\ 1550 & & $0.04\pm0.02$ & & \\
& \SiIV\ 1393 & & $0.045\pm0.007$ & & $12.69\pm0.06$ \\
& \SiIV\ 1402 & & $0.023\pm0.005$ & & \\
& \CII\ 1334 & & & & $<13$ \\ 
%\CIV$^{\rm a}$ & $5.8770\pm0.0001$ &  $16\pm10$ & $13.22\pm0.07$ \\
22 & \CIV\ 1548s & $5.8770\pm0.0001$ & $0.09\pm0.02$ & $16\pm10$ & $13.22\pm0.07$ \\
& \CIV\ 1550 & & $ 0.031\pm0.009$ & & \\
& \SiIV & & & & $< 12.5$ \\
& \CII\ 1334 & $5.87644\pm0.00002$ & & $11.2\pm0.8$ & $13.81\pm0.03$ \\
\hline
\end{tabular}
Notes. b: Line blended with another absorption line; m: multiple
components; s: line contaminated by a sky emission line; l: feature below $3\,\sigma$ detection.
\end{minipage}
\end{center}
\end{table*}

\begin{table*}
\begin{center}
\caption{\CIV\ absorption systems in the spectrum of SDSS J0836+0054. In the first column, the letters denote the single
  velocity components which was not possible to deblend in the
  computation of the equivalent width.}
\begin{minipage}{120mm}
\label{tab_CIV_0836}
\begin{tabular}{@{} l l c c c c}
%\hline  
%\multicolumn{4}{c}{SDSS J0836+0054} \\
\hline
System & Ion &  $z$ & $W_0$ & $b$ & $\log N$   \\
& & & (\AA) & (km s$^{-1}$) & (cm$^{-2}$) \\
\hline
1 & \CIV\ 1548 & $4.53013\pm0.00003$ & $0.048\pm0.004$ & $11\pm3$ &  $13.08\pm0.03$ \\
& \CIV\ 1550b & & $<0.046\pm0.004$ & & \\
%\CIV$^{\rm a}$ &  $4.61195\pm0.00007$  &  $18\pm7$ &   $12.82\pm0.08$ \\
%\CIV$^{\rm a}$ &  $4.66840\pm0.00005$  &  $10\pm5$ &   $12.82\pm0.06$ \\
2 & \CIV\ 1548 &  $4.61195\pm0.00007$  & $0.030\pm0.006$ & $18\pm7$ &   $12.82\pm0.08$ \\
& \CIV\ 1550 & & $0.017\pm0.006$ & & \\
3 & \CIV\ 1548 &  $4.66840\pm0.00005$  & $0.028\pm0.005$ & $10\pm5$ &   $12.82\pm0.06$ \\
& \CIV\ 1550 & & $0.016\pm0.005$ & & \\
%\hline
4 & \CIV\ 1548mb & $4.6855$ & $0.398\pm0.008$ & & \\
& \CIV\ 1550m & & $0.189\pm0.009$ & & \\
4a & \CIV & $4.68427\pm 0.00004$  &&  $29\pm3$ &   $13.46\pm0.04$ \\
4b & \CIV & $4.68651\pm0.00004$  &&  $44\pm5$ &   $13.58\pm0.03$  \\
%\hline
%\CIV$^{\rm a,b}$ & $4.6985\pm0.0001$   &  15 &  $12.80\pm0.07$ \\
%\CIV$^{\rm a}$ &  $4.7733\pm0.0001$   &  $40\pm9$ &   $12.96\pm0.06$ \\ 
%\CIV$^{\rm a}$ &  $4.99048\pm0.00004$  &  $26\pm 4$ & $13.06\pm 0.04$ \\ 
5 & \CIV\ 1548b & $4.6985\pm0.0001$   &  $<0.040\pm0.007$ & 15 &  $12.80\pm0.07$ \\
& \CIV\ 1550b & & $>0.021\pm0.004$ & & \\
6 & \CIV\ 1548 &  $4.7733\pm0.0001$   & $0.041\pm0.006$ & $40\pm9$ &   $12.96\pm0.06$ \\ 
& \CIV\ 1550 & & $0.034\pm0.006$ & & \\
7 & \CIV\ 1548 &  $4.99048\pm0.00004$  & $0.051\pm0.006$ & $26\pm 4$ & $13.06\pm 0.04$ \\ 
& \CIV\ 1550 & & $0.024\pm0.006$ & & \\
& \SiIV\ 1393 & & $0.015\pm0.003$ & $17\pm2$ & $12.14\pm0.08$ \\
& \SiIV\ 1402l & & $0.005\pm0.003$ & & \\
%\CIV$^{\rm a}$ &  $4.99264\pm0.00003$  &  $28\pm 3$ & $13.10\pm0.04$  \\
8 & \CIV\ 1548 &  $4.99264\pm0.00003$  & $0.060\pm0.006$ & $28\pm 3$ & $13.10\pm0.04$  \\
& \CIV\ 1550 & & $0.029\pm0.004$ && \\
& \SiIV\ 1393 & & $0.052\pm0.004$ & $18\pm2$ & $12.71\pm0.03$ \\
& \SiIV\ 1402 & & $0.028\pm0.004$ & & \\
%\CIV$^{\rm a}$ &  $4.99421\pm0.00009$  &  $29\pm10$ & $12.90\pm 0.10$\\
9 & \CIV\ 1548 &  $4.99421\pm0.00009$  & $0.044\pm0.005$ & $29\pm10$ & $12.90\pm 0.10$\\
& \CIV\ 1550 & & $0.022\pm0.005$ & & \\
& \SiIV\ 1393 & & $0.011\pm0.005$ & $19\pm6$ & $12.0\pm0.15$ \\
& \SiIV\ 1402 & & $0.006\pm0.003$ & & \\
%\hline
10 & \CIV\ 1548m & $4.9965$ & $0.208\pm0.006$ & & \\
& \CIV\ 1550m & & $0.130\pm0.007$ & & \\
& \SiIV\ 1393m & & $0.083\pm0.004$ & & \\
& \SiIV\ 1402m & & $0.066\pm0.005$ & & \\
10a & \CIV & $4.99634\pm0.00006$  & & $52\pm4$ &  $13.58\pm0.03$ \\
& \SiIV & & & $34\pm3$ & $12.63\pm0.05$ \\
%\CIV$^{\rm b}$ & $4.99693\pm0.00002$  &  10 &   $13.29\pm0.05$ \\
%\SiIV$^{\rm b}$ & & 10 & $12.73\pm0.04$ \\
10b & \CIV & $4.99693\pm0.00002$  &  10 &   $13.29\pm0.05$ \\
& \SiIV & & 10 & $12.73\pm0.04$ \\
%\hline
11 & \CIV\ 1548b & $5.12487\pm0.00003$  & $<0.211\pm0.006$ & 33    &   $<13.60$ \\
& \CIV\ 1550b & & $<0.234\pm0.005$ & & \\
& \SiIV\ 1393 & & $0.104\pm0.004$ & $33\pm2$ & $13.04\pm0.02$ \\
& \SiIV\ 1402 & & $0.047\pm0.005$ & & \\
12 & \CIV\ 1548b & $5.12712\pm0.00003$  &  $<0.112\pm0.007$ & 20 &   $<13.30$ \\
& \CIV\ 1550b & & $<0.074\pm0.004$ & & \\
& \SiIV\ 1393 & & $0.073\pm0.005$ & $20\pm2$ & $12.84\pm0.03$ \\
& \SiIV\ 1402 & & $0.019\pm0.004$ & & \\
%\CIV$^{\rm a}$ & $5.32280\pm0.00004$  &  $33\pm4$ &   $13.58\pm0.04$\\
13 & \CIV\ 1548 & $5.32277\pm0.00004$  & $0.17\pm0.02$ &  $31\pm4$ &   $13.65\pm0.04$\\
& \CIV\ 1550 & & $0.10\pm0.01$ & & \\
& \SiIV\ 1393 & & $0.050\pm0.005$ & & $12.77\pm0.04$ \\
& \SiIV\ 1402s & & $0.005\pm0.006$ & & \\
& \CII\ 1334 & & & & $<13.00$ \\
\hline
\end{tabular}
Notes. b: Line blended with another absorption line; m: multiple
components; s: line contaminated by a sky emission line; l: feature below $3\,\sigma$ detection.
\end{minipage}
\end{center}
\end{table*}

\begin{table*}
\begin{center}
\caption{\CIV\ absorption systems in the spectrum of SDSS J1030+0524. In the first column, the letters denote the single
  velocity components which was not possible to deblend in the
  computation of the equivalent width.}
\begin{minipage}{120mm}
\label{tab_CIV_1030}
\begin{tabular}{@{}l l c c  c c}
%\hline  
%\multicolumn{4}{c}{SDSS J1030+0524} \\
\hline
System & Ion & $z$ & $W_0$ & $b$ & $\log N$   \\
& & & (\AA) & (km s$^{-1}$) & (cm$^{-2}$) \\
\hline
1 & \CIV\ 1548 & $4.76671\pm0.00004$  & $0.063\pm0.006$ & $36\pm  4$  & $13.13\pm 0.03$  \\ 
& \CIV\ 1550b & & $<0.076\pm0.005$ & & \\
2 & \CIV\ 1548b & $4.7966\pm0.0001$ & $<0.097\pm0.006$ & $94\pm 12$  &    $13.30\pm 0.04$  \\ 
& \CIV\ 1550l & & $0.008\pm0.004$ & & \\
3 & \CIV\ 1548b & $4.79931\pm0.00001$ & $<0.128\pm0.005$ & $13\pm2$ & $13.37\pm0.02$  \\ 
& \CIV\ 1550 & & $0.072\pm0.004$ & & \\
4 & \CIV\ 1548 & $4.80107\pm 0.00001$ & $0.149\pm0.005$ & $19\pm2$  & $13.46\pm 0.01$  \\ 
& \CIV\ 1550b & & $<0.377\pm0.005$ & & \\
5 &\CIV\ 1548 & $4.89066\pm0.00003$  & $0.086\pm0.007$ & $30\pm4$ & $13.21\pm 0.02$  \\ 
& \CIV\ 1550 & & $0.033\pm0.007$ & & \\
%\hline
6 & \CIV\ 1548m & $4.9482$ & $0.31\pm0.01$ & & \\
 & \CIV\ 1550m & & $0.18\pm0.01$ & & \\
6a & \CIV & $4.94709\pm  0.00004$  &  & $12\pm 4$  &    $13.22\pm 0.04$  \\ 
6b & \CIV & $4.94849\pm  0.00002$  &  & $29\pm 2$  &    $13.77\pm 0.01$  \\ 
%\hline
7 & \CIV\ 1548m & $5.5172$ & $0.39\pm0.06$ & & \\
 & \CIV\ 1550ms & & $0.22\pm0.07$ & & \\
 & \SiIV\ 1393m & $5.5164$ & $0.233\pm0.006$ & & \\
 & \SiIV\ 1402m & & $0.093\pm0.007$ & & \\
%\CIV$^{\rm a}$ & $5.51553\pm0.00003$  &   34 &    $13.49\pm 0.14$  \\ 
7a & \CIV & $5.51553\pm0.00002$  &  & 34 &    $13.4\pm 0.2$  \\ 
& \SiIV & & & $34\pm2$ & $13.10\pm0.01$ \\
%\CIV$^{\rm a}$ & $5.51731\pm0.00001$ & $21\pm 1$ & $13.57\pm0.16$  \\ 
7b & \CIV & $5.5178\pm0.0001$ & &$51\pm 7$ & $13.92\pm0.05$  \\ 
& \SiIV & $5.51730\pm0.00001$ & & $20\pm1$ & $13.13\pm0.01$ \\
%\CIV$^{\rm a}$ & $5.5184\pm0.0002$ & $32\pm14$ & $13.63\pm0.13$  \\ 
%\hline
8 & \CIV\ 1548 & $5.72419\pm0.0001$ & $0.75\pm0.06$ & $47\pm8$ & $14.52\pm 0.08$  \\ 
& \CIV\ 1550 & & $0.49\pm0.05$ & & \\
& \SiIV & & & & $<13.7$ \\ 
& \CII\ 1334 & & & & $<13.5$ \\
%\CIV$^{\rm a}$ & $5.74116\pm0.00004$ & $29\pm 3$  & $13.76\pm0.12$  \\ 
%\hline
9 & \CIV\ 1548m &  $5.7428$ & $0.54\pm0.05$ & & \\
  & \CIV\ 1550m & &  $0.25\pm0.06$ & & \\
 & \SiIV\ 1393m & $5.7429$ & $0.31\pm0.02$ & & \\
 & \SiIV\ 1402mb & & $<0.71\pm0.03$ & & \\
9a & \CIV & $5.74116\pm0.00004$ & & $29\pm 3$  & $13.8\pm0.1$  \\ 
& \SiIV & & & & $13.20\pm0.03$ \\  
& \CII\ 1334 & $5.74097\pm0.00001$ & & $25.2\pm 0.6$ & $14.40\pm0.01$ \\
%\CIV$^{\rm a}$ & $5.7440\pm0.0002$ & $29\pm3$  & $13.89\pm0.09$  \\ 
9b & \CIV & $5.7440\pm0.0002$ & & $29\pm3$  & $13.89\pm0.09$  \\ 
& \SiIV & $5.74425\pm0.00004$ & & $29\pm3$ & $13.34\pm0.03$ \\
& \CII\ 1334 & $5.74399\pm0.00004$ & & $50.6\pm0.8$ & $14.34\pm0.01$ \\
%\hline
%\CIV$^{\rm a,b}$ & $5.9757\pm0.0004$   &   15 &   $13.12\pm 0.26$  \\ 
10 & \CIV\ 1548l & $5.9757\pm0.0004$ & $0.05\pm0.02$ & 15 &   $13.1\pm 0.3$  \\ 
   & \CIV\ 1550l & & $0.02\pm0.02$ & & \\
   & \SiIV & & & & $<12.7$ \\
   & \CII\ 1334  & & & & $<13.5$ \\
%\CIV$^{\rm a,b}$ & $5.9784\pm0.0002$   &   15 &  $13.42\pm 0.17$ \\
%\SiIV$^{\rm a,b}$ & $5.97896\pm0.00009$ & 15 & $12.6\pm0.1$ \\
11 & \CIV\ 1548l & $5.9784\pm0.0002$ & $0.07\pm0.03$ & 15 &  $13.4\pm 0.2$ \\
 & \CIV\ 1550l & & $0.08\pm0.02$ & & \\
& \SiIV\ 1393l & $5.97896\pm0.00009$ & $0.03\pm0.01$ & 15 & $12.6\pm0.1$ \\
& \SiIV\ 1402l & & $0.01\pm0.01$ & & \\
   & \CII\ 1334  & & & & $<13.5$ \\
\hline
\end{tabular}
Notes. b: Line blended with another absorption line; m: multiple
components; s: line contaminated by a sky emission line; l: feature
below $3\,\sigma$ detection.
\end{minipage}
\end{center}
\end{table*}

\begin{table*}
\begin{center}
\caption{\CIV\ Absorption systems in the spectrum of SDSS J1306+0356. In the first column, the letters denote the single
  velocity components which was not possible to deblend in the
  computation of the equivalent width.}
\begin{minipage}{120mm}
\label{tab_CIV_1306}
\begin{tabular}{@{}l l c c c c}
%\hline  
%\multicolumn{4}{c}{SDSS J1306+0356} \\
\hline
System & Ion & $z$ & $W_0$ & $b$ & $\log N$   \\
& & & (\AA) & (km s$^{-1}$) & (cm$^{-2}$) \\
\hline
1 & \CIV\ 1548m & $4.5292$ & $ 0.090\pm0.004$ & & \\
 & \CIV\ 1550mb & & $>0.021\pm0.003$ & & \\
1a & \CIV & $4.52897\pm0.00003$ & & $32\pm 3$ &  $13.15\pm 0.02$ \\ 
%\CIV$^{\rm a,b}$ & $4.53040\pm0.00007$ &  15 &  $12.44\pm 0.07$ \\
%\CIV$^{\rm a}$ &  $4.58045\pm0.00007$ &  $16\pm 7$ &  $12.49\pm 0.08$ \\
%\CIV$^{\rm a}$ &  $4.61252 \pm0.00007$ &  $32\pm 7$ &  $12.81\pm 0.06$ \\
1b & \CIV & $4.53040\pm0.00007$ & & 15 &  $12.44\pm 0.07$ \\
%\hline
2 & \CIV\ 1548 &  $4.58045\pm0.00007$ & $0.019\pm0.003$ & $16\pm 7$ &  $12.49\pm 0.08$ \\
  & \CIV\ 1550s & & $0.001\pm0.002$ & & \\
%\hline
3 & \CIV\ 1548m & $4.6149$ & $ 0.384\pm0.006$ & & \\
 & \CIV\ 1550m & & $0.196\pm0.006$ & & \\
3a & \CIV &  $4.61252 \pm0.00007$ & & $32\pm 7$ &  $12.81\pm 0.06$ \\
3b & \CIV & $4.61464\pm0.00002$ & & $35\pm 2$ &  $13.79\pm 0.02$ \\  
3c & \CIV & $4.61591\pm0.00003$ & & $25\pm 2$ &  $13.44\pm 0.03$ \\
%\hline 
%\CIV$^{\rm a}$ & $4.65361 \pm0.00003$ &  $16\pm 3$ &  $13.07\pm 0.04$ \\ 
%\CIV$^{\rm a}$ & $4.6671 \pm0.0001$  &  $22\pm 9$ &  $12.68\pm 0.25$ \\
4 & \CIV\ 1548 & $4.65361 \pm0.00003$ & $0.054\pm0.005$ & $16\pm 3$ &  $13.07\pm 0.04$ \\ 
 & \CIV\ 1550 & & $0.016\pm0.007$ & & \\
%\hline
5 & \CIV\ 1548m & $4.6685$ & $0.42\pm0.01$ & & \\
 & \CIV\ 1550m & & $0.234\pm0.009$ & & \\
5a  & \CIV & $4.6671 \pm0.0001$  & & $22\pm 9$ &  $12.68\pm 0.25$ \\
5b  & \CIV & $4.66818\pm0.00001$ & & $12\pm 2$ &  $13.84\pm 0.04$ \\ 
5c  & \CIV & $4.6688\pm0.0001$  & & $50\pm 6$ &  $13.82\pm 0.07$ \\
%\hline
6 & \CIV\ 1548 & $4.71113\pm 0.00007$ & $0.085\pm0.008$ & $48\pm 6$ &  $13.27\pm 0.04$ \\ 
 & \CIV\ 1550b & & $<0.046\pm0.008$ & & \\
7 & \CIV\ 1548b & $4.72315\pm 0.00005$ & $<0.083\pm0.007$ & $37\pm 4$ &  $13.26\pm 0.04$ \\
  & \CIV\ 1550 & & $0.034\pm0.006$ & & \\
8 & \CIV\ 1548 & $4.82048\pm 0.00008$ & $0.092\pm0.008$ & $51\pm 7$ &  $13.27\pm 0.04$ \\
  & \CIV\ 1550 & & $0.040\pm0.006$ & & \\
%\CIV$^{\rm a}$ & $4.8591\pm 0.0003$  &  50 &  $13.14\pm 0.07$ \\
%\hline
9 & \CIV\ 1548m & $4.864$ & $1.58\pm0.01$ & & \\
  & \CIV\ 1550m & & $1.125\pm0.01$ & & \\ 
9a & \CIV & $4.8591\pm 0.0003$   & &  50 &  $13.14\pm 0.07$ \\
9b & \CIV & $4.86043\pm 0.00003$ & &  $18\pm 3$ &  $13.76\pm 0.03$ \\   
9c & \CIV & $4.86116\pm 0.00005$ & &  $10\pm 0$ &  $13.33\pm 0.07$ \\   
9d & \CIV & $4.86238\pm 0.00004$ & &  $33\pm 4$ &  $13.97\pm 0.04$ \\
9e & \CIV & $4.86341\pm 0.00003$ & &  $15\pm 2$ &  $13.91\pm 0.04$ \\ 
9f & \CIV & $4.86459\pm 0.00003$ & &  $27\pm 3$ &  $14.12\pm 0.03$ \\ 
9g & \CIV & $4.86560\pm 0.00004$ & &  $17\pm 2$ &  $13.72\pm 0.06$ \\  
9h & \CIV & $4.86686\pm 0.00001$ & &  $25\pm 1$ &  $14.02\pm 0.01$ \\  
9i & \CIV & $4.86911\pm 0.00006$ & &  $52\pm 5$ &  $13.60\pm 0.03$ \\  
%\hline
%\CIV$^{\rm a,b}$ & $4.87912\pm0.00004$  &  10 &  $13.00\pm 0.05$ \\        
10 & \CIV\ 1548m & $4.8806$ & $0.51\pm0.01$ & & \\
 & \CIV\ 1550m & & $0.264\pm0.008$ & & \\
10a & \CIV & $4.87912\pm0.00004$  & & 10 &  $13.00\pm 0.05$ \\        
10b & \CIV & $4.88013\pm0.00003$  & & $17\pm 2$ &  $13.77\pm 0.03$ \\  
10c & \CIV & $4.88126 \pm0.00004$ & & $27\pm 3$ &  $13.80\pm 0.03$ \\ 
%\hline
%\CIV$^{\rm a}$ & $4.88340 \pm0.00008$ &  $41\pm 7$ &  $13.10\pm 0.05$ \\
%\CIV$^{\rm a}$ & $4.88694 \pm0.00004$ &  $33\pm 3$ &  $13.31\pm 0.03$ \\
11 & \CIV\ 1548b & $4.88340 \pm0.00008$ & $<0.057\pm0.006$ & $41\pm 7$ &  $13.10\pm 0.05$ \\
 & \CIV\ 1550 & & $0.019\pm0.005$ & & \\
12 & \CIV\ 1548 & $4.88694 \pm0.00004$ & $0.092\pm0.006$ & $33\pm 3$ &  $13.31\pm 0.03$ \\
   & \CIV\ 1550b & & $<0.072\pm0.006$ & & \\
\hline
\end{tabular}
Notes. b: Line blended with another absorption line; m: multiple
components; s: line contaminated by a sky emission line; l: feature below $3\,\sigma$ detection. 
\end{minipage}
\end{center}
\end{table*}

\begin{table*}
\begin{center}
\caption{\CIV\ Absorption systems in the spectrum of ULAS J1319+0950. In the first column, the letters denote the single
  velocity components which was not possible to deblend in the
  computation of the equivalent width.}
\begin{minipage}{120mm}
\label{tab_CIV_1319}
\begin{tabular}{@{} l l c c  c c}
%\hline  
%\multicolumn{4}{c}{ULAS J1319+0950} \\
\hline
System & Ion & $z$ & $W_0$ & $b$ & $\log N$   \\
& & & (\AA) & (km s$^{-1}$) & (cm$^{-2}$) \\
\hline
1 & \CIV\ 1548 & $4.61273\pm 0.00003$ & $0.078\pm0.004$ & $29\pm  2$ &  $13.24\pm 0.02$ \\ 
 & \CIV\ 1550 & & $0.044\pm0.004$ & & \\
%\CIV$^{\rm a}$ & $4.62931\pm 0.00008$ &  $36\pm  6$ &  $12.86\pm 0.06$ \\
%\CIV$^{\rm a}$ & $4.64478\pm 0.00004$ &   $7\pm  4$ &  $12.72\pm 0.05$ \\
2 & \CIV\ 1548 & $4.62931\pm 0.00008$ & $0.035\pm0.004$ & $36\pm  6$ &  $12.86\pm 0.06$ \\
 & \CIV\ 1550b & & $< 0.069\pm0.004$ & & \\
3 & \CIV\ 1548 & $4.64478\pm 0.00004$ & $0.023\pm0.004$ &  $7\pm  4$ &  $12.72\pm 0.05$ \\
 & \CIV\ 1550b & & $<0.017\pm0.002$ & & \\
4 & \CIV\ 1548b & $4.65317\pm 0.00005$ & $0.043\pm0.003$ & $34\pm  5$
&  $13.01\pm 0.04$ \\ 
 & \CIV\ 1550b & & $< 0.120\pm0.004$ & & \\
%\CIV$^{\rm a,b}$ & $4.66010\pm 0.00003$ &  10 &  $12.89\pm 0.05$ \\
%\hline
5 & \CIV\ 1548mb & $4.663$ & $0.34\pm0.01$ & & \\
 & \CIV\ 1550mb & & $<0.189\pm0.008$ & & \\
5a & \CIV & $4.66010\pm 0.00003$ &  &  10 &  $12.89\pm 0.05$ \\
5b & \CIV & $4.66127 \pm  0.00005$ & & $29\pm  4$ &  $13.12\pm 0.05$ \\
5c & \CIV & $4.66294 \pm  0.00003$ & & $18\pm  3$ &  $13.09\pm 0.05$ \\
5d & \CIV & $4.66482 \pm  0.00006$ & & $68\pm  6$ &  $13.62\pm 0.03$ \\   
%\hline
6 & \CIV\ 1548s & $4.70325 \pm  0.00002$ & $0.05\pm0.01$ & $10\pm  3$
&  $13.50\pm 0.03$ \\  
 & \CIV\ 1550 & & $0.061\pm0.003$ & & \\
%\hline
7 & \CIV\ 1548m & $4.7167$ & $ 0.155\pm0.008$ & & \\
 & \CIV\ 1550m & & $0.111\pm0.007$ & & \\
7a & \CIV & $4.71659 \pm   0.00002$ & & $13\pm  3$ &  $13.27\pm 0.05$ \\
7b & \CIV & $4.7169\pm 0.0001$  & & $74\pm  9$ &  $13.37\pm 0.05$ \\ 
%\hline
%\CIV$^{\rm a}$ & $5.2627\pm 0.0001$  &  $57\pm10$ &  $12.83\pm 0.14$ \\ 
8 & \CIV\ 1548m & $5.2640$ & $0.090\pm0.009$ & & \\
 & \CIV\ 1550ms & & $0.06\pm0.01$ & & \\
 & \SiIV\ 1393m & & $0.089\pm0.005$ & & \\
 & \SiIV\ 1402mb & & $<0.105\pm0.004$ & & \\
8a & \CIV & $5.2627\pm 0.0001$  & & $57\pm10$ &  $12.83\pm 0.14$ \\ 
& \SiIV &   &  & &$12.69\pm0.06$ \\
%\CIV$^{\rm a}$ & $5.26452\pm 0.00006$ &  $24\pm 3$ &  $13.20\pm 0.05$\\ 
8b & \CIV & $5.26452\pm 0.00006$ & & $24\pm 3$ &  $13.20\pm 0.05$\\ 
& \SiIV & &  & &$12.61\pm0.05$ \\
%\hline
9 & \CIV\ 1548 & $5.37488\pm 0.00004$ & $0.17\pm0.01$ & $32\pm  4$ &  $13.59\pm 0.03$ \\
 & \CIV\ 1550 & & $0.08\pm0.01$ & & \\
 & \SiIV\ 1393m & & $0.111\pm0.004$ & & \\
 & \SiIV\ 1402m & & $0.074\pm0.003$ & & \\
9a & \SiIV & $5.37427\pm0.00005$ & &$11\pm6$ & $12.32\pm0.07$ \\
9b & \SiIV & $5.37508\pm0.00001$ & & $10$ & $13.17\pm0.02$ \\
%\hline
%\CIV$^{\rm a}$ & $5.57049 \pm0.00003$  &  $34\pm3$ &  $13.97\pm 0.10$\\ 
10 & \CIV\ 1548s & $5.57049 \pm0.00003$  & $0.4\pm0.1$ & $34\pm3$ &  $13.97\pm 0.10$\\ 
 & \CIV\ 1550s & & $0.28\pm0.07$ & & \\
 & \SiIV\ 1393 &  & $0.108\pm0.006$ & & $13.04\pm 0.02$ \\
 & \SiIV\ 1402 & & $0.062\pm0.005$ & & \\
 & \CII\ 1334 & & & & $<13.5$ \\
%\CIV$^{\rm a}$ & $5.5740\pm 0.0001$  &  $36\pm 11$ &  $14.09\pm 0.08$\\ 
%\hline
11 & \CIV\ 1548s & $5.5740\pm 0.0001$  & $0.35\pm0.06$ & $36\pm 11$ &  $14.09\pm 0.08$\\ 
   & \CIV\ 1550s & & $0.29\pm0.05$ & & \\
   & \SiIV\ 1393m & $5.5739$ & $0.238\pm0.006$ && \\
   & \SiIV\ 1402m & & $0.178\pm0.005$ & & \\
   & \CII\ 1334 & $5.57372\pm0.00001$ & & $12\pm1$ & $13.88\pm0.02$ \\
11a & \SiIV & $5.57358\pm 0.00001$ & & $20\pm1$ & $13.52\pm0.01$ \\
11b & \SiIV & $5.57457\pm 0.00003$ & & $10$ & $12.71\pm0.05$ \\
\hline
\end{tabular}
Notes. b: Line blended with another absorption line; m: multiple
components; s: line contaminated by a sky emission line; l: feature
below $3\,\sigma$ detection.
\end{minipage}
\end{center}
\end{table*}

\begin{table*}
\begin{center}
\caption{\CIV\ absorption systems in the spectrum of CFHQS
  J1509-1749. In the first column, the letters denote the single
  velocity components which was not possible to deblend in the
  computation of the equivalent width.}
\begin{minipage}{120mm}
\label{tab_CIV_1509}
\begin{tabular}{@{} l l c c c c}
%\hline  
%\multicolumn{4}{c}{CFHQS J1509-1749} \\
\hline
System & Ion & $z$ & $W_0$ & $b$ & $\log N$   \\
& & & (\AA) & (km s$^{-1}$) & (cm$^{-2}$) \\
\hline
1 & \CIV\ 1548 & $4.61086\pm0.00004$ & $0.072\pm0.007$ & $34\pm5$ & $13.31\pm0.03$ \\
 & \CIV\ 1550 & & $0.034\pm0.007$ & & \\
2 & \CIV\ 1548$^{\rm a}$ & $4.64188\pm0.00003$ & $0.105\pm0.008$ &
$19\pm3$ & $13.39\pm0.03$\\
 & \CIV\ 1550b & & $>0.085\pm0.007$ & & \\
3 & \CIV\ 1548b$^{\rm a}$ & $4.6501\pm0.0002$ & $>0.033\pm0.005$ & $68\pm15$ & $13.15\pm0.08$ \\
 & \CIV\ 1550s & & $0.03\pm0.01$ & & \\
4 & \CIV\ 1548$^{\rm a}$ & $4.65508\pm0.00006$ & $0.040\pm0.005$ &
$16\pm8$ & $12.93\pm0.06$ \\
 & \CIV\ 1550s & & $0.04\pm0.01$ & & \\
6 & \CIV\ 1548$^{\rm a}$ & $4.66629\pm0.00004$ & $0.08\pm0.01$ &
$24\pm4$ & $13.21\pm0.04$ \\
 & \CIV\ 1550 & & $0.043\pm0.006$ && \\
%\CIV$^{\rm a}$ & $4.7690\pm0.0001$  & $44\pm10$ & $13.13\pm0.08$ \\
7 & \CIV\ 1548 & $4.7690\pm0.0001$ & $0.06\pm0.01$ & $44\pm10$ & $13.13\pm0.08$ \\
& \CIV\ 1550s & & $0.014\pm0.006$ & & \\
8 & \CIV\ 1548 & $4.79182\pm0.00005$ & $0.14\pm0.01$ & $46\pm4$ & $13.51\pm0.03$ \\
 & \CIV\ 1550 & & $0.073\pm0.009$ & & \\
9 & \CIV\ 1548s$^{\rm b}$ & $4.81568\pm0.00003$ & $0.24\pm0.02$ &
$39\pm3$ & $13.98\pm0.02$ \\
 & \CIV\ 1550 & & $0.20\pm0.01$ & & \\
10 & \CIV\ 1548s$^{\rm b}$ & $5.91572\pm0.00006$ & $0.31\pm0.06$ & $18\pm5$ & $14.11\pm0.14$ \\
 & \CIV\ 1550s & & $0.22\pm0.06$ & & \\
 & \SiIV\ 1393 & & $0.09\pm0.01$ & $18\pm5$ & $13.04\pm0.06$ \\
 & \SiIV\ 1402s & & $ 0.09\pm0.04$ & & \\
 & \CII\ 1334 & & & & $<13.4$ \\
\hline
\end{tabular} 
% (a) Features below $3\,\sigma$ detection;  
Notes. b: Line blended with another absorption line; m: multiple
components; s: line contaminated by a sky emission line; l: feature
below $3\,\sigma$ detection. 
$^{\rm a}$We found an alternative identification for this line: it could be
\AlIII\ at $z_{\rm abs}\simeq3.71$. $^{\rm b}$The line could be saturated
\end{minipage}
\end{center}
\end{table*}

\FloatBarrier

\section{Other Absorption Systems}

\subsection{SDSS J0818+1722}

\begin{figure*}
\begin{center}
\includegraphics[width=8cm]{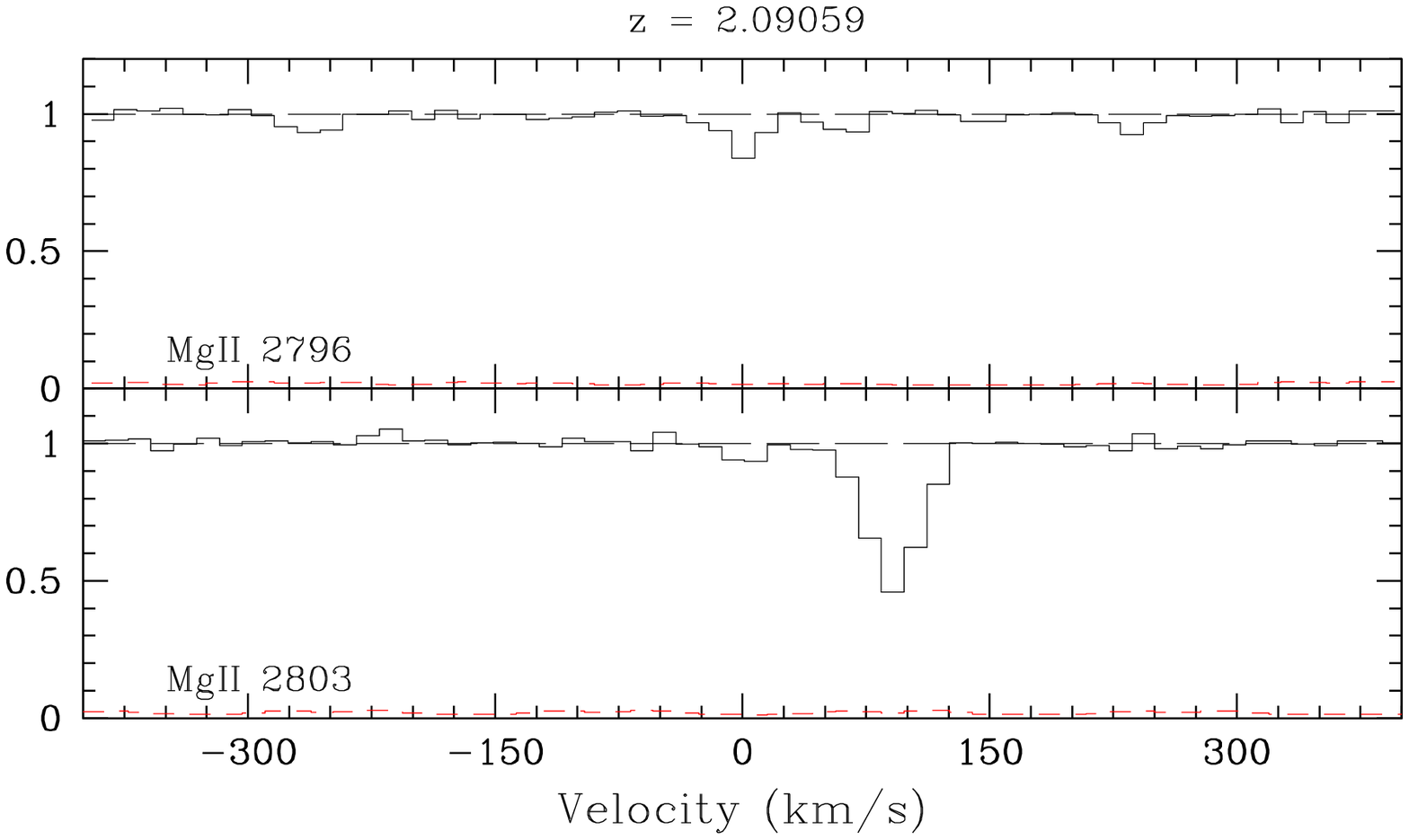}
\includegraphics[width=8cm]{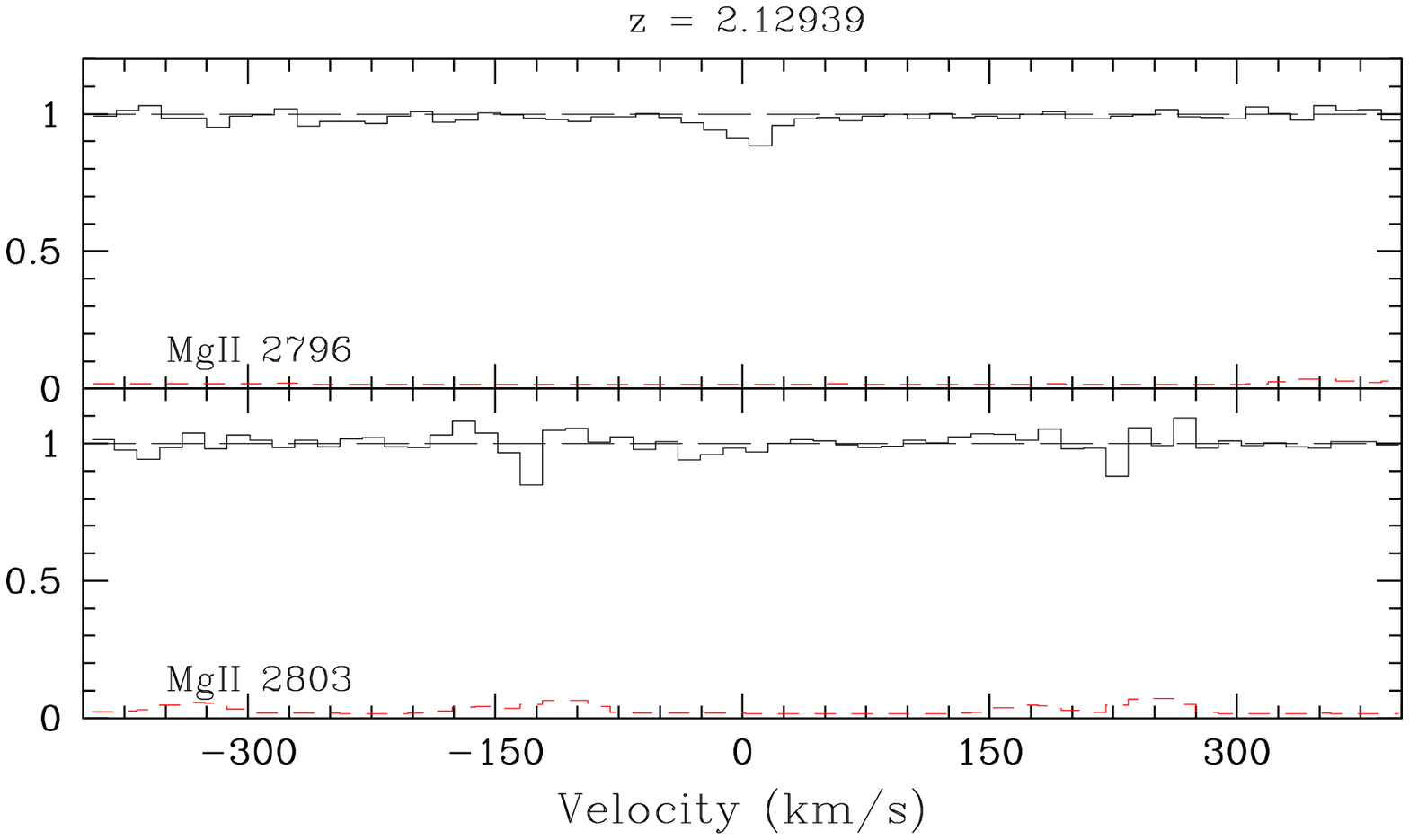}
\includegraphics[width=8cm]{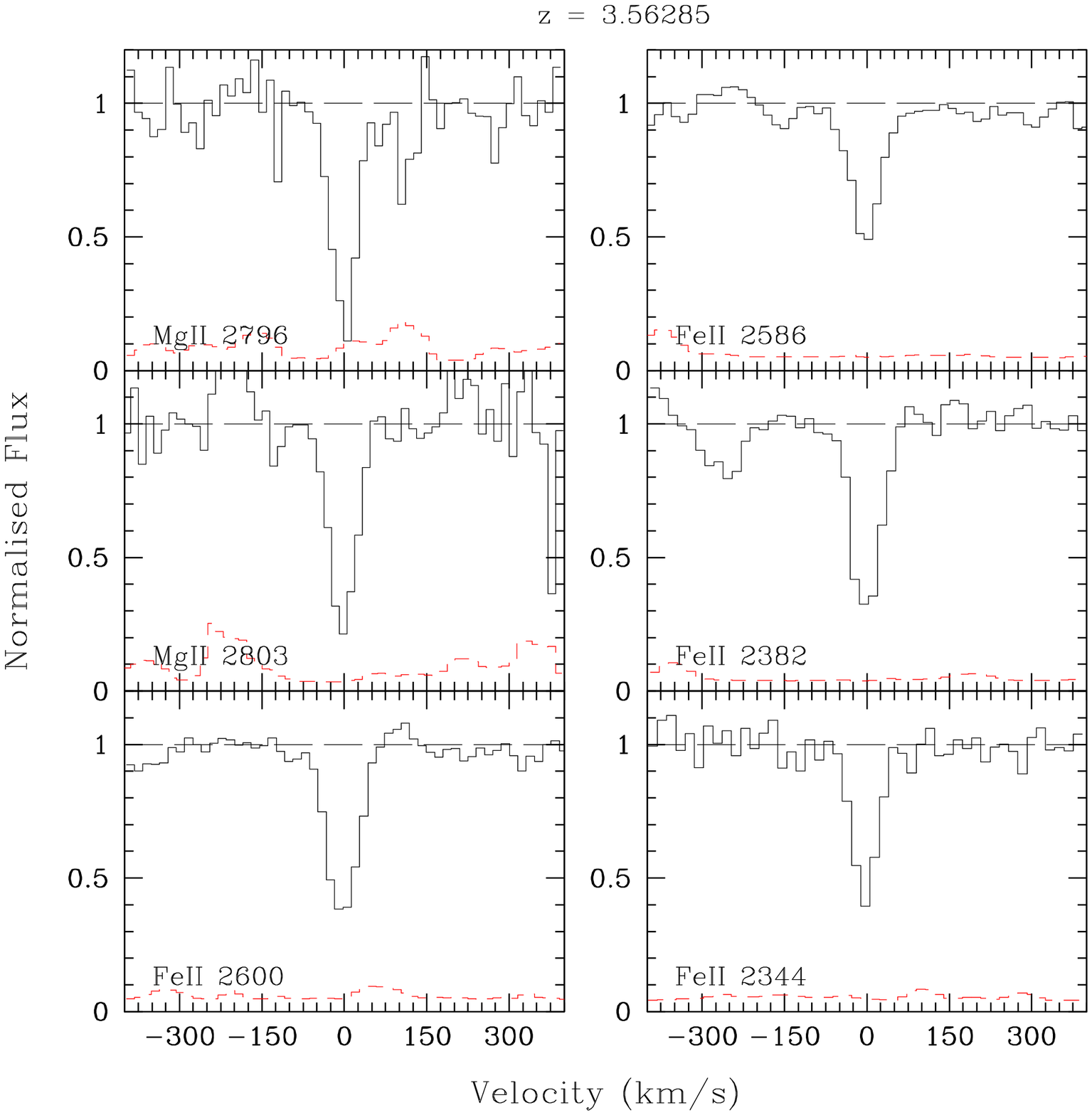}
\caption{\MgII\ systems in the spectrum of SDSS J0818+1722. }   
\label{J0818_mgII}
\end{center}
\end{figure*}

\subsection{SDSS J0836+0054}

\begin{figure*}
\begin{center}
\includegraphics[width=8cm]{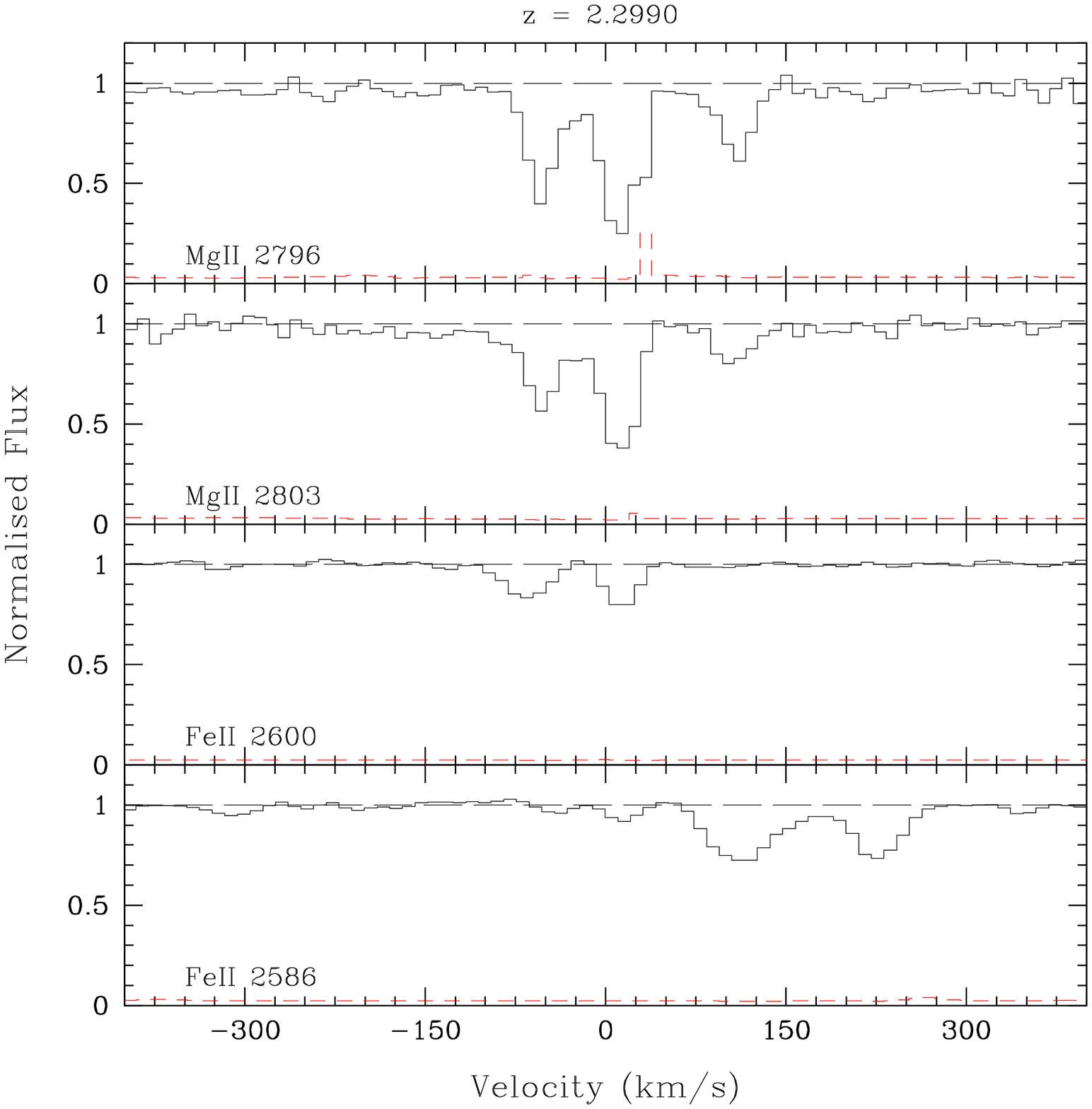}
\includegraphics[width=8cm]{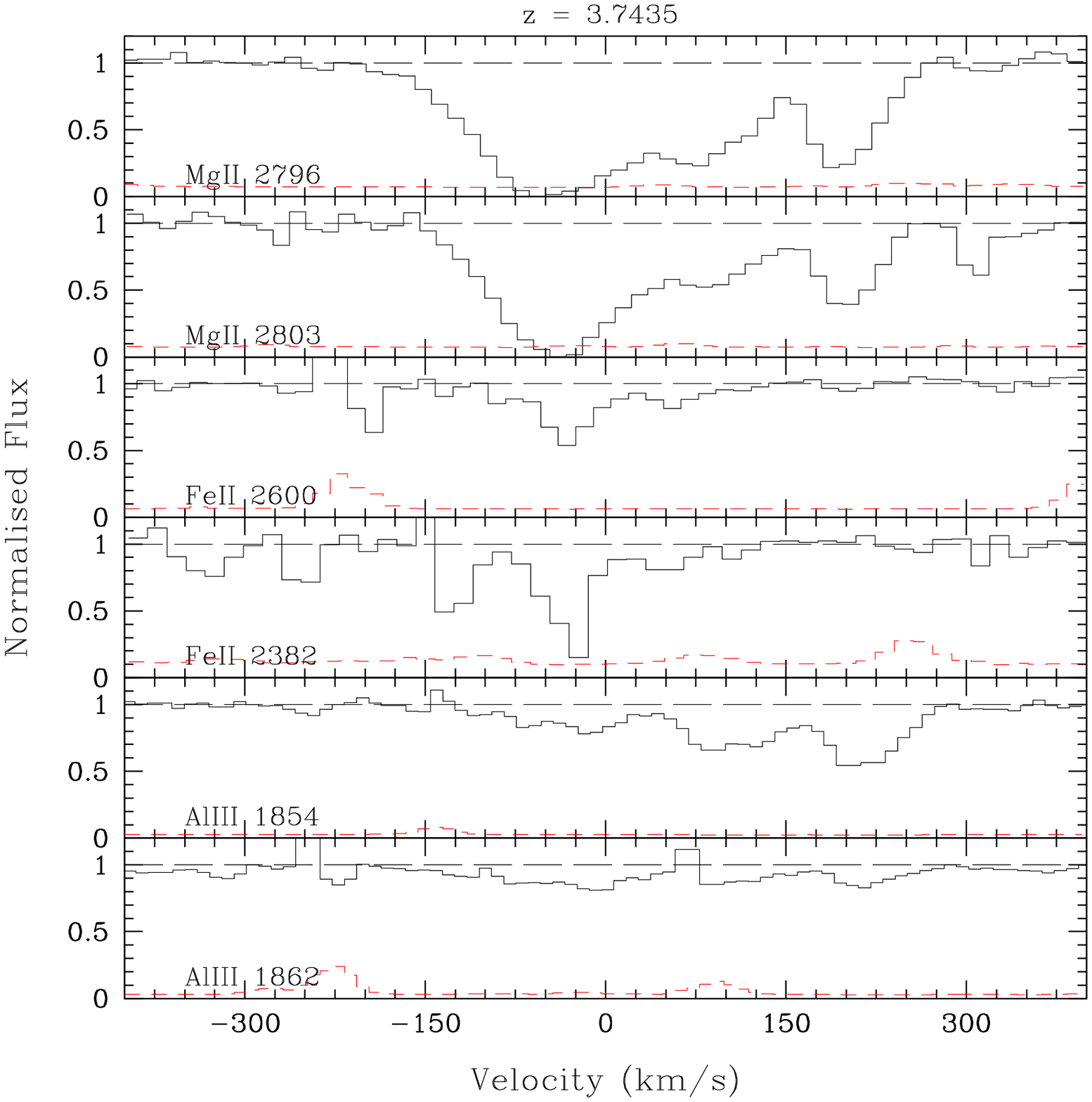}
\caption{\MgII\ systems in the spectrum of SDSS J0836+0054. }   
\label{J0836_mgII}
\end{center}
\end{figure*}

\subsection{SDSS J1030+0524}

\begin{figure*}
\begin{center}
\includegraphics[width=8cm]{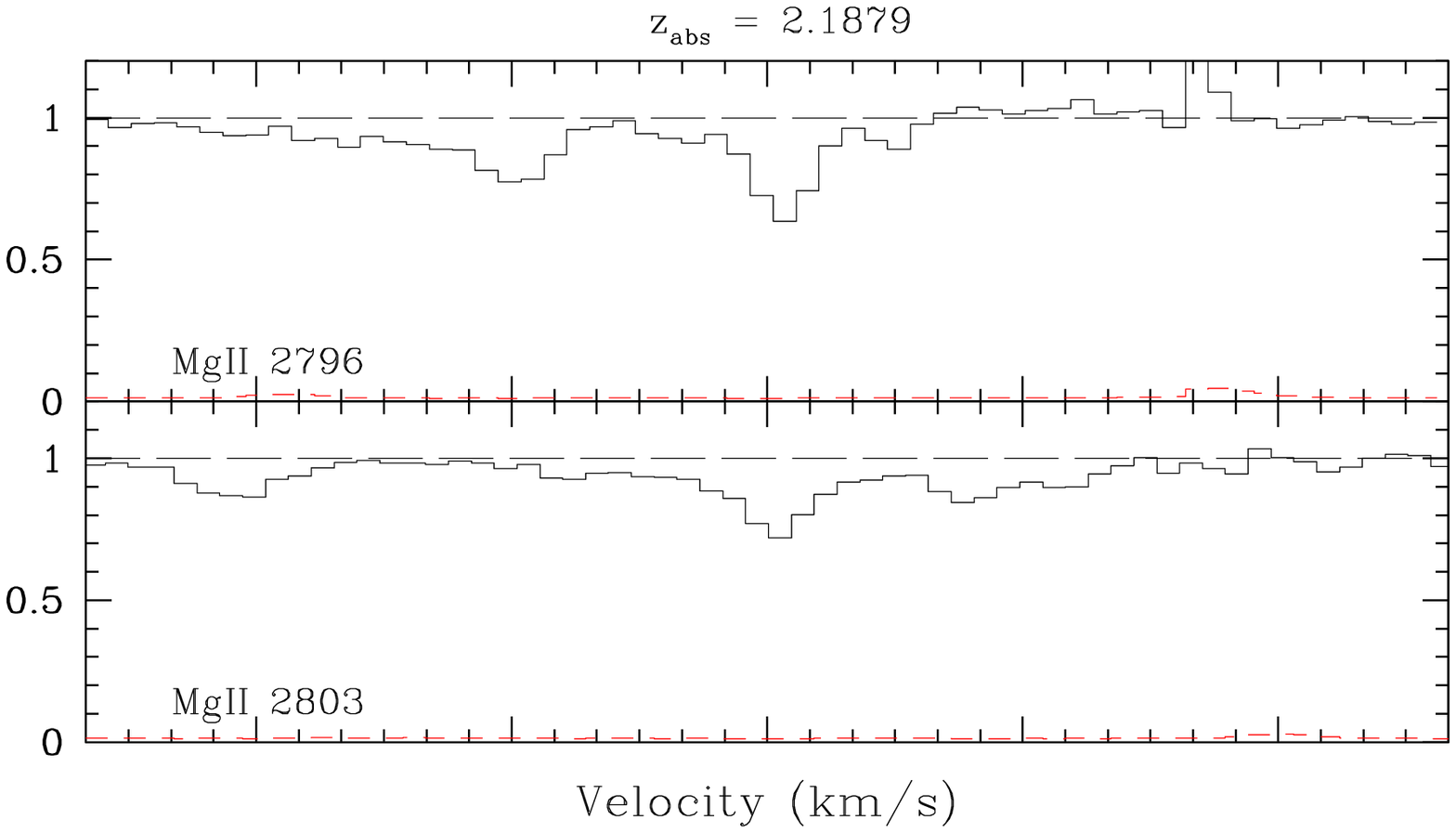}
\includegraphics[width=8cm]{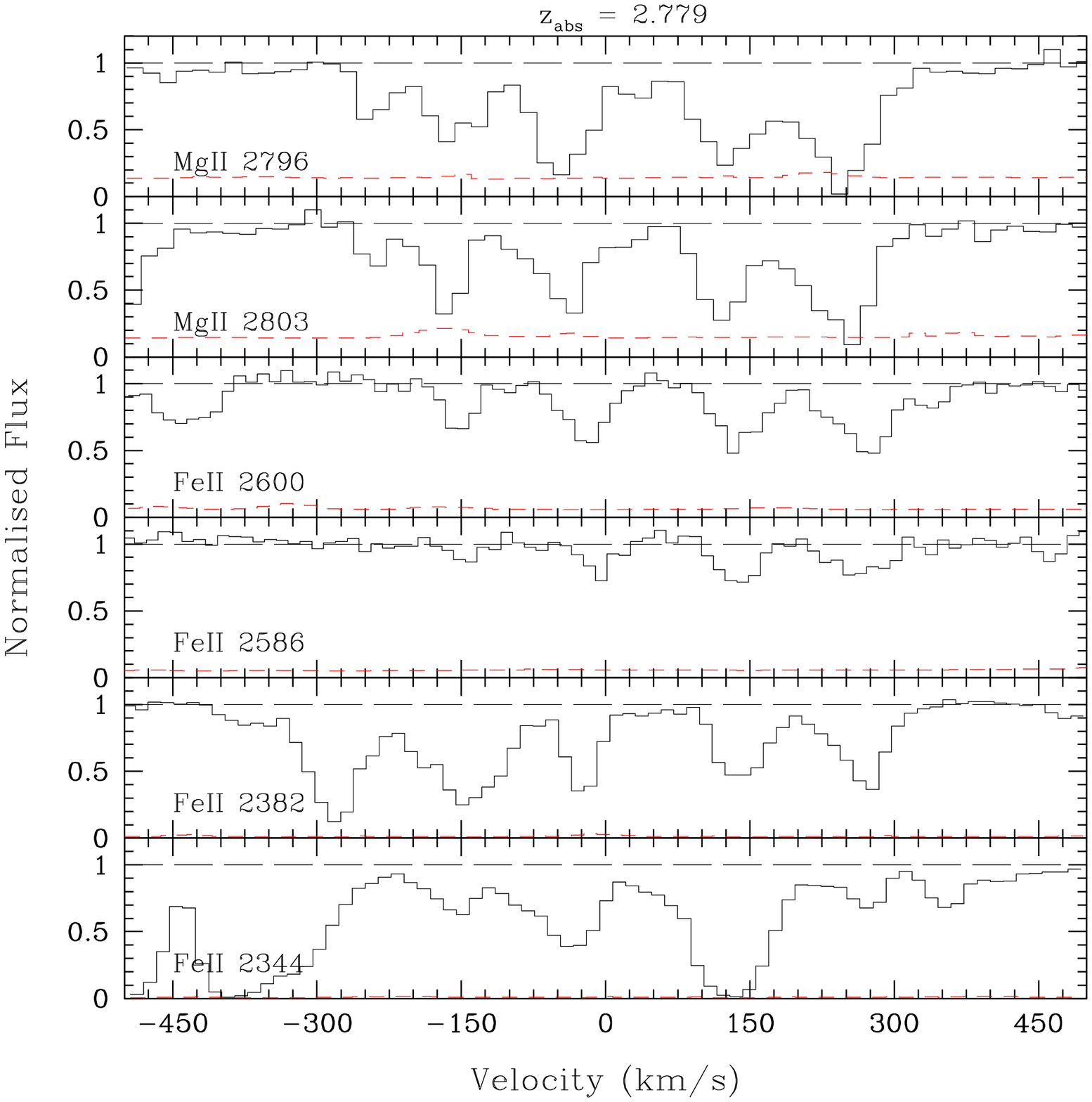}
\includegraphics[width=8cm]{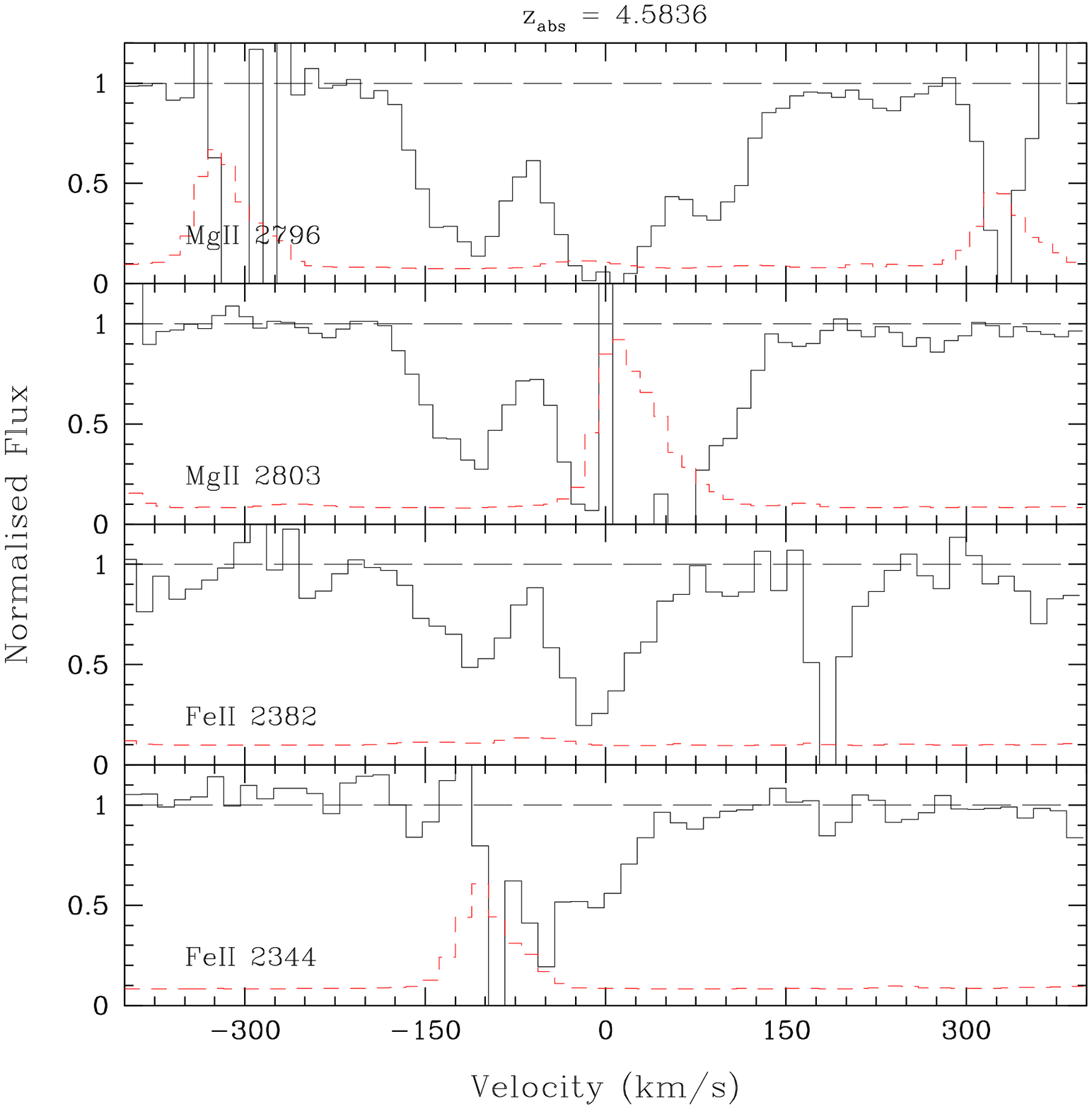}
\caption{\MgII\ systems in the spectrum of SDSS J1030+0524. }   
\label{J1030_mgII_1}
\end{center}
\end{figure*}

\begin{figure*}
\begin{center}
\includegraphics[width=8cm]{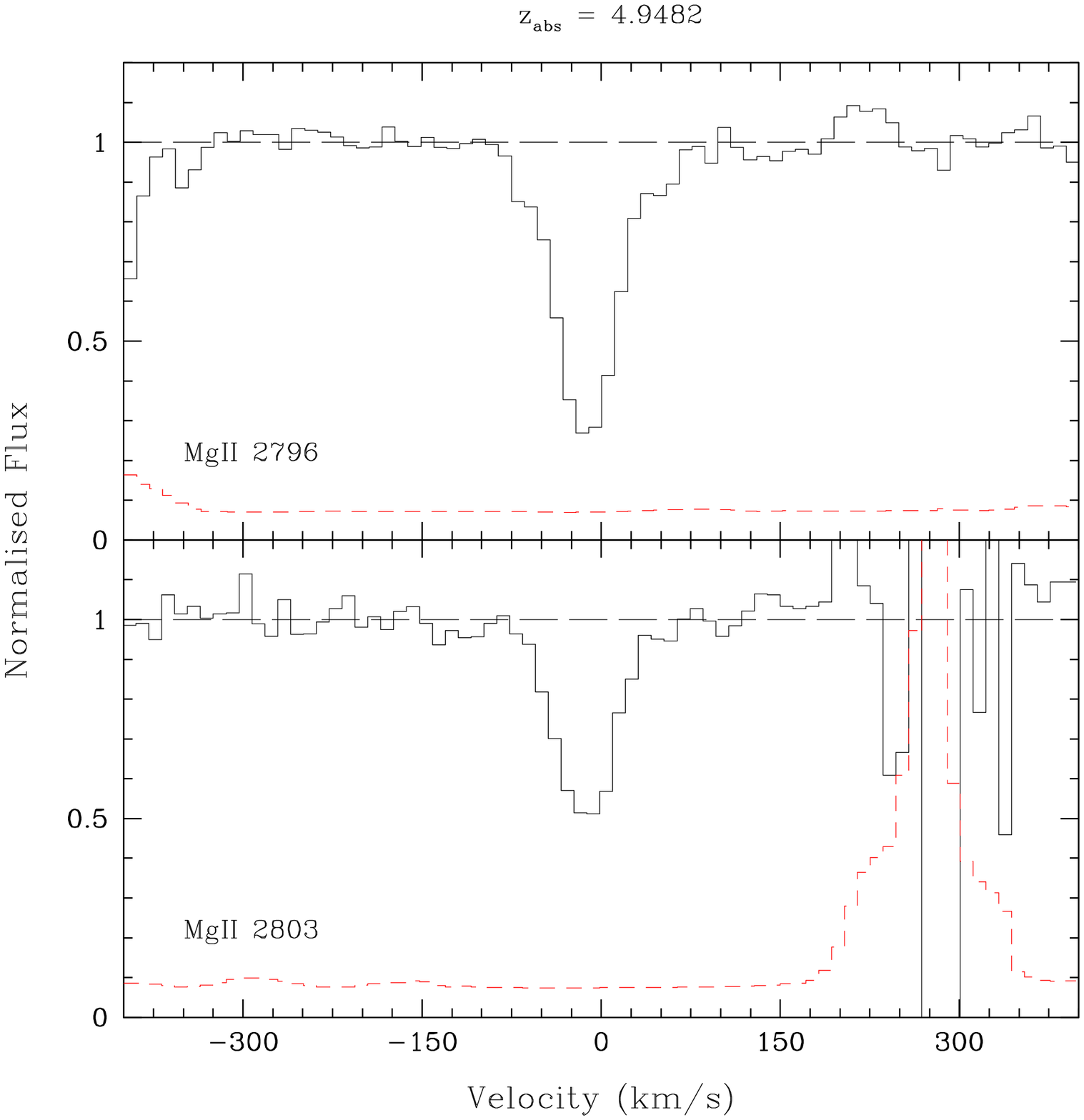}
\includegraphics[width=8cm]{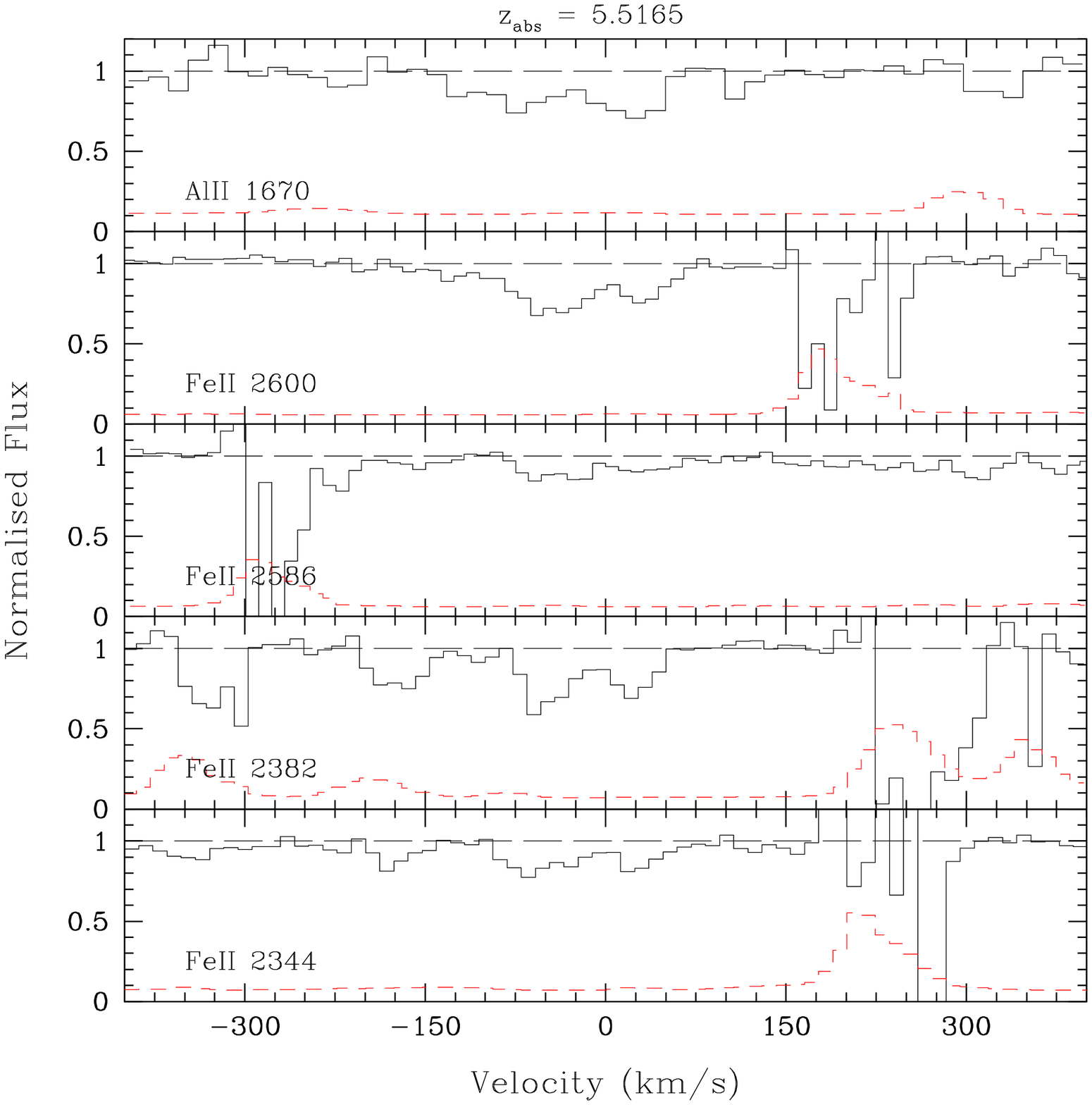}
\includegraphics[width=8cm]{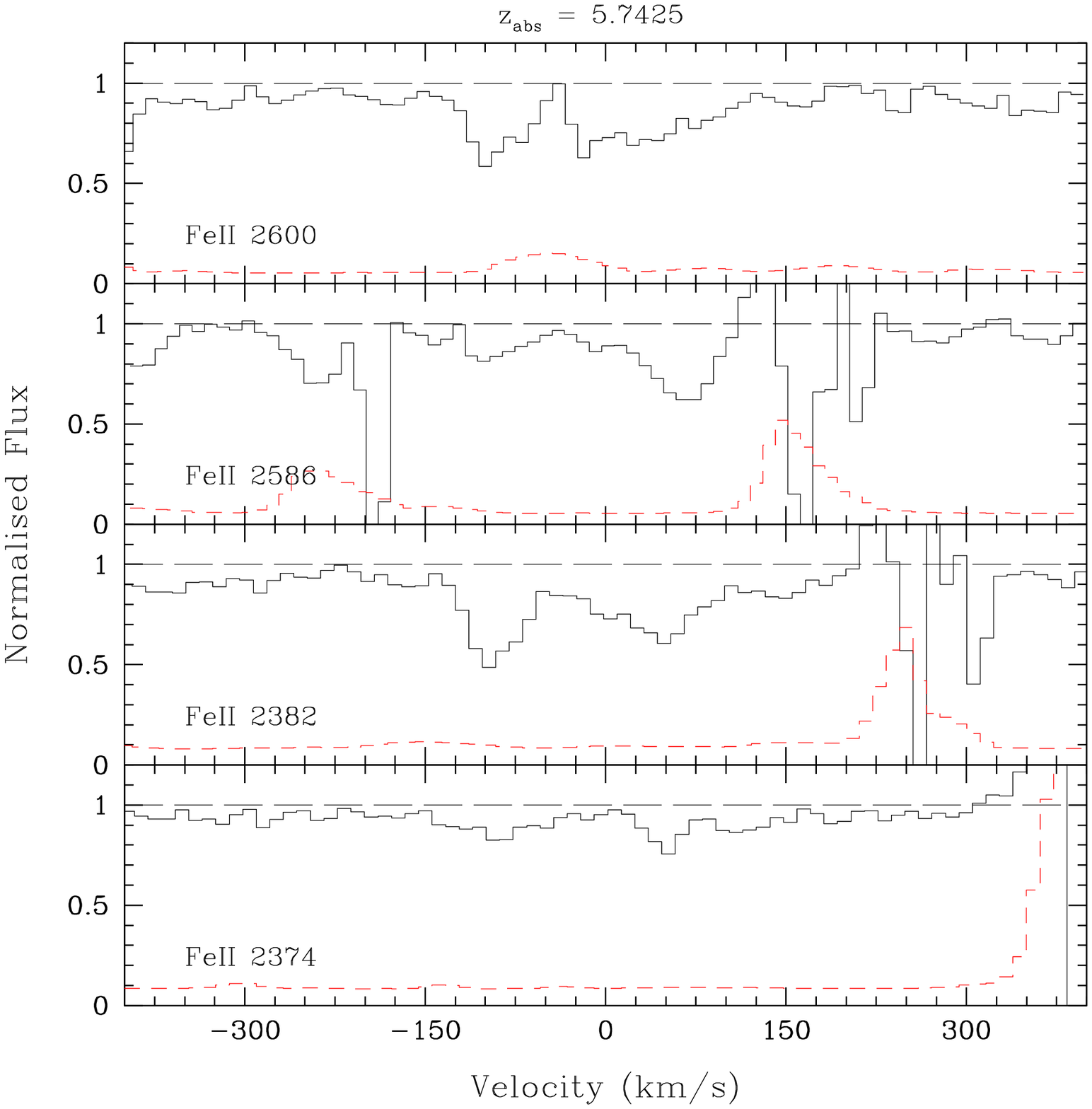}
\caption{\MgII\ systems in the spectrum of SDSS J1030+0524
  (continuation of Fig.~\ref{J1030_mgII_1}). }   
\label{J1030_mgII_2}
\end{center}
\end{figure*}

\subsection{SDSS J1306+0356}

\begin{figure*}
\begin{center}
\includegraphics[width=8cm]{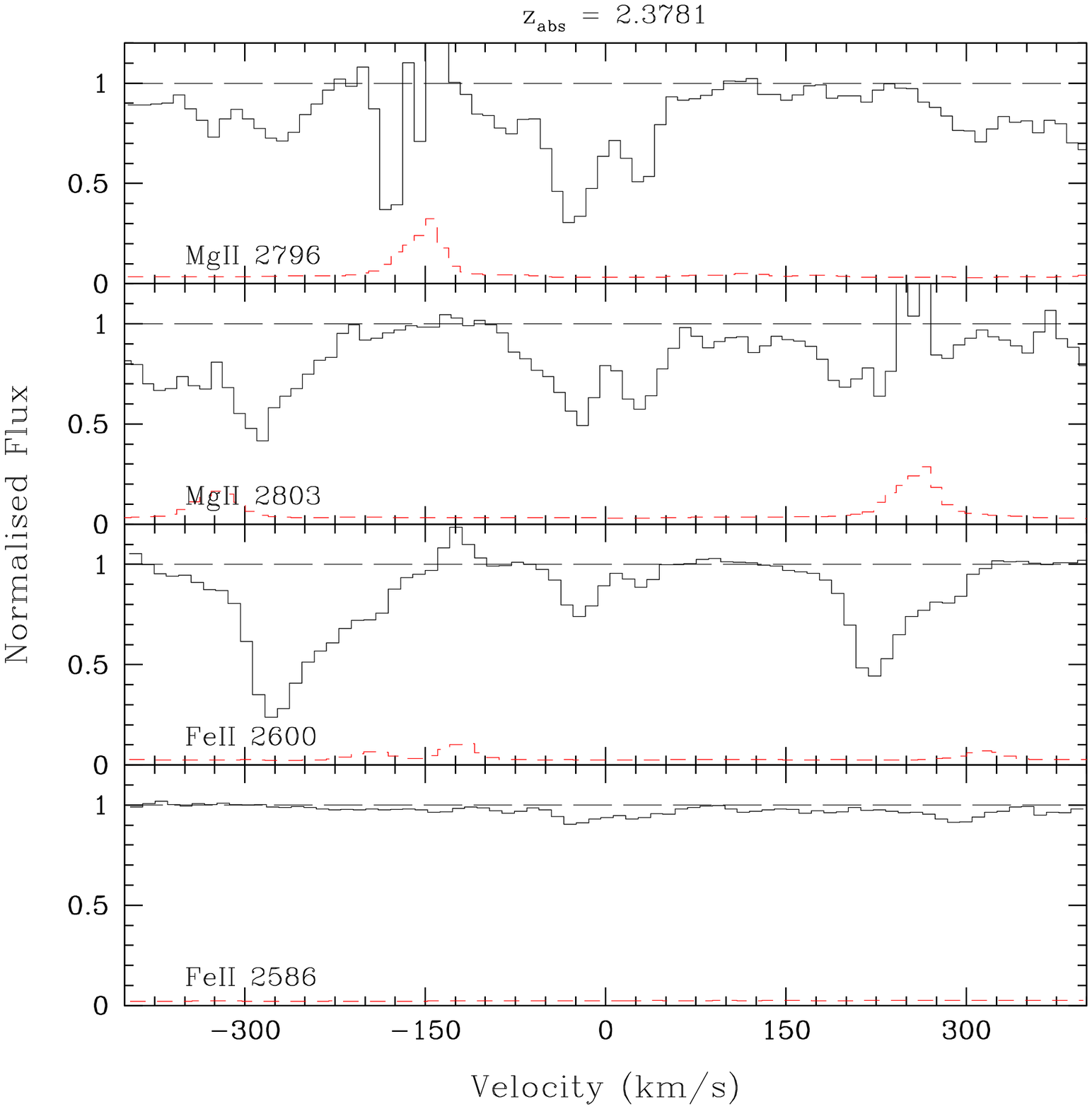}
\includegraphics[width=8cm]{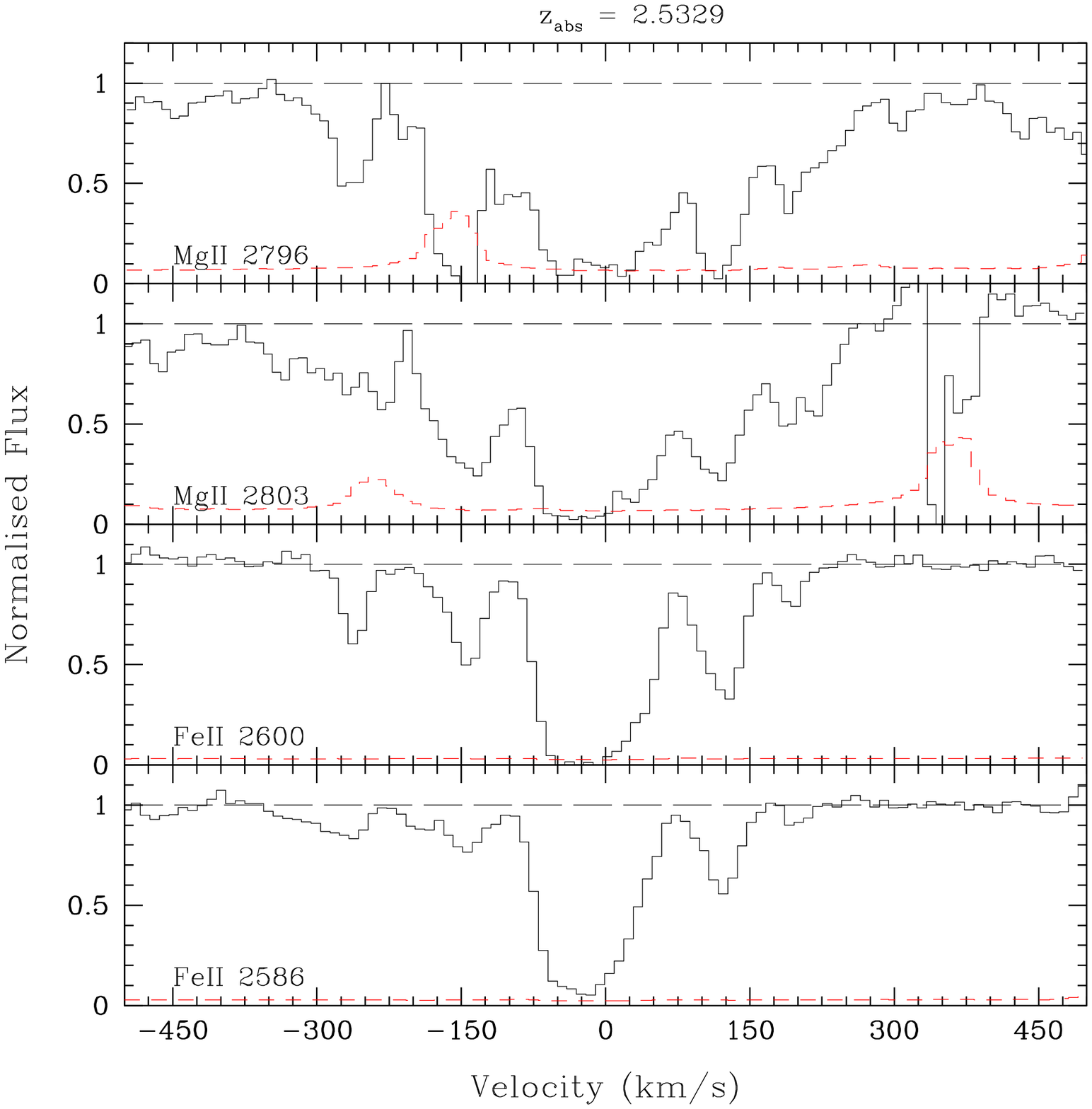}
\includegraphics[width=8cm]{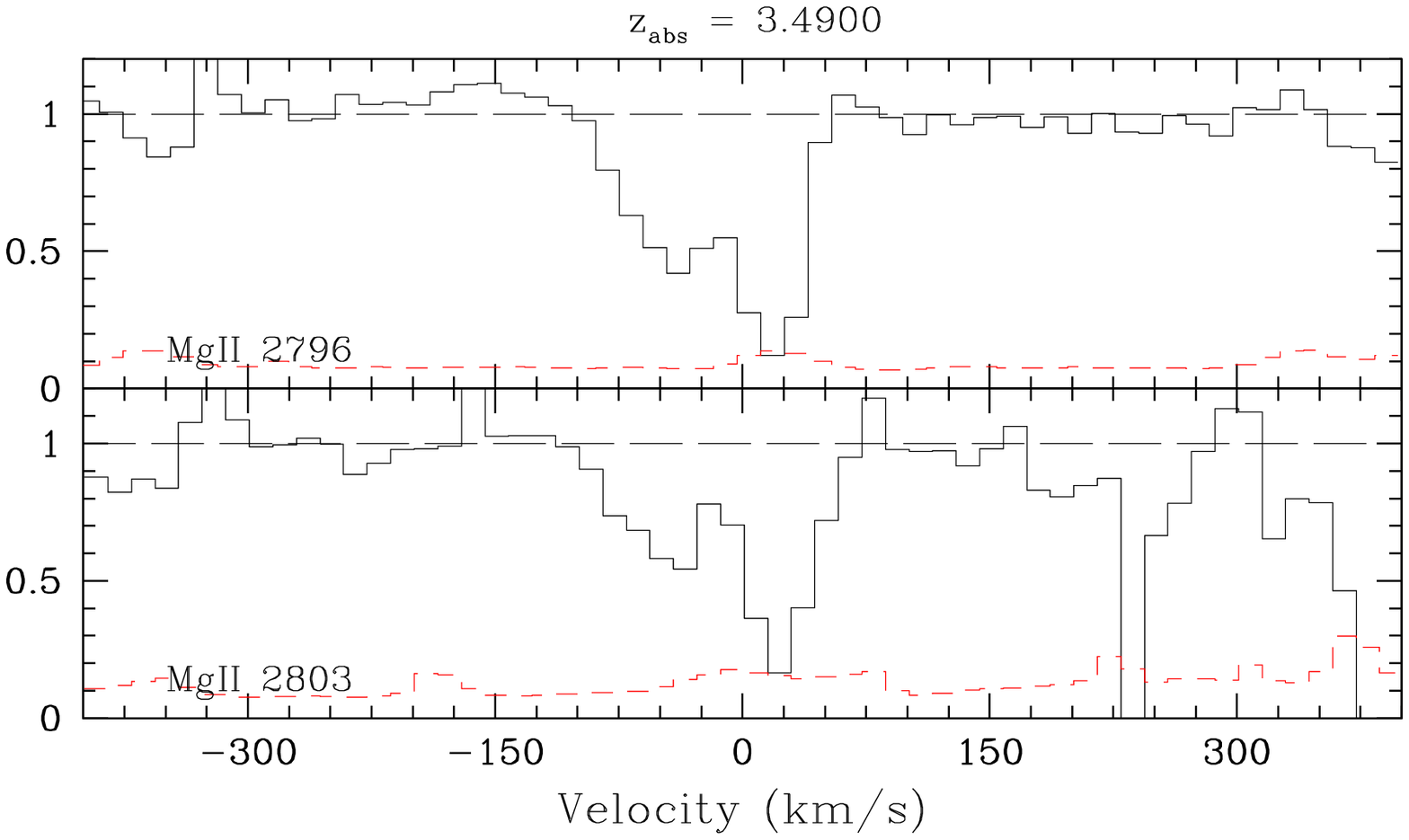}
\caption{\MgII systems in the spectrum of SDSS J1306+0356. }   
\label{J1306_mgII}
\end{center}
\end{figure*}

\begin{figure*}
\begin{center}
\includegraphics[width=8cm]{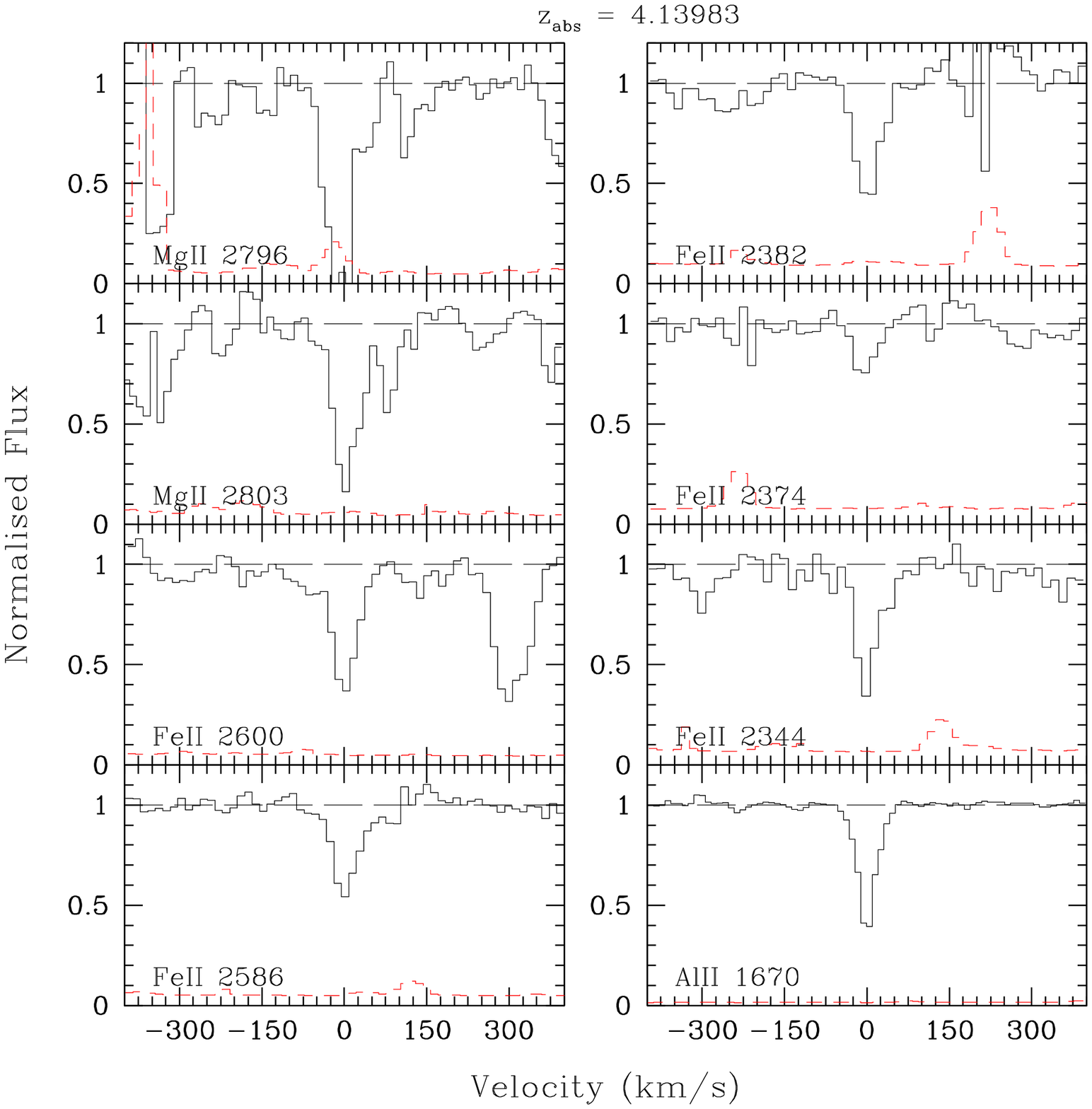}
\includegraphics[width=8cm]{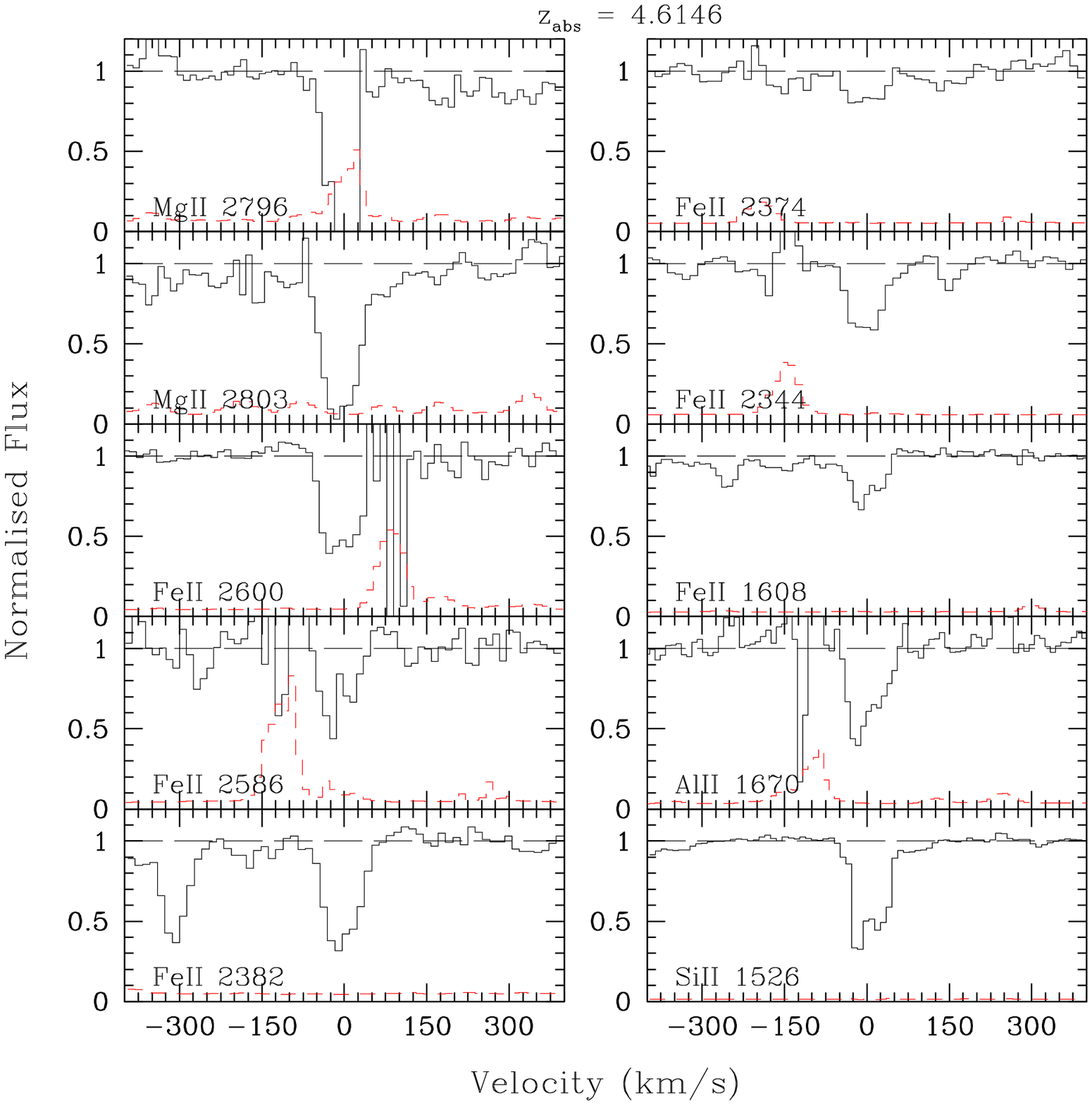}
\includegraphics[width=8cm]{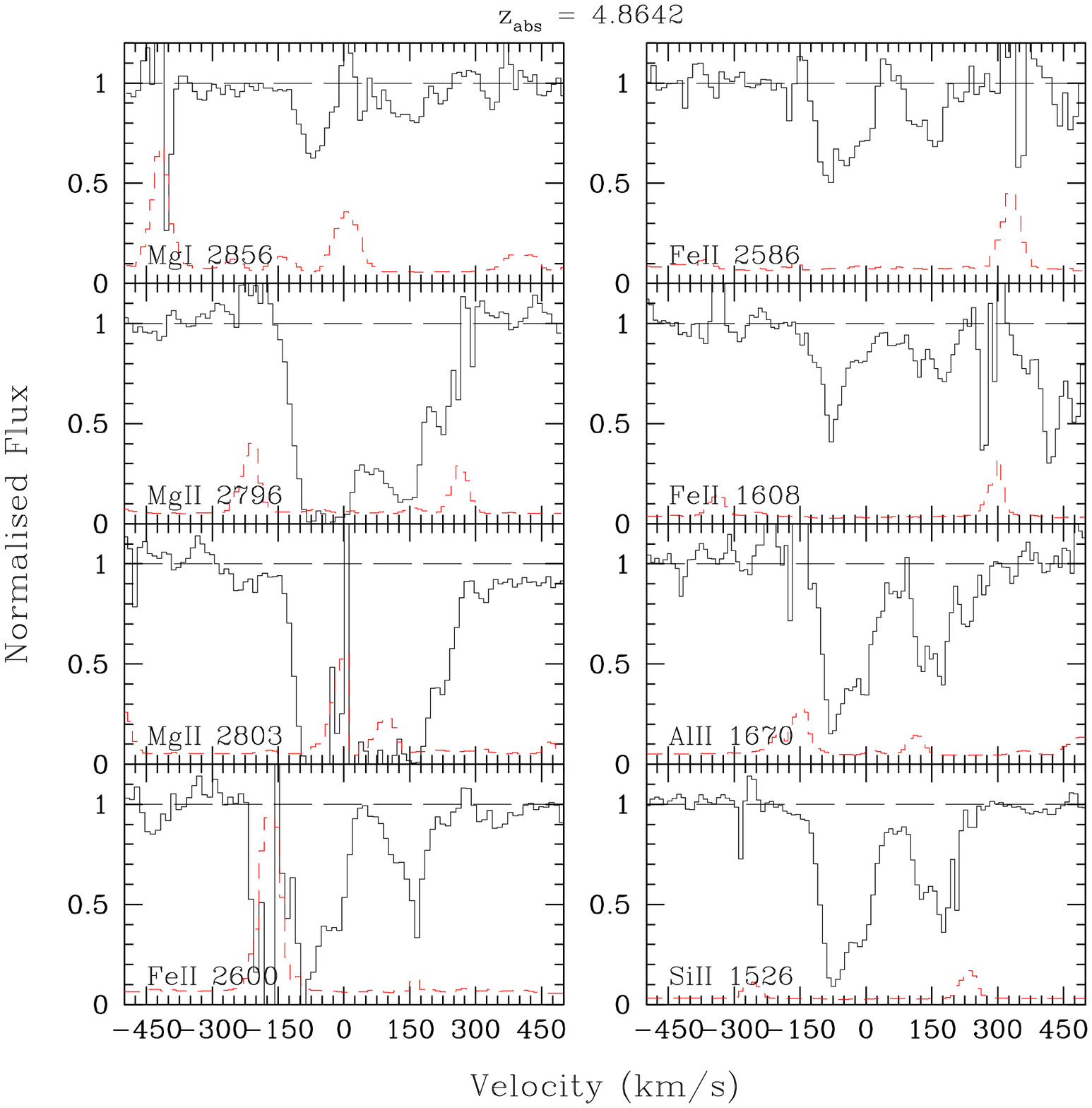}
\includegraphics[width=8cm]{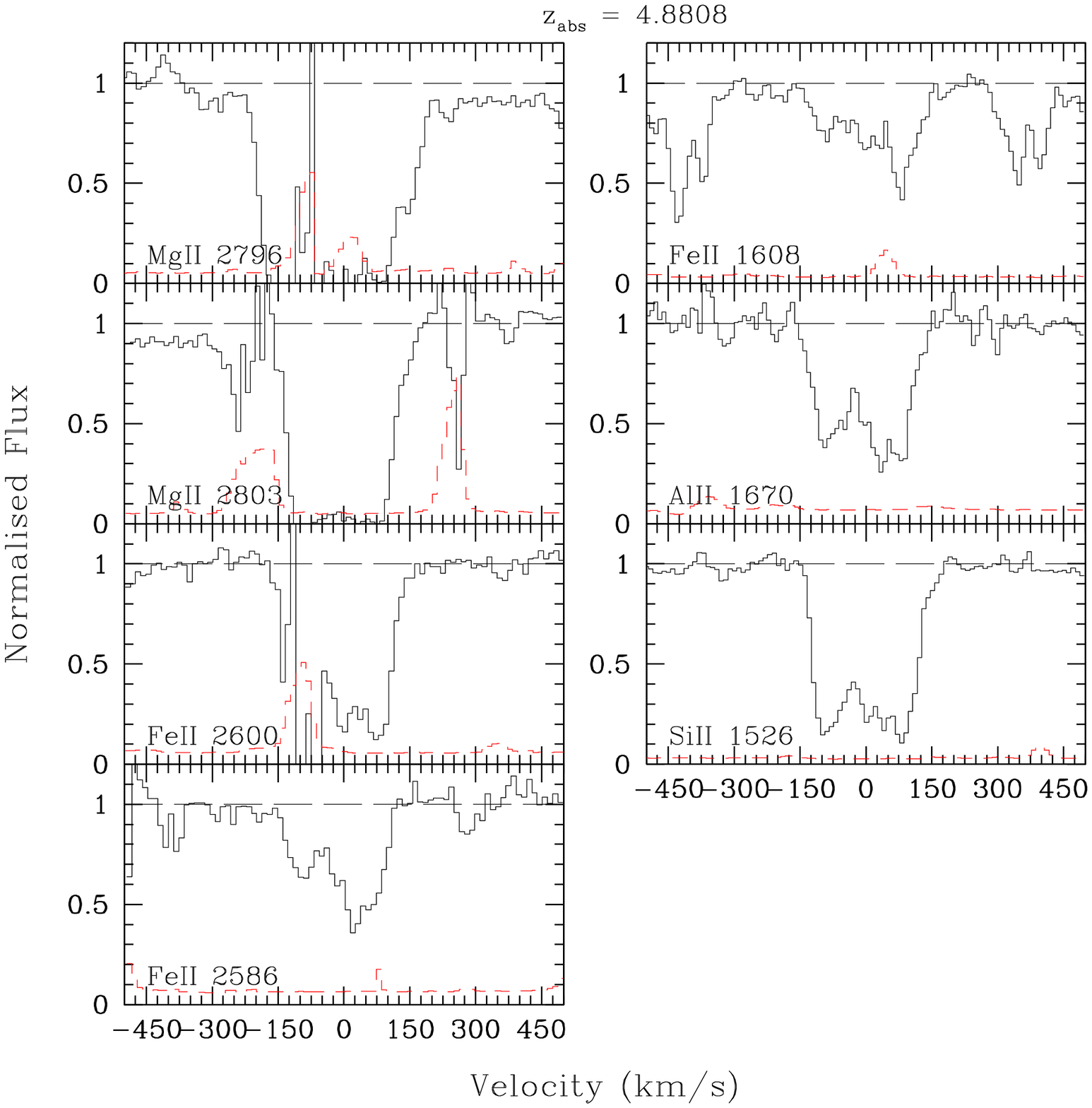}
\caption{\MgII\ systems in the spectrum of SDSS J1306+0356
  (continuation of Fig.~\ref{J1306_mgII}. }   
\label{J1306_mgII_2}
\end{center}
\end{figure*}

\subsection{ULAS J1319+0950}

\begin{figure*}
\begin{center}
\includegraphics[width=8cm]{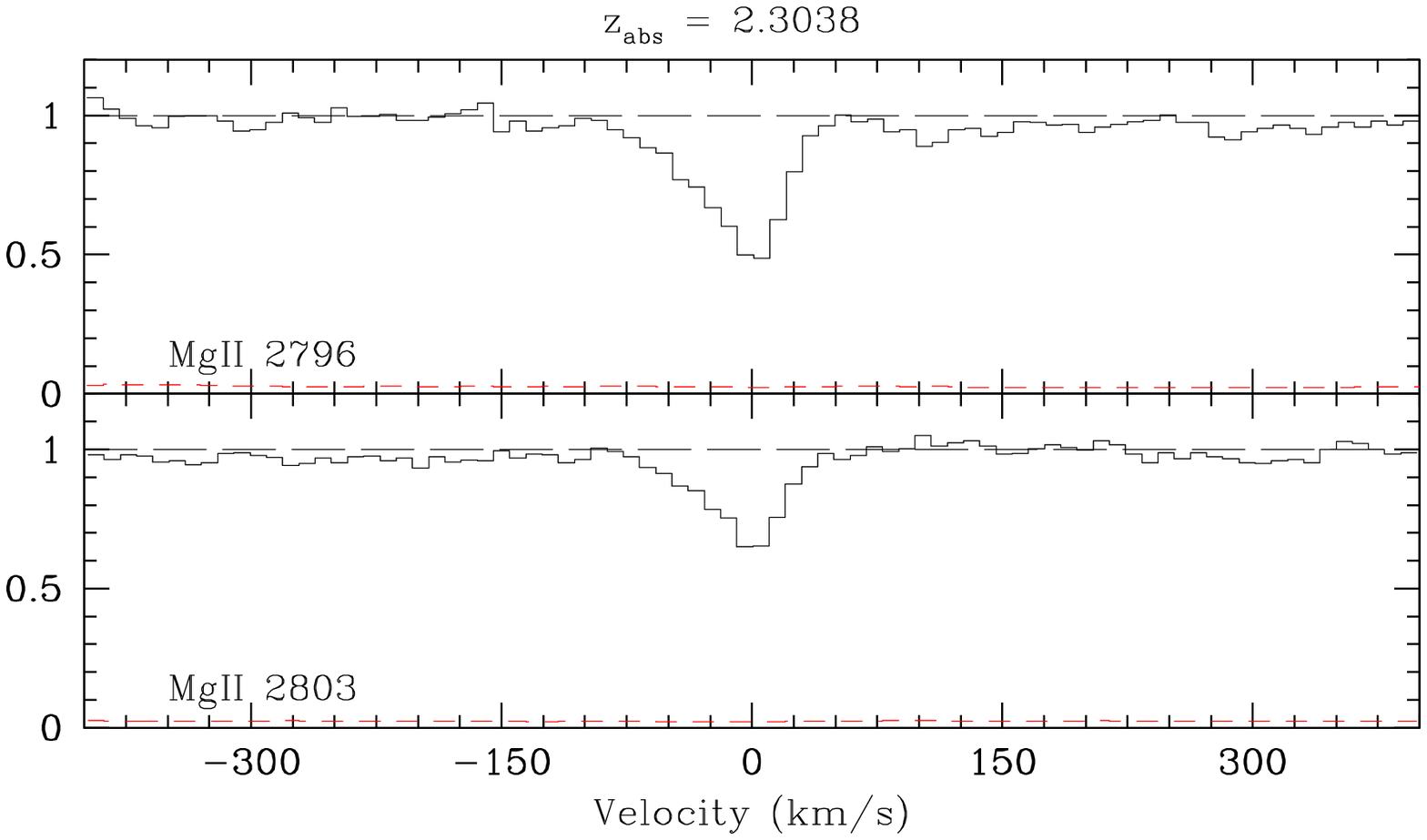}
\includegraphics[width=8cm]{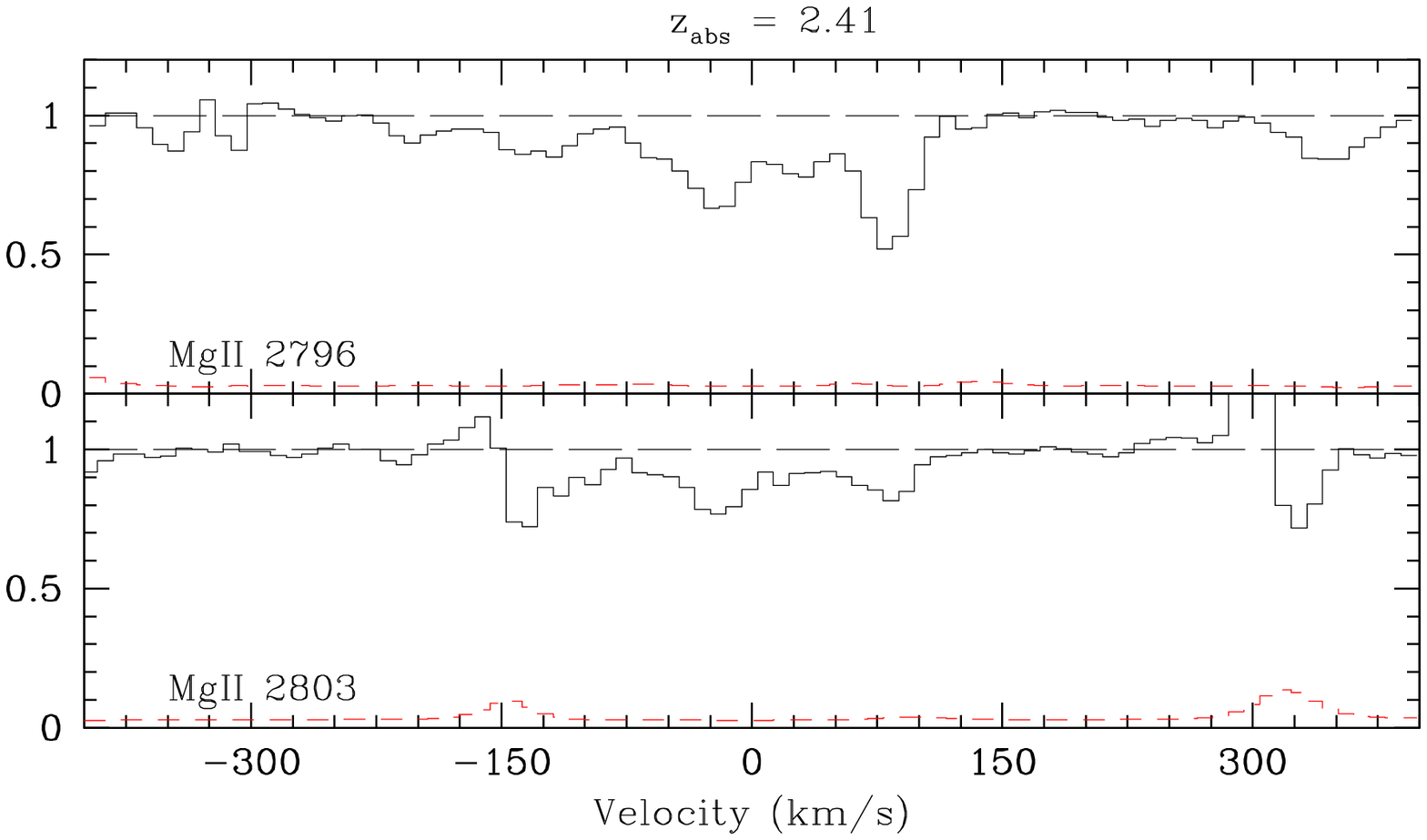}
\includegraphics[width=8cm]{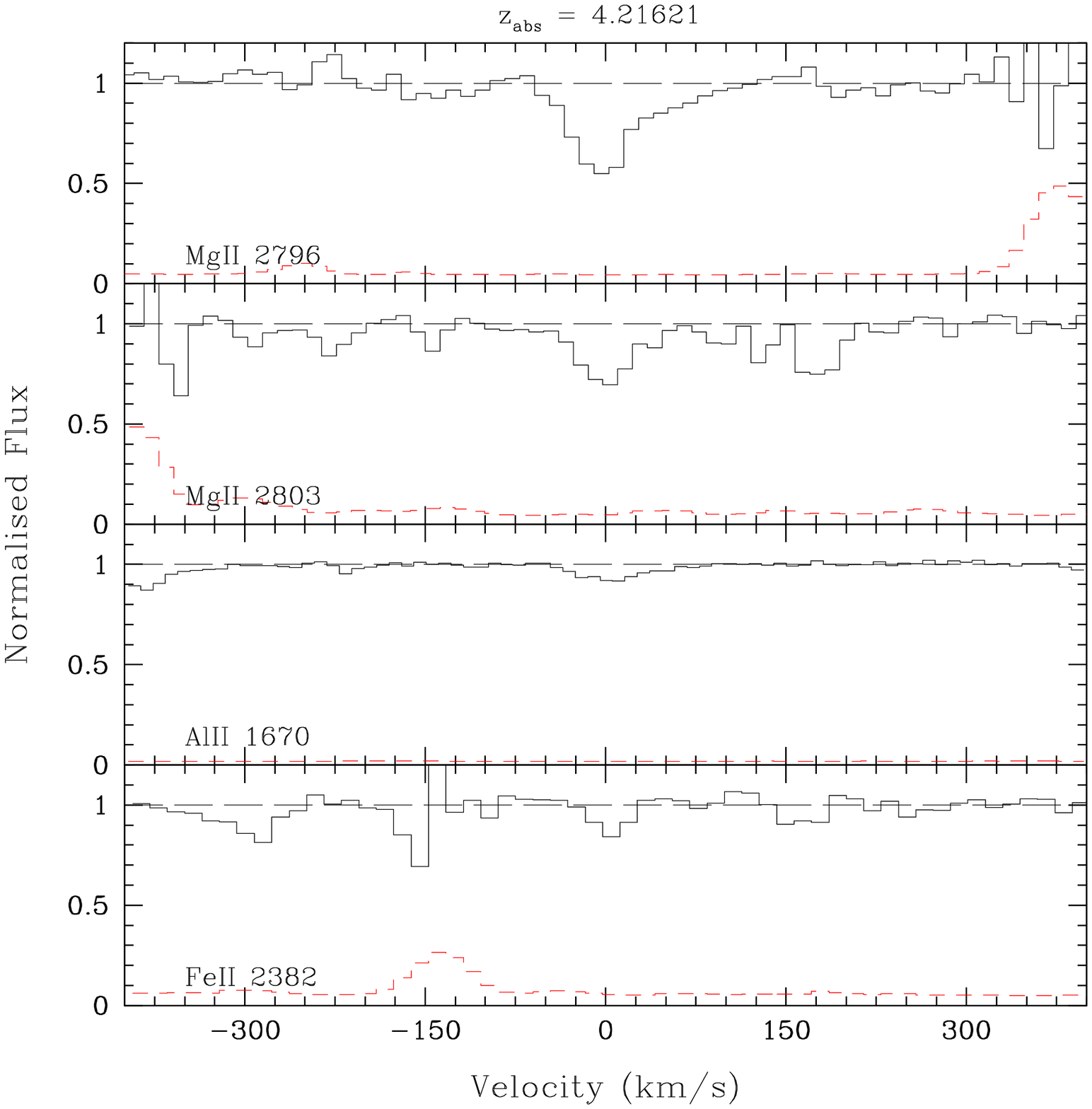}
\includegraphics[width=8cm]{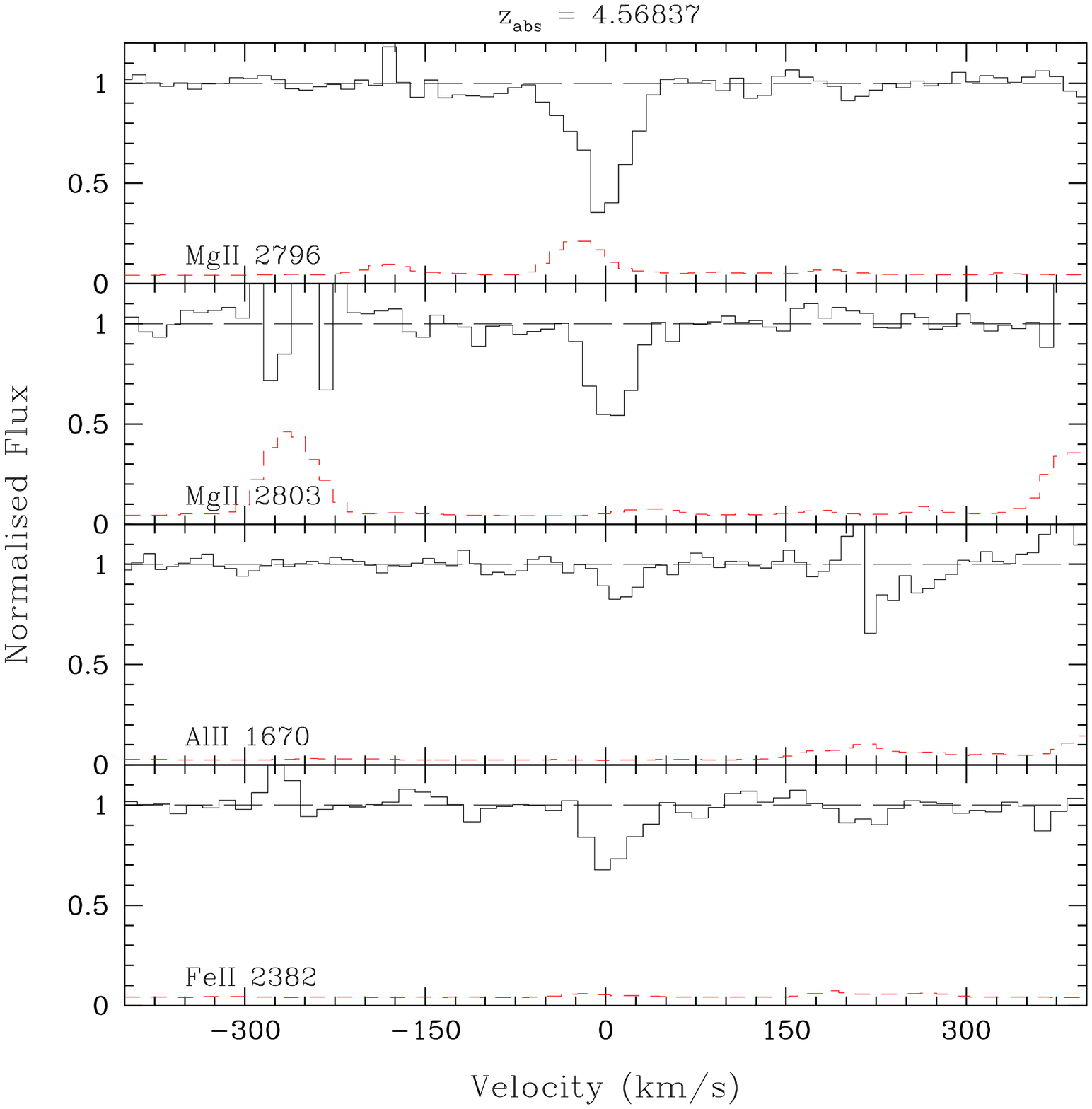}
\caption{\MgII\ systems in the spectrum of ULAS J1319+0950. }   
\label{J1319_mgII}
\end{center}
\end{figure*}

\subsection{CFHQS J1509-1749}

\begin{figure*}
\begin{center}
\includegraphics[width=8cm]{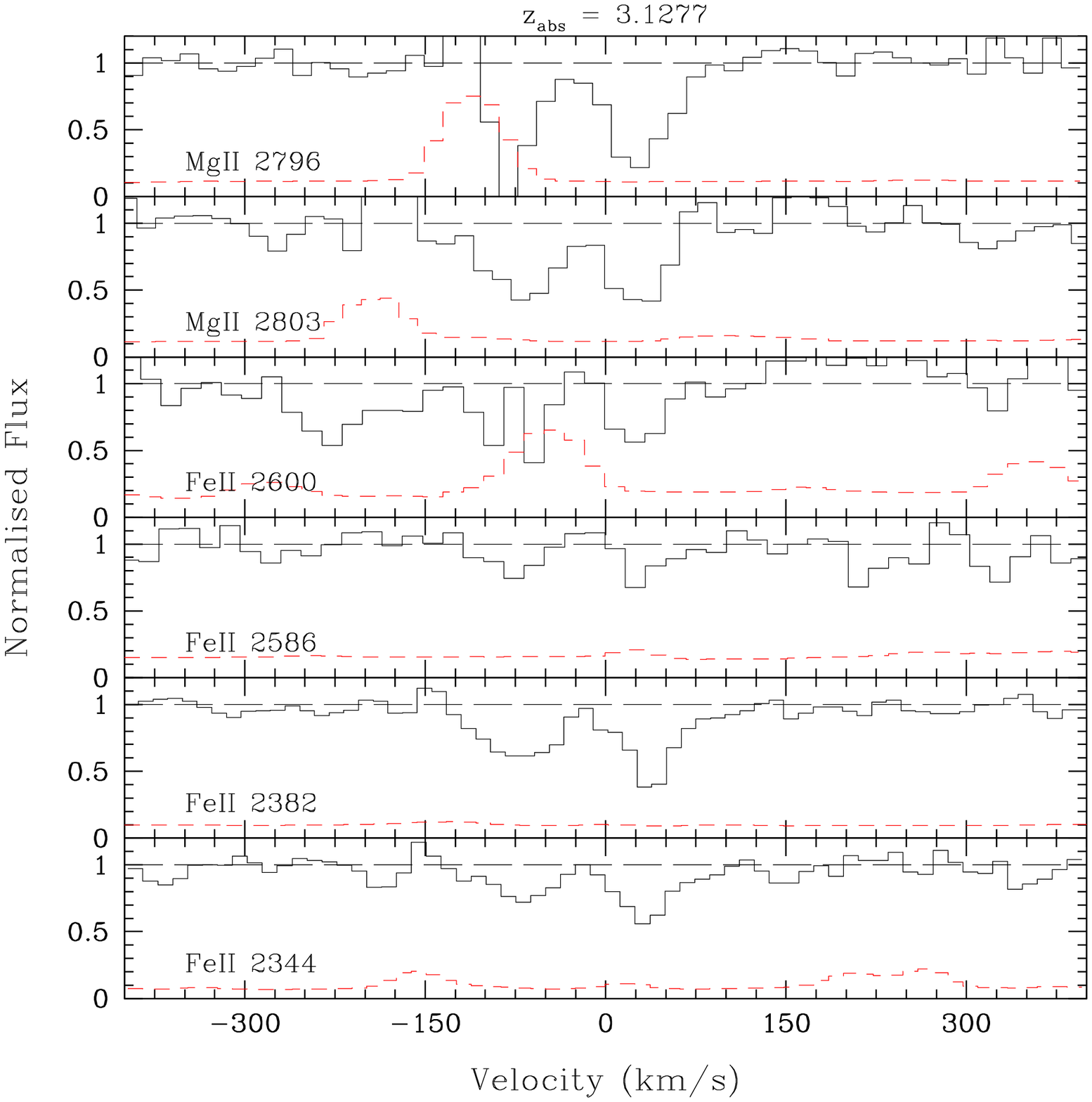}
\includegraphics[width=8cm]{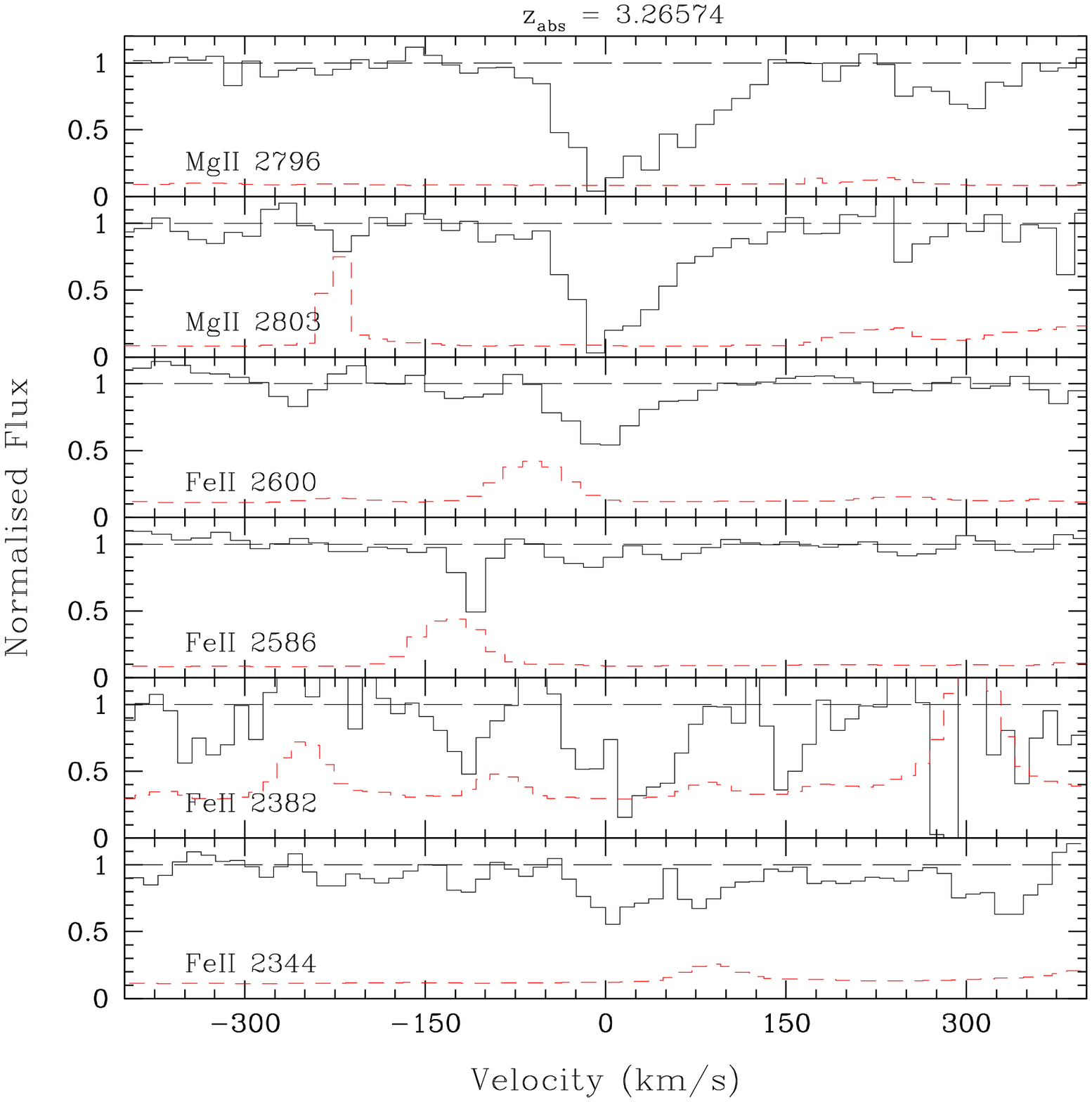}
\includegraphics[width=8cm]{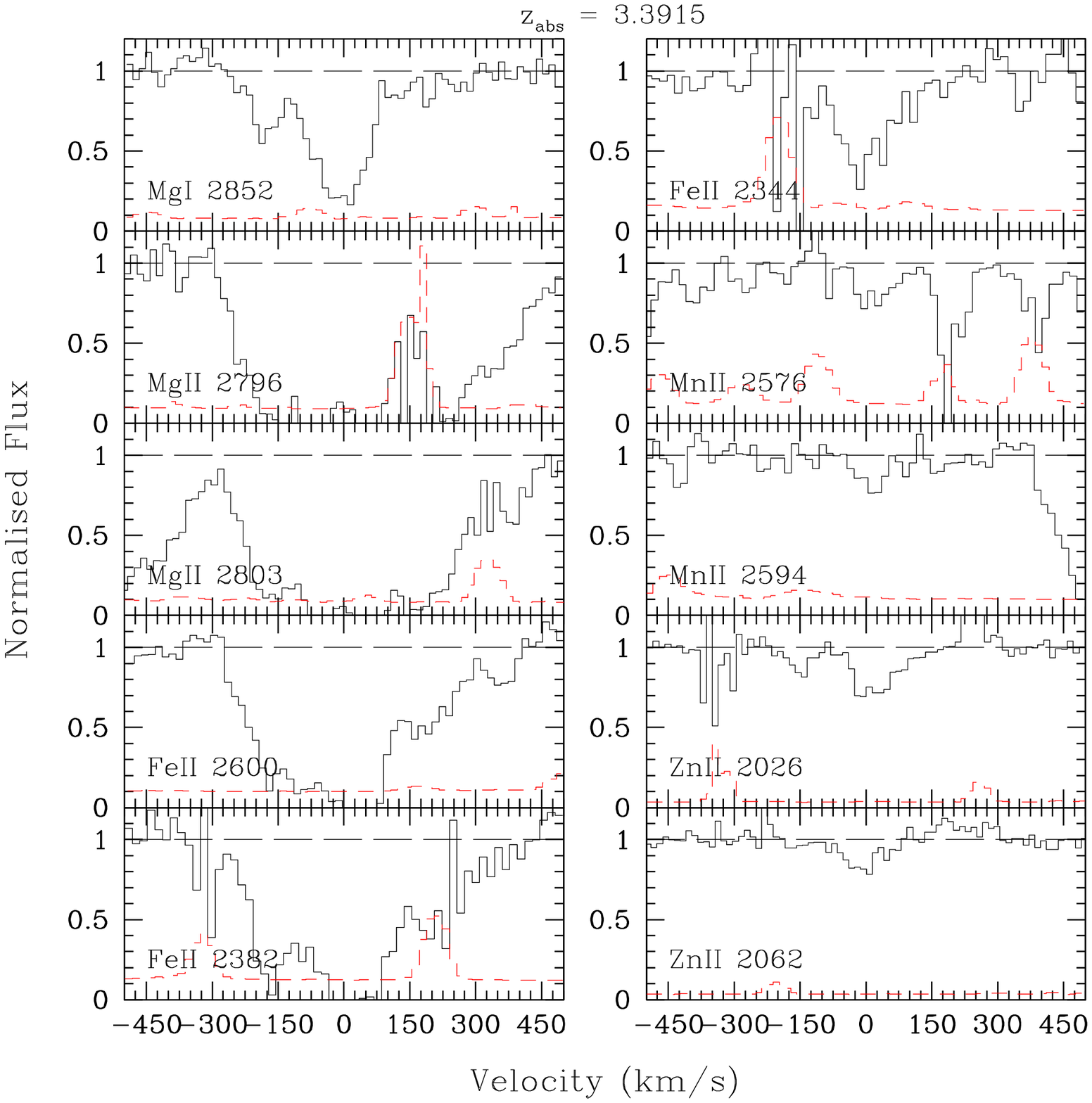}
\includegraphics[width=8cm]{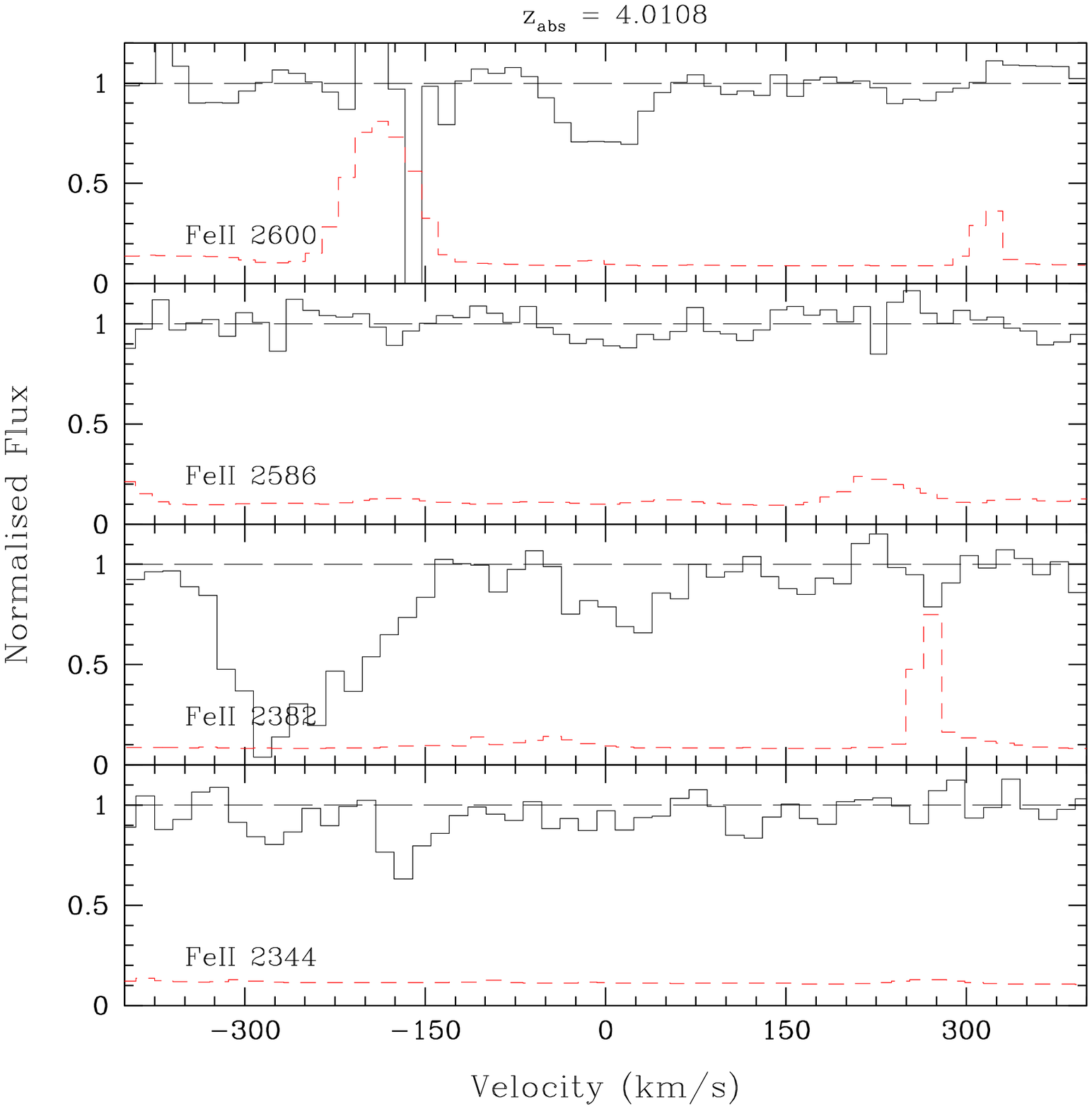}
\caption{\MgII\ systems in the spectrum of CFHQS J1509-1749. }   
\label{J1509_mgII}
\end{center}
\end{figure*}

\bsp

\label{lastpage}

\end{document}